\documentclass[
aps,
prd,
amsmath,amssymb,
superscriptaddress,
onecolumn,
nofootinbib,
floatfix,
noeprint,
11pt
]{revtex4-2}

\usepackage[english]{babel}
\usepackage[utf8]{inputenc}
\usepackage[T1]{fontenc}
\usepackage{physics}
\usepackage{mathrsfs}
\usepackage{bbm}
\usepackage{bm}
\usepackage{enumerate}
\usepackage{graphicx}
\usepackage[dvipsnames]{xcolor}
\usepackage{float}
\usepackage{subdepth}
\usepackage{makecell}
\usepackage{fnpct}
\usepackage{booktabs}
\usepackage{multirow}
\usepackage{setspace}
\usepackage[math]{cellspace}
\usepackage{subfigure}

\usepackage[colorlinks,
linkcolor=BrickRed,
citecolor=MidnightBlue,
urlcolor=MidnightBlue,
bookmarks=true,
bookmarksopen=true,
bookmarksnumbered=true,
]{hyperref}

\newcommand{\eref}[1]{(\ref{#1})}
\newcommand{\dob}{\begin{tabular}{c}}
\newcommand{\edob}{\end{tabular} }
\newcommand{\unit}{1\!\!1}
\newcommand{\one}{1\!\!1}
\newcommand{\rA}{\mathrm{A}}
\newcommand{\rB}{\mathrm{B}}
\newcommand{\rC}{\mathrm{C}}
\newcommand{\rD}{\mathrm{D}}
\renewcommand{\Im}{\rm Im}
\renewcommand{\Re}{\rm Re}
\newcommand{\nn}{\nonumber}

\renewcommand{\i}{\mathrm{i}}
\renewcommand{\(}{\left(}
\renewcommand{\)}{\right)}
\newcommand{\id}{\mathbbm{1}}

\newcommand{\sT}{\mathcal{T}}
\newcommand{\sC}{\mathcal{C}}
\newcommand{\sP}{\mathcal{P}}

\newcommand{\sO}{\mathcal{O}}
\newcommand{\sQ}{\mathcal{Q}}
\newcommand{\sU}{\mathcal{U}}

\newcommand{\sL}{\mathcal{L}}

\newcommand{\z}{\zeta}

\newcommand{\be}{\begin{eqnarray}}
\newcommand{\bea}{\begin{eqnarray}}
\newcommand{\eea}{\end{eqnarray}}
\newcommand{\beq}{\begin{equation}}
\newcommand{\ee}{\end{eqnarray}}
\newcommand{\eeq}{\end{equation}}
\renewcommand{\mod}{\mathrm{\,mod\,}}

\newcounter{ls}

\newcounter{amg}

\newcounter{jvc}

\newcounter{yc}

\begin{document}
\title{Toward a classification of PT-symmetric quantum systems: From dissipative dynamics to topology and wormholes}

\author{Antonio M. Garc\'\i a-Garc\'\i a}
\email{amgg@sjtu.edu.cn}
\affiliation{Shanghai Center for Complex Physics,
	School of Physics and Astronomy, Shanghai Jiao Tong
	University, Shanghai 200240, China}

\author{Lucas S\'a}
\email{ld710@cam.ac.uk}
\affiliation{TCM Group, Cavendish Laboratory, University of Cambridge, JJ Thomson Avenue, Cambridge CB3 0HE, UK}
\affiliation{CeFEMA, Instituto Superior T\'ecnico, Universidade de Lisboa, Av.\ Rovisco Pais, 1049-001 Lisboa, Portugal}

\author{Jacobus J. M. Verbaarschot}
\email{jacobus.verbaarschot@stonybrook.edu}

\affiliation{Center for Nuclear Theory and Department of Physics and Astronomy, Stony Brook University, Stony Brook, New York 11794, USA}
\author{Can Yin}
\email{yin_can@sjtu.edu.cn}
\affiliation{Shanghai Center for Complex Physics,
	School of Physics and Astronomy, Shanghai Jiao Tong
	University, Shanghai 200240, China}

\date{\today}

\begin{abstract}
	\vspace{+1em}
	Studies of many-body non-Hermitian parity-time (PT)-symmetric quantum systems are attracting a lot of interest due to their relevance in research areas ranging from quantum optics and continuously monitored dynamics to Euclidean wormholes in quantum gravity and dissipative quantum chaos.
	While a symmetry classification of non-Hermitian systems leads to 38 universality classes, we show that, under certain conditions, PT-symmetric systems are grouped into 24 universality classes. We identify 14 of them in a coupled two-site Sachdev-Ye-Kitaev (SYK) model and confirm the classification by spectral analysis using exact diagonalization techniques.
	Intriguingly, in 4 of these 14 universality classes, AIII$_\nu$, BDI$^\dagger_\nu$, BDI$_{++\nu}$, and CI$_{--\nu}$, we identify a basis in which the SYK Hamiltonian has a block structure in which some blocks are rectangular, with $\nu \in \mathbb{N}$ the difference between the number of rows and columns. We show analytically that this feature leads to the existence of $\nu$ robust purely \emph{real} eigenvalues, whose level statistics follow the predictions of Hermitian random matrix theory for classes A, AI, BDI, and CI, respectively. We have recently found that this $\nu$ is a topological invariant, so these classes are topological.
	By contrast, nontopological real eigenvalues display a crossover between Hermitian and non-Hermitian level statistics.
	Similarly to the case of Lindbladian dynamics, the reduction of universality classes leads to unexpected results, such as the absence of Kramers degeneracy in a given sector of the theory.
    Another novel feature of the classification scheme is
    that different sectors of the PT-symmetric Hamiltonian may have different symmetries.
\end{abstract}

\maketitle
\newpage

{
\setstretch{1.1}
\hypersetup{linkcolor=black}
\tableofcontents
}

\clearpage

\section{Introduction}
One of the most striking features of quantum chaotic systems is the robust universality of their dynamics.
For sufficiently long times and in the absence of localization effects~\cite{anderson1958},
quantum chaotic systems relax to an ergodic state that depends only on global symmetries.
For Hermitian systems, the number of universality classes labeled by these global symmetries is limited to ten~\cite{altland1997,verbaarschot1994} and depends on the existence or not of time-reversal, particle-hole, and chiral symmetries.
A powerful tool to probe the late stages of the dynamics is the study of level statistics due to the celebrated Bohigas-Giannoni-Schmit (BGS) conjecture~\cite{bohigas1984} that states that spectral correlations of quantum chaotic systems are described by random matrix theory (RMT)~\cite{wigner1951,dyson1962a,dyson1962b,dyson1962c,dyson1962d,dyson1972}. 

The search for RMT fingerprints in the spectrum of single- and many-body quantum systems,
and the identification of these ten universality classes, the so-called tenfold way~\cite{altland1997,verbaarschot1993a,verbaarschot1994}, has been intensively investigated for more than fifty years in completely different fields.
Initial applications of RMT were focused on nuclear physics, where it was employed to model excitations of nuclei~\cite{wigner1951}. Starting from the establishment of the BGS conjecture~\cite{bohigas1984}, research then focused on studies of noninteracting single-particle quantum chaotic~\cite{berry1985,grempel1984} and disordered systems~\cite{anderson1958,altshuler1988,efetov1983}.
Later, in the nineties, these ideas found a fertile ground in quantum chromodynamics (QCD) where it was found that the spectrum of the Euclidean QCD Dirac operator was correlated according to chiral RMT~\cite{verbaarschot1993a}, with the global symmetry depending on the number of colors and the representation of the gauge fields~\cite{verbaarschot1994,verbaarschot1994a}.

In recent years, quantum chaos and RMT have experienced a phenomenal revival of interest due to novel applications in the context of quantum information and quantum gravity~\cite{maldacena2015,kitaev2015}, starting with the proposal
of a bound on quantum chaos, which is characterized by a maximal value of the Lyapunov exponent that controls the exponential growth of certain out-of-time-order correlation functions for short times (of the order of the Eherenfest time).
This bound may be saturated by field theories with a gravity dual, for example, the low-temperature limit of the Sachdev-Ye-Kitaev (SYK) model~\cite{kitaev2015,bohigas1971,french1970,french1971,bohigas1971a,sachdev1993,benet2001,maldacena2016}, consisting of $N$ Majoranas in zero spatial dimension with all-to-all $q$-body random interactions in Fock space. It was quickly realized that the SYK model is also quantum chaotic for late times, with level statistics in agreement with RMT~\cite{garcia2016}, and that, by tuning $q$ and $N$~\cite{you2016,garcia2016,Cotler:2016fpe,altland:2017eao,li2017,kanazawa2017,garcia2018a,beri2019,beri2020,sun2020} and considering supersymmetric~\cite{fu2018} and chiral~\cite{sa2022b} extensions of the model, all ten universality classes could be reproduced.
Moreover, recent results --- showing that the dual field theory of certain near anti-de Sitter (AdS) configurations in two dimensions is a random matrix with broken time reversal invariance ~\cite{saad2019} and that their dynamics is quantum chaotic~\cite{garciagarcia2019} --- indicate that the tenfold way
found in the SYK model could be employed to classify quantum black holes.
The relation of the SYK model with quantum gravity is also not restricted to black-hole configurations. For instance,
a two-dimensional traversable wormhole~\cite{gao2016} is related to the low-temperature limit of a two-site SYK model with a weak intersite coupling~\cite{maldacena2018}. A recent symmetry classification of traversable wormholes in the SYK model~\cite{garcia2023b} has
revealed that, unlike the ten RMT symmetry classes, only six universality classes are allowed.

So far, we have focused our discussion on Hermitian systems. However, in the past few years, there has been a major boost of interest in the dynamics of open and dissipative quantum systems encompassing several fields: quantum optics~\cite{guo2010}, QCD at finite chemical potential~\cite{Stephanov:1996ki,Janik:1996va,Halasz:1997fc,Osborn:2004rf,Osborn:2005ss,kanazawa2021},
quantum information~\cite{kawabata2017,chen2020replica,Luo:2017bno,Jafferis:2022crx}, dissipative quantum chaos~\cite{magan2018,xu2019,campo2020,denisov2018,can2019,can2019a,sa2019,sa2020,xu2020,zhai2020a,li2021a,cornelius2022,garcia2023a,costa2023,weinstein2022,zhang2022a,roubeas2023a,roubeas2023b}, cold atoms~\cite{li2019}, superconductivity~\cite{kawabata2018}, sensitivity enhancement of particle detectors~\cite{wiersig2014}, and different aspects of quantum gravity, from the information paradox~\cite{penington2020,almheiri2020} to wormhole physics~\cite{garcia2021,garcia2022e}.
Not surprisingly, the late-time quantum chaotic dynamics of these systems have been connected to non-Hermitian RMT~\cite{ginibre1965,grobe1988,grobe1989,fyodorov1997,fyodorov2003}, depending on global symmetries~\cite{hamazaki2019,akemann2019,sa2019PRX}. In this case, there are 38 non-Hermitian universality classes~\cite{bernard2002,ueda2019,zhou2019}. A single-site SYK model with complex couplings and its extensions has been shown to reproduce 19 of these classes~\cite{garcia2022d}.
Despite this success, the exact degree of the universality of the quantum dynamics and its precise relation to RMT still remains much less understood than in the Hermitian case~\cite{garcia2023a}.

Contrary to Hermitian systems, not all non-Hermitian systems have sensible thermodynamical properties or a (nonunitary)
evolution still consistent with the postulates of quantum mechanics. As such, the 38-fold classification of non-Hermitian matrices includes classes describing unphysical dynamics. Therefore, it would be desirable to have a symmetry classification of non-Hermitian quantum systems that describes physically relevant situations only. A step in this direction has recently be given by proposing a symmetry classification of Liouvillians~\cite{sa2022a,kawabata2022a} that describes quantum many-body systems coupled to a Markovian environment (see also Refs.~\cite{lieu2020PRL,kawasaki2022PRB,altland2021PRX}).
This is not, however, the only non-Hermitian quantum system consistent with the standard postulates of quantum mechanics. It is believed~\cite{bender1998,bender1999} that a minimal condition for nonpathological quantum dynamics and thermodynamics, despite the non-Hermiticity of the Hamiltonian, is the existence of parity-time (PT) symmetry. More specifically, if the spectrum is purely real, such that PT symmetry is unbroken, it is believed that the resulting quantum dynamics is unitary, namely, that probability is conserved. In general, this is not the case if the PT symmetry is spontaneously broken, leading to a complex spectrum with complex-conjugation symmetry. However, thermodynamic properties can, in some cases~\cite{garcia2022a}, still be consistent with those of closed systems.
In this paper, we initiate a symmetry classification of these systems. Assuming some restrictions on the structure of the Hamiltonian, we propose a classification scheme resulting in 24 universality classes.

Our main goal is to explicitly implement this symmetry classification in a simple non-Hermitian, but PT-symmetric, two-site SYK model with complex couplings and a weak intersite interaction. This model was recently introduced in the study of transitions between Euclidean and traversable wormholes~\cite{garcia2022e}.
Interestingly, we identify a total of 14 universality classes by tuning the different parameters of the model.
A subset of them was already investigated in Ref.~\cite{kawabata2022a} in the context of a classification scheme of Lindbladian quantum chaos, so our classification is a generalization of it. Moreover, it also enlarges the results of Ref.~\cite{kawabata2022a}, since we have found that some symmetry classes were missed in Ref.~\cite{kawabata2022a} because it was not noticed that different sectors of the theory (i.e., blocks of the Hamiltonian) may have different symmetries.

Another interesting feature of our results is the observation in certain universality classes of purely real eigenvalues whose number $\nu$ is robust to changes in the intersite couplings and other parameters of the model.
We have shown that the origin of these real modes is traced back to the existence of a basis in which the many-body SYK Hamiltonian has rectangular blocks. We have recently found \cite{garcia2023c} that the difference in size between the number of rows and columns is a topological index, closely related to the index of the
Wilson Dirac operator in QCD \cite{Itoh:1987iy,Gattringer:1997ci,Damgaard:2010cz,Akemann:2010em,Kieburg:2011uf,Akemann:2011kj,Kieburg:2013xta,Kieburg:2015vqa}.
  Intriguingly, the level statistics of these real eigenvalues is different from the one expected~\cite{Shindou2022} for the universality class corresponding to square blocks ($\nu=0$). In the context of RMT,
some of these ensembles with rectangular blocks and purely real eigenvalues were previously studied for QCD at nonzero chemical potential
\cite{Osborn:2004rf,Akemann:2004dr,Kanazawa:2009en}
  and for random matrix models of the
  Wilson Dirac operator~\cite{Damgaard:2010cz,Akemann:2010em,Kieburg:2011uf,Akemann:2011kj,Kieburg:2013xta,Kieburg:2015vqa}.

 Our results are of direct relevance not only in dissipative quantum chaos but also in quantum information, as this SYK setting is also related to the dynamics of systems under continuous monitoring~\cite{jacobs2006,jacobs2014,wiseman2009}.
Quantum gravity is yet another field that can benefit from our classification as different types of wormhole configurations~\cite{maldacena2018,garcia2021,garcia2022e,penington2020} in near (A)dS$_2$ backgrounds~\cite{jackiw1985,teitelboim1983,almheiri2015,maldacena2016a} have been related to the low-temperature limit of the SYK setting that we investigate. Although the aforementioned spectral properties require a comparison with RMT results and therefore implicitly assume dissipative quantum chaotic dynamics, the proposed symmetry classification is by no means restricted to these systems. For Hermitian systems,
a similar symmetry classification for topological insulators and superconductors~\cite{ryu2010}, or systems at the metal-insulator transition~\cite{garcia2000, garcia2002,evers2008}, exists without any reference to quantum chaos or quantum ergodicity requirements. Therefore, we envisage that our classification could also be relevant for problems where the dynamics is not necessarily quantum chaotic.

The remainder of the paper is organized as follows. We establish general results
for fermionic PT-symmetric systems in Sec.~\ref{sec:general_classification} and restrict our attention to the PT-symmetric SYK Hamiltonian in Sec.~\ref{sec:2siteSYK}. Its classification is then performed in Sec.~\ref{sec:classification}, in all regimes of the model. In Sec.~\ref{topo}, we discuss the consequences of the rectangular structure, recently related to topology \cite{garcia2023c}, of the SYK Hamiltonian classified in the previous section. In Sec.~\ref{sec:level_statistics} we study the quantum chaos properties of the model and confirm our symmetry classification through a detailed analysis of level statistics, which allows us to unambiguously distinguish the identified symmetry classes. In Sec.~\ref{sec:examples} we discuss examples of applications of our model, namely to gravitational wormholes in Sec.~\ref{sec:examples_wormholes}, and open quantum systems (coupled to a bath or subject to continuous monitoring) in Sec.~\ref{sec:examples_open}. Finally, we present our conclusions and outlook in Sec.~\ref{sec:conclusion}.

\section{General classification of fermionic PT-symmetric quantum systems}
\label{sec:general_classification}

A Hamiltonian $H$ is PT symmetric if it commutes with an anti-unitary operator $\sT_+$ that squares to $\sT_+^2=+1$~\cite{bender2002JPhysA}. This symmetry allows us~\cite{benderbook} to see the system as composed of two copies, left (L) and right (R), where the R copy is time-reversed (complex-conjugated) with respect to the L copy, and assumes a left-right symmetric interaction, in such a way that a gain (loss) of probability in L is exactly balanced by a loss (gain) in R. The existence of this anti-unitary symmetry ensures that the eigenvalues of $H$ are either real or come in complex-conjugated pairs. The former case allows us to define meaningful quantum dynamics despite $H$ being non-Hermitian. As mentioned earlier, for complex-conjugated pairs, probability is not in general conserved with the standard definition of scalar product in quantum mechanics, but thermodynamic properties may still resemble in many cases those of closed systems.
In this paper, we consider PT-symmetric systems composed of $2N$ Majorana fermions. We first derive the general form of a Majorana PT-symmetric non-Hermitian Hamiltonian and, then discuss its classification.

\subsection{Majorana fermions and their symmetries}
\label{sec:Majorana}

With the above picture in mind, we partition the $2N$ Majoranas into two sets, one with $N$ left Majoranas $\psi^L_i$ and one with $N$ right Majoranas $\psi^R_i$, where, for concreteness, we assume $N$ to be even.\footnote{A straightforward, albeit tedious, extension of the contents of this paper to odd $N$ is possible. Since no new interesting results arise in this case, we will not consider it any further.} They satisfy the anticommutation relation $\{ \psi_{i}^A, \psi_{j}^B \} =\delta_{AB}\delta_{ij}$ ($i,j=1,\dots, N$ and $A,B=L,R$).
We choose a representation where the left Majoranas are real and symmetric and the right Majoranas are purely imaginary and antisymmetric. The complex-conjugation operator, $K$, which performs the T operation (``time reversal'') in PT symmetry, thus acts on Majoranas as
\begin{align}
\label{eq:action_K_psi}
K \psi_i^L K^{-1} = \psi_i^L,
\qquad
K \psi_i^R K^{-1} = -\psi_i^R.
\end{align}
We further introduce the unitary operator (exponential of the spin),
\begin{equation}
\label{eq:def_Q}
Q=\exp\left\{-\frac{\pi}{4}\sum_{i=1}^N\psi_i^L\psi_i^R\right\}
=\prod_{i=1}^N\frac{1}{\sqrt{2}}\(1-2\psi_i^L\psi_i^R\),
\end{equation}
which exchanges the left and right copies up to a sign,
\be
\label{eq:action_Q_psi}
Q\psi_i^L Q^{-1} = \psi_i^R, \qquad Q\psi_i^R Q^{-1} = -\psi_i^L,
\ee
and, therefore, acts as the P operation (``parity'') in PT symmetry.
Their joint action, implemented by the anti-unitary PT symmetry $\sT_+=QK$, is, therefore, an exchange symmetry of left and right Majoranas,
\be
\label{eq:PTsymm_psi}
\sT_+\psi_i^L \sT_+^{-1} = \psi_i^R, \qquad \sT_+\psi_i^R \sT_+^{-1} = \psi_i^L.
\ee
Moreover, using Eqs.~(\ref{eq:action_K_psi}) and (\ref{eq:def_Q}), we find $\sT_+^2=+1$, see also Eq.~(\ref{eq:comm_K_S_Q}) below, as required~\cite{bender2002JPhysA}. While the PT symmetry is implemented by the anti-unitary combination $QK$, the unitary operator $Q$ itself will play a pivotal role in our classification.

We also note that all Hamiltonians $H$ must be even in Majoranas (bosonic), meaning that they commute with the total parity
\begin{equation}
\label{eq:def_S}
S=S_LS_R,
\end{equation}
where
\begin{equation}
\label{eq:def_SLR}
S_L=\i^{N(N-1)/2}\prod_{i=1}^N\sqrt{2}\psi_i^L
\qquad \text{and}
\qquad
S_R=\i^{N(N+1)/2}\prod_{i=1}^N\sqrt{2}\i\psi_i^R
\end{equation}
are the left and right parities. They act on Majorana operators as
\begin{align}
\label{eq:action_SL_psi}
&S_L \psi_i^L S_L^{-1} = -\psi_i^L,
&&S_L \psi_i^R S_L^{-1} = \psi_i^R,
\\
\label{eq:action_SR_psi}
&S_R \psi_i^L S_R^{-1} = \psi_i^L,
&&S_R \psi_i^R S_R^{-1} = -\psi_i^R,
\\
\label{eq:action_S_psi}
&S \psi_i^L S^{-1} = -\psi_i^L,
&&S \psi_i^R S^{-1} = -\psi_i^R.
\end{align}
Some specific Hamiltonians may also conserve $S_L$ and $S_R$ individually (we will construct such examples below).

The unitary operators $S$, $S_L$, $S_R$, and $Q$ and the complex conjugation operator $K$ play a central role in the symmetry classification we develop. We collect here their properties, for later use.
As mentioned above, we choose a representation where the left Majoranas are real and symmetric and the right Majoranas are purely imaginary and antisymmetric. With this representation, the unitary operators $S_{L,R}$ are antidiagonal, $Q$ and $S$ are diagonal, and they satisfy the reality conditions:
\begin{align}
 \label{eq:comm_K_SLSR}
&K S_L= (-1)^{N/2}S_L K,
&&K S_R =(-1)^{N/2}S_R K,
\\
\label{eq:comm_K_S_Q}
&K S = S K,
&&K Q=Q^{-1} K=S QK.
\end{align}
Moreover, they square to
\begin{align}
\label{eq:square_unitaries}
S_L^2=S_R^2=S^2=+1,
\qquad
Q^2=S,
\end{align}
and, hence, have eigenvalues $s_{L,R}=\pm1$, $s=\pm1$ and $k=\pm1,\pm\i$, respectively.
They satisfy the commutation relations
\begin{align}
\label{eq:comm_SLSR}
&S_L S_R = S_R S_L,
\\
\label{eq:comm_QSLR}
&Q S_{L,R} = S_{L,R} Q^{-1}= S S_{L,R} Q,
\\
\label{eq:comm_QS}
&QS=SQ.
\end{align}

\subsection{PT-symmetric Majorana Hamiltonians}
We now write down the most general PT-symmetric Majorana Hamiltonian.
A basis of operators on the L$+$R space of Majoranas is $\{\Gamma_I^L\Gamma^R_{J}\}_{I,J}$, where $I,J=0,1,\dots,2^{N}-1$, and $\Gamma_I$ are all possible products of Majorana fermions, i.e., $\Gamma_0=\id$, $\Gamma_1=\psi_1,\ \dots,\ \Gamma_{N}=\psi_N$, $\Gamma_{N+1}=\psi_1\psi_2,\ \dots,\  \Gamma_{2^{N}-1}=\psi_1\cdots \psi_N$. Note that a
conserved parity restricts the number of independent basis operators.
Denoting the number of Majoranas in $\Gamma^{L,R}_I$ by $|I|$, we have that
\begin{equation}
\label{eq:Gamma_comm}
\Gamma^L_I \Gamma^R_J = (-1)^{|I||J|}\Gamma^R_J \Gamma^L_I.
\end{equation}
In this basis, we can write down the most general non-Hermitian Hamiltonian as
\begin{equation}
\label{eq:general_fermionic}
H=\sum_{I,J=0}^{2^N-1}h_{IJ} \Gamma^L_I \Gamma^R_{J},
\end{equation}
where $h_{IJ}$ are arbitrary complex coefficients. The even fermionic parity of $H$ ($[H,S]=0$) implies that $|I|$ and $|J|$ must have the same parity in each term of the sum.

Under the PT operation, we have that
\begin{equation}
\sT_+ \Gamma^L_I \sT_+^{-1}=\Gamma^R_I,
\qquad
\sT_+ \Gamma^R_I \sT_+^{-1}=\Gamma^L_I,
\end{equation}
and, hence, $H$ transforms as
\begin{equation}
\begin{split}
\sT_+ H\sT_+^{-1}
&=\sum_{I,J} h_{IJ}^* \Gamma^R_I \Gamma^L_{J}
\\
&=\sum_{I,J} h_{IJ}^* (-1)^{|I||J|}\Gamma^L_{J} \Gamma^R_I
\\
&=\sum_{I,J} h_{JI}^* (-1)^{|I|}\Gamma^L_I \Gamma^R_{J},
\end{split}
\end{equation}
where in the last line we used that $|I|$ and $|J|$ have the same parity.
To ensure that $H$ is PT-symmetric, $\sT_+ H\sT_+^{-1}=H$, we must, therefore, only require that
\begin{equation}
h_{IJ}=(-1)^{|I|}h_{JI}^*,
\end{equation}
or, equivalently, that the $2^{N}$-dimensional matrix $\tilde{h}$ defined by $\tilde{h}_{IJ}=(-i)^{|I|}h_{IJ}$ is Hermitian.

To conclude this discussion, we show that we can write a PT-symmetric Hamiltonian in a canonical form, reminiscent of general Hermiticity-preserving quantum master equations~\cite{hall2014} governing the dynamics of certain open quantum systems.
This relation will be used in Sec.~\ref{sec:examples}.
Because $\tilde{h}$ is Hermitian, it can be diagonalized as $\tilde{h}=U D U^\dagger$, where $U$ is unitary and $D$ real diagonal. It is convenient to separate the $I=0$ and $J=0$ terms from the remaining ones: The $I=J=0$ term is just a constant (proportional to the identity operator) and can be dropped; the terms with $I\neq0$, $J=0$ ($|I|$ must then be even) and $I=0$, $J\neq0$ (even $|J|$) give two uncoupled complex-conjugated single-site Hamiltonians $H_L$ and $H_R$; and the terms with $I\neq 0$ and $J\neq0$ describe the coupling between the two sites, $H_I$. We arrive at a canonical form:
\begin{align}
\label{eq:H_general_PT}
&H=H_L+H_R+H_I,
\\
&H_L=\sum_{I=1}^{2^N-1}h_{I0}\Gamma_L^I,
\qquad
H_R=\sum_{I=1}^{2^N-1}h_{I0}^*\Gamma_R^{I},
\\
\label{eq:H_general_PT_HI}
&H_I=\sum_{K=1}^{2^N-1} D_K O_L^KO_R^K,
\qquad
O_L^K=\sum_{I=1}^{2^N-1} U_{IK}\Gamma_L^I,
\qquad
O_R^K=\sum_{I=1}^{2^N-1} i^{|I|} U_{IK}^*\Gamma_R^{I}.
\end{align}

Having established the general form of a fermionic PT-symmetric Hamiltonian, we now
turn to its symmetry classification. We start by reviewing the classification of arbitrary non-Hermitian operators, then we establish the classification of PT-symmetric models of 0d single-flavor Majorana fermions, and finally we arrive at a
conjecture for the classification of general fermionic PT-symmetric models.

\subsection{Symmetry classification of non-Hermitian Hamiltonians}
\label{sec:non-Hermitian_classes}

The non-Hermitian symmetry classification follows from the behavior of the irreducible blocks of a non-Hermitian Hamiltonian under involutive anti-unitary and unitary symmetries, summarized in Table~\ref{eq:nHsym}. There are four types of anti-unitary symmetries ($\sT_\pm$ and $\sC_\pm$) that can square to either $+1$ or $-1$ and three types of unitary involutions ($\sP$ and $\sQ_\pm$) that always square to $+1$. The fourth column of Table~\ref{eq:nHsym} denotes the pairing of the complex eigenvalues, which follows from considering the secular equation (and whose associated eigenstates are connected by the symmetry), which will prove important in Sec.~\ref{sec:level_statistics}.

More precisely, if there is a unitary $\sU$ that commutes with the Hamiltonian $H$,
\begin{equation}
	\sU H\sU^{-1}=H,
\end{equation}
we can block diagonalize (reduce) $H$ into blocks (sectors) of fixed eigenvalue of $\sU$. Since different sectors are independent, we consider a single one. Inside this block, the unitary symmetry $\sU$ has a fixed eigenvalue and, therefore, acts trivially as the identity (up to a phase). If no further unitary symmetries exist, the block is irreducible.
The symmetry class of \emph{an irreducible block of} $H$ is determined by the anti-unitary symmetries and unitary involutions that either map that block into itself or its adjoint (up to a sign).

\begin{table}[tbp]
	\caption{Anti-unitary and involutive symmetries of non-Hermitian Hamiltonians. The first column gives the transformation relation of the Hamiltonian under the symmetry, the second column, its square, and the third specifies whether it is unitary or anti-unitary. The last column indicates the spectral symmetry implied by each symmetry transformations.}
	\label{eq:nHsym}
	\begin{tabular}{llll}
		$\sT_+ H \sT_+^{-1} = +H$,\hspace*{1cm}
		&$\sT_+^2=\pm 1$, \hspace*{1cm} &
		$\sT_+ \i\, \sT_+^{-1}=-\i$,\hspace*{1cm}
		& $\zeta_k, \zeta_k^*$,
		\\
		$\sT_- H \sT_-^{-1} = -H$ ,
		& $\sT_-^2=\pm 1$,
		& $ \sT_- \i\, \sT_-^{-1}=-\i$,
		&$\zeta_k, -\zeta_k^*$,
		\\
		$\sC_+ H^\dagger \sC_+^{-1} = +H$,
		&$\sC_+^2=\pm 1$,
		&$\sC_+ \i\, \sC_+^{-1}=-\i$,
		&$\zeta_k$,
		\\
		$\sC_- H^\dagger \sC_-^{-1} = -H$,
		&$\sC_-^2=\pm 1$,
		&$\sC_- \i\, \sC_-^{-1}=-\i$,
		&$\zeta_k,-\zeta_k$,
		\\
		$\sP H \sP^{-1} = -H$,
		&$\sP^2=+1$,
		&$\sP \i\, \sP^{-1}=+\i$,
		&$\zeta_k,-\zeta_k$,
		\\
		$\sQ_+ H^\dagger \sQ_+^{-1} = +H$,
		&$\sQ_+^2=+ 1$,
		&$\sQ_+ \i\, \sQ_+^{-1}=+\i$,
		&$\zeta_k, \zeta_k^*$,
		\\
		$\sQ_- H^\dagger \sQ_-^{-1} = -H$,
		&$\sQ_-^2=+ 1$,
		&$\sQ_- \i\, \sQ_-^{-1}=+\i$,
		& $\zeta_k, -\zeta_k^*$.
	\end{tabular}
\end{table}

We look for the existence of anti-unitary operators $\sT_\pm$, such that $H$ satisfies
\begin{alignat}{99}
	\label{eq:nHsym_Tp_new}
	&\sT_+ H \sT_+^{-1} = +H,\qquad
	&&\sT_+^2=\pm 1,
	\\
	\label{eq:nHsym_Tm_new}
	&\sT_- H \sT_-^{-1} = -H,\qquad
	&&\sT_-^2=\pm 1.
\end{alignat}
Since $H$ is non-Hermitian, it can also be related to its adjoint $H^\dagger\neq H$ through anti-unitary operators. To this end, we look for the existence of anti-unitaries $\sC_\pm$ implementing:
\begin{alignat}{99}
	\label{eq:nHsym_Cp_new}
	&\sC_+ H^\dagger \sC_+^{-1} = +H,\qquad
	&&\sC_+^2=\pm 1,
	\\
	\label{eq:nHsym_Cm_new}
	&\sC_- H^\dagger\sC_-^{-1} = -H,\qquad
	&&\sC_-^2=\pm 1.
\end{alignat}
In the presence of a commuting unitary symmetry that block diagonalizes $H$, the transformations in Table~\ref{eq:nHsym} must act within a single block to define a symmetry class. As such, only one transformation of each type can exist, as the composition of two of the same type gives a commuting unitary symmetry of the Hamiltonian, leading to further block diagonalization. Moreover, any three anti-unitary symmetries determine the fourth, and hence, a class has either zero, one, two, or four antinunitary symmetries.

The combined action of anti-unitaries of different types gives rise to unitary involutions: the composition of $\sT_+$ and $\sT_-$ (or $\sC_+$ and $\sC_-$) is a unitary transformation that anticommutes with $H$ (chiral symmetry); the composition of $\sT_+$ and $\sC_+$ (or $\sT_- $ and $\sC_-$) is a unitary similarity transformation between $H$ and $H^\dagger$ (pseudo-Hermiticity); and the composition of $\sT_+$ and $\sC_-$ (or $\sT_-$ and $\sC_+$) unitarily maps $H$ to $-H^\dagger$ (antipseudo-Hermiticity).
In the absence of anti-unitary symmetries, these unitary involutions can still act on their own and we look for unitary operators $\sP$ and $\sQ_\pm$, such that $H$ transforms as
\begin{alignat}{99}
	\label{eq:nHsym_P_new}
	&\sP H \sP^{-1} = -H,\qquad
	&&\sP^2=1,
	\\
	\label{eq:nHsym_Qp_new}
	&\sQ_+ H^\dagger \sQ_+^{-1} =  +H,\qquad
	&&\sQ_+^2=1,
	\\
	\label{eq:nHsym_Qm_new}
	&\sQ_- H^\dagger \sQ_-^{-1} = - H,\qquad
	&&\sQ_-^2=1.
\end{alignat}
Similarly to before, any two unitary involutions determine the third, and, hence, we have classes with either zero, one, or three unitary involutions.

In principle, besides the square of its symmetries, the labeling of a symmetry class can depend on their commutation relations. In the case of anti-unitary symmetries, it is possible to show~\cite{sa2023thesis} that this information is already contained in their squares. Indeed, we can always choose one of the four anti-unitary symmetries to commute with the other three, while the remaining commutation relations are fixed by the squares of the symmetries.
Finally, in the absence of anti-unitary symmetries, we consider unitary involutions, in which case one commutation relation is an independent label. Taking the independent unitary involutions to be $\sP$ and $\sQ_+$, we find that they either commute or anticommute~\cite{sa2023thesis}.

\begin{table}[t!]
	\centering
	\caption{Classification of non-Hermitian Hamiltonians by anti-unitary symmetries  $\sT_\pm$ and $\sC_\pm$. The columns of the table give the square of each symmetry operator and the name of the class introduced in Ref.~\cite{ueda2019}.
    The subscript $\pm$ of the classes is the product $\sC_+^2\sC_-^2$ and/or $\sT_+^2\sT_-^2$. The value of
          these products can also be related to
          commutation relations of the $\sP$ symmetry and the anti-unitary symmetries,
          see Ref.~\cite{ueda2019}.  In case the class does not have a standard name,
          we use the name under
          $H\sim \i H$ (see Table~\ref{tab:correlations_classes_equiv}). Other classes have two names, both of which we have included for completeness
          (see Ref.~\cite{ueda2019}).}
	\label{tab:correlations_classes_anti}
	\begingroup
	\renewcommand*{\arraystretch}{1.0375}
	\begin{tabular}[t]{@{}Sl Sc Sc Sc Sc Sl@{}}
		\toprule
		\#  & $\sT_+^2$ & $\sC_-^2$ & $\sC_+^2$ & $\sT_-^2$ & Class                            \\ \midrule
		1  & ---        & ---        & ---        & ---         & A                                  \\
		2  & $+1$       & ---        & ---        & ---         & AI                                \\
		3  & $-1$       & ---        & ---        & ---         & AII                               \\
		4  & ---        & $+1$       & ---        & ---         & D                                 \\
		5  & ---        & $-1$       & ---        & ---         & C                                 \\
		6  & ---        & ---        & $+1$       & ---         & AI$^\dagger$                      \\
		7  & ---        & ---        & $-1$       & ---         & AII$^\dagger$                     \\
		8  & ---        & ---        & ---        & $+1$        & D$^\dagger$                       \\
		9  & ---        & ---        & ---        & $-1$        & C$^\dagger$                       \\
		10 & $+1$       & $+1$       & ---        & ---         & BDI                               \\
		11 & $+1$       & $-1$       & ---        & ---         & CI                                \\
		12 & $-1$       & $+1$       & ---        & ---         & DIII                              \\
		13 & $-1$       & $-1$       & ---        & ---         & CII                               \\
		14 & $+1$       & ---        & $+1$       & ---         & no name ($\equiv$ BDI$^\dagger$)  \\
		15 & $+1$       & ---        & $-1$       & ---         & no name ($\equiv$ DIII$^\dagger$) \\
		16 & $-1$       & ---        & $+1$       & ---         & no name ($\equiv$ CI$^\dagger$)   \\
		17 & $-1$       & ---        & $-1$       & ---         & no name ($\equiv$ CII$^\dagger$)  \\
		18 & $+1$       & ---        & ---        & $+1$        & AI$_+$ / D$^\dagger_+$            \\
		19 & $+1$       & ---        & ---        & $-1$        & AI$_-$ / C$^\dagger_-$            \\
		20 & $-1$       & ---        & ---        & $+1$        & AII$_-$ / D$^\dagger_-$           \\
		21 & $-1$       & ---        & ---        & $-1$        & AII$_+$ / C$^\dagger_+$           \\
		22 & ---        & $+1$       & $+1$       & ---         & D$_+$ / AI$^\dagger_+$            \\
		23 & ---        & $+1$       & $-1$       & ---         & D$_-$ / AII$^\dagger_-$           \\
		24 & ---        & $-1$       & $+1$       & ---         & C$_-$ / AI$^\dagger_-$            \\
		25 & ---        & $-1$       & $-1$       & ---         & C$_+$ / AII$^\dagger_+$           \\ \bottomrule
	\end{tabular}
	\endgroup
	\hspace{+2ex}
	\begin{tabular}[t]{@{}Sl Sc Sc Sc Sc Sl@{}}
		\toprule
		\#  & $\sT_+^2$ & $\sC_-^2$ & $\sC_+^2$ & $\sT_-^2$ & Class                             \\ \midrule
		26 & ---        & $+1$       & ---        & $+1$        & no name ($\equiv$ BDI)            \\
		27 & ---        & $+1$       & ---        & $-1$        & no name ($\equiv$ DIII)           \\
		28 & ---        & $-1$       & ---        & $+1$        & no name ($\equiv$ CI)             \\
		29 & ---        & $-1$       & ---        & $-1$        & no name ($\equiv$ CII)            \\
		30 & ---        & ---        & $+1$       & $+1$        & BDI$^\dagger$                     \\
		31 & ---        & ---        & $+1$       & $-1$        & CI$^\dagger$                      \\
		32 & ---        & ---        & $-1$       & $+1$        & DIII$^\dagger$                    \\
		33 & ---        & ---        & $-1$       & $-1$        & CII$^\dagger$                      \\
		34 & $+1$       & $+1$       & $+1$       & $+1$        & BDI$_{++}$ / BDI$^\dagger_{++}$   \\
		35 & $+1$       & $+1$       & $+1$       & $-1$        & BDI$_{-+}$ / CI$^\dagger_{+-}$   \\
		36 & $+1$       & $+1$       & $-1$       & $+1$        & BDI$_{+-}$ / DIII$^\dagger_{-+}$  \\
		37 & $+1$       & $+1$       & $-1$       & $-1$        & BDI$_{--}$ / CII$^\dagger_{--}$   \\
		38 & $+1$       & $-1$       & $+1$       & $+1$        & CI$_{+-}$ / BDI$^\dagger_{-+}$    \\
		39 & $+1$       & $-1$       & $+1$       & $-1$        & CI$_{--}$ / CI$^\dagger_{--}$     \\
		40 & $+1$       & $-1$       & $-1$       & $+1$        & CI$_{++}$ / DIII$^\dagger_{++}$   \\
		41 & $+1$       & $-1$       & $-1$       & $-1$        & CI$_{-+}$ / CII$^\dagger_{+-}$    \\
		42 & $-1$       & $+1$       & $+1$       & $+1$        & DIII$_{-+}$ / BDI$^\dagger_{+-}$  \\
		43 & $-1$       & $+1$       & $+1$       & $-1$        & DIII$_{++}$ / CI$^\dagger_{++}$   \\
		44 & $-1$       & $+1$       & $-1$       & $+1$        & DIII$_{--}$ / DIII$^\dagger_{--}$ \\
		45 & $-1$       & $+1$       & $-1$       & $-1$        & DIII$_{+-}$ / CII$^\dagger_{-+}$  \\
		46 & $-1$       & $-1$       & $+1$       & $+1$        & CII$_{--}$ / BDI$^\dagger_{--}$   \\
		47 & $-1$       & $-1$       & $+1$       & $-1$        & CII$_{+-}$ / CI$^\dagger_{-+}$    \\
		48 & $-1$       & $-1$       & $-1$       & $+1$        & CII$_{-+}$ / DIII$^\dagger_{+-}$  \\
		49 & $-1$       & $-1$       & $-1$       & $-1$        & CII$_{++}$ / CII$^\dagger_{++}$   \\ \bottomrule
	\end{tabular}
	\vspace{+10pt}
	\caption{The classification of non-Hermitian Hamiltonians without anti-unitary symmetries by the unitary involutions $\sP$ and $\sQ_\pm$.
        We list the square of each of them as well as the value of $\epsilon_{\sP\!\sQ_+}$ in
        $\sP\sQ_+ -  \epsilon_{\sP\!\sQ_+}\sQ_+ \sP =0$, and
        the name of the class adopted in Ref.~\cite{ueda2019}. Class 52 has no name, but it is equivalent
        to AIII under multiplication by $\i$
(see Table~\ref{tab:correlations_classes_equiv}).}
	\label{tab:correlations_classes_noanti}
	\vspace{+8pt}
	\begin{tabular}{@{}Sl Sc Sc Sc Sc Sl@{}}
		\toprule
		\#  & $\sP^2$ & $\sQ_+^2$ & $\sQ_-^2$ & $\epsilon_{\sP\!\sQ_+}$ & Class                   \\ \midrule
		50  & $+1$     & ---        & ---        & ---                       & AIII$^\dagger$          \\
		51  & ---      & $+1$       & ---        & ---                       & AIII                    \\
		52  & ---      & ---        & $+1$       & ---                       & no name ($\equiv$ AIII) \\
		53  & $+1$     & $+1$       & $+1$       & $+1$                      & AIII$_+$                \\
		54  & $+1$     & $+1$       & $+1$       & $-1$                      & AIII$_-$                \\ \bottomrule
	\end{tabular}
\end{table}

The non-Hermitian classes are determined by  the anti-unitary symmetries and unitary involutions.
Following closely the discussion in~\cite{sa2023thesis},
we have the following possibilities:
\begin{itemize}
	\item \textit{No symmetries.} One class.
	\item \textit{One anti-unitary symmetry.} The four possible anti-unitary symmetries are
$\sT_\pm$ and $\sC_\pm$, each squaring to $\pm 1$.
This gives  $4\times2=8$ classes.
\item \textit{Two anti-unitary symmetries.}
  The four anti-unitary symmetry operators  give 6 combinations with
  each of the anti-unitary operators  squaring to $\pm 1$. This results in $6\times2^2=24$ classes.
\item \textit{Three anti-unitary symmetries.} This is not possible because the product of three anti-unitary
  symmetries is another anti-unitary symmetry, resulting in four anti-unitary symmetries.
\item \textit{Four anti-unitary symmetries.} This gives only one choice for the operators, but each of them squares to $\pm1$, adding $2^4=16$ classes.
\item \textit{One unitary involution.} Each of the $\sP$ and $\sQ_{\pm}$ can be the unitary involution. They
  can only square to the identity resulting in three classes.
\item \textit{Two unitary involutions.} The combination of two different unitary involutions gives the third one, so that this case is not possible.
\item \textit{Three unitary involutions.}
  Because $\sP \sQ_+ \sQ_- = \alpha \one$ with $|\alpha|=1 $, using that the operators
  square to the identity, we can derive $(\sP \sQ_+)^2 =\alpha^2 \one$. This gives $\sP = \alpha^2\sQ_+ \sP \sQ_+$. Since $ \sP^2=\sQ_+^2=1$, it follows that $\alpha^2 = \pm 1$, and $\sP$ and $\sQ_+$ either commute or anticommute. The same argument can
  be applied to  $\sP$ and $\sQ_-$, and to  $\sQ_+$ and $\sQ_-$. From
  $\sP \sQ_+ \sQ_- = \alpha \one$ it also follows that the three pairs have to either all commute or
  all anticommute. Therefore, we can have only two classes with three unitary involutions.
\end{itemize}
As was already observed in Refs.~\cite{liu2019PRB,ashida2020}, the total number of classes adds up to $1+8+24+16+3+2=54$.
Tables~\ref{tab:correlations_classes_anti} and \ref{tab:correlations_classes_noanti} show the classes with
and without anti-unitary symmetry (note that our notation is slightly different from the nomenclature in~\cite{ueda2019}). Other equivalent possibilities to label the classes have appeared in the literature,
see Refs.~\cite{bernard2002,ueda2019,zhou2019,ashida2020}. More discussion of the classification
can be found in Ref.~\cite{sa2023thesis}.

\begin{table}[t]
	\centering
	\caption{This table  shows the 16 classes  classes that are equivalent
		to another class under $H \sim \i H$. Therefore we have 38 inequivalent classes.
		The numbering of the classes is as in Tables~\ref{tab:correlations_classes_anti} and \ref{tab:correlations_classes_noanti}. This and the previous two tables are taken from the thesis~\cite{sa2023thesis}.}
	\vspace{+8pt}
	\label{tab:correlations_classes_equiv}
	\begin{tabular}{@{}ll@{}}
		\toprule
		Equivalence  & Classes                                                                \\ \midrule
		2 $\equiv$ 8   & AI $\equiv$ D$^\dagger$                                                  \\
		3 $\equiv$ 9   & AII $\equiv$ C$^\dagger$                                                 \\
		10 $\equiv$ 26 & BDI                                                                    \\
		11 $\equiv$ 28 & CI                                                                     \\
		12 $\equiv$ 26 & DIII                                                                   \\
		13 $\equiv$ 27 & CII                                                                    \\
		14 $\equiv$ 30 & BDI$^\dagger$                                                          \\
		15 $\equiv$ 32 & DIII$^\dagger$                                                         \\
		16 $\equiv$ 31 & CI$^\dagger$                                                           \\
		17 $\equiv$ 33 & CII$^\dagger$                                                           \\
		19 $\equiv$ 20 & AI$_-$ / C$_-^\dagger$ $\equiv$ AII$_-$ / D$^\dagger_-$                    \\
		35 $\equiv$ 42 & BDI$_{-+}$ / CI$_{+-}^\dagger\equiv$ DIII$_{-+}$ / BDI$^\dagger_{+-}$   \\
		37 $\equiv$ 44 & BDI$_{--}$ / CII$_{--}^\dagger$ $\equiv$ DIII$_{--}$ / DIII$^\dagger_{--}$ \\
		39 $\equiv$ 46 & CI$_{--}$ / CI$_{--}^\dagger$ $\equiv$ CII$_{--}$ / BDI$^\dagger_{--}$     \\
		41 $\equiv$ 48 & CI$_{-+}$ / CII$_{+-}^\dagger$ $\equiv$ CII$_{-+}$ / DIII$^\dagger_{+-}$   \\
		51 $\equiv$ 52 & AIII                                                                   \\ \bottomrule
	\end{tabular}
\end{table}

For our purposes, it proves convenient to identify the matrices $H$ and $\i H$, which only changes the
eigenvalues by $\i$ but does not affect the eigenvectors.  This multiplication interchanges the role of
$\sT_+$ and $\sT_-$ and of $\sQ_+$ and $\sQ_-$ but does not change the action of the other symmetry
operators. As is shown in Table~\ref{tab:correlations_classes_equiv} we find 16 equivalence relations between classes, reducing
the total number of classes to $54-16=38$~\cite{ueda2019}, the Bernard-LeClair (BL) classes.

The previous considerations are completely general. For a Majorana system as the one described in Sec.~\ref{sec:Majorana}, the unitary symmetry operators are given by compositions of the available unitary operators ($Q$, $S$, and $S_{L,R}$), while the anti-unitary symmetries further include a product with the complex conjugation operator $K$. The precise expressions depend on the details of the system, and will be considered in depth in what follows. A natural question is how the 38-fold classification of arbitrary non-Hermitian Hamiltonians is affected by the presence of PT symmetry, which we examine next.

\subsection{Symmetry classification of 0d PT-symmetric Majorana Hamiltonians}
\label{sec:class_0dPT}

As discussed above, PT-symmetric systems have an anti-unitary symmetry $\sT_+=QK$, satisfying $\sT_+^2=+1$, and their eigenvalues come in complex-conjugated pairs. Nevertheless, PT symmetry of the Hamiltonian can be broken, either explicitly (in which case not all its irreducible blocks are PT-symmetric) or spontaneously (in which case some or all eigenstates are not PT-symmetric). Regarding the first, $H$ always admits one or more unitary symmetries, which can, in some cases, break PT symmetry explicitly. That is, while the full Hamiltonian $H$ is PT-symmetric, it is not guaranteed that individual parity sectors, or blocks of the Hamiltonian, are also PT-symmetric, because $\sT_+$ can connect different irreducible blocks of $H$ instead of acting inside a single one. Even if the PT symmetry is not explicitly broken, the eigenstates of $H$ are not necessarily invariant under $\sT_+$. Indeed, if there is at least one complex-conjugated (nonreal) eigenvalue pair, PT symmetry is spontaneously broken.
If, instead, all eigenstates are invariant under $\sT_+$, then all eigenvalues are real and PT symmetry is unbroken. We note that the ground state of the Hamiltonian, corresponding to the steady state of the evolution of the density matrix by means of the Lindblad equation (see Sec.~\ref{sec:examples_open}), never spontaneously breaks PT symmetry. The phase transition between the PT-broken and unbroken phases was studied in Ref.~\cite{garcia2022c} for a four-body non-Hermitian two-site SYK model for which the transition occurs at a large
intersite coupling.

From the preceding discussion, it is clear that the existence of PT symmetry does not preclude or enforce the presence (and the sign of the square) of any anti-unitary symmetry of \emph{an irreducible block} of the Hamiltonian. Nevertheless, we conjecture that classes with $\sT_+^2=-1$ and $\sC_+^2=-1$ do not occur in PT-symmetric systems.
Starting with the 54 classes without the identification $H\sim\i H$~\cite{liu2019PRB,ashida2020}, this restriction excludes 25 classes, leaving 29 allowed ones. Performing the identification $H\sim\i H$ leads to a set of 24 admissible classes out of the 38 of the BL classification. These classes are conventionally named A, AI, D, C, AI$^\dagger$, BDI, CI, DIII, CII, BDI$^\dagger$, CI$^\dagger$, CII$^\dagger$, AI$_+$, AI$_-$, D$_+$, C$_-$, BDI$_{++}$, BDI$_{-+}$, CI$_{+-}$, CI$_{--}$, AIII$^\dagger$, AIII, AIII$_+$, and AIII$_-$, see Tables~\ref{tab:correlations_classes_anti} and \ref{tab:correlations_classes_noanti}. This 24-fold classification constitutes our main statement on the general classification of PT-symmetric systems and we proceed by giving arguments for
its validity.

In the case of 0d single-flavor Majorana fermions (where the Majorana fermions have no additional indices besides $i=1,\dots,N$ and $A=L,R$, in particular, no spatial or flavor indices), and which constitutes the main focus of the present paper, the statement above can be checked explicitly. Indeed, in this case, there are no other symmetry operations available besides $K$, $Q$, $S$, $S_{L,R}$ and combinations of them. Of these, as follows from Eqs.~(\ref{eq:comm_K_SLSR})--(\ref{eq:comm_QS}), the only anti-unitaries that can square to $-1$ are $S_L K$ and $Q S_L K$. However, using the results of Sec.~\ref{sec:2siteSYK}, these operators only act as $\sT_-$ or $\sC_-$ on irreducible blocks of $H$. We, therefore, cannot construct a block with $\sT_+^2=-1$ and $\sC_+^2=-1$. In Sec.~\ref{sec:2siteSYK} we perform this calculation in detail for a paradigmatic two-site SYK model. The more general result follows from combining
several of these SYK models.

\subsection{Toward a symmetry classification of arbitrary PT-symmetric Hamiltonians}
\label{sec:class_general_PT}

The absence of commuting anti-unitary symmetries that square to $-1$ is true for any PT-symmetric systems where all L operators involved in the classification commute with all R operators,\footnote{This includes, obviously, bosonic and spin systems. It includes, moreover, some fermionic systems, e.g., those with intersite couplings quadratic in L and R fermions, since a product of two L fermions commutes with a product of two R fermions.} which can be mapped (see Sec.~\ref{sec:examples_open}) to the Liouvillians of Ref.~\cite{sa2022a}. There, it was shown that $\sC_+^2=-1$ and $\sT_+^2=-1$ symmetries are excluded, as these symmetries are not possible inside a single block of $H$. The argument consists of (i) showing that a $\sC_+^2=-1$ (or $\sT_+^2=-1$) symmetry of the full Hamiltonian can exist; but (ii) because of PT symmetry, it always induces an additional commuting unitary symmetry, with respect to which $H$ has to be reduced; and (iii) the $\sC_+^2=-1$ (or $\sT_+^2=-1$) symmetry always connects different blocks with respect to the induced symmetry. Therefore, the $\sC_+^2=-1$ (or $\sT_+^2=-1$) symmetry of $H$ does not define a symmetry class of the blocks of $H$ with $\sC_+^2=-1$ (or $\sT_+^2=-1$).

Fermionic systems in higher dimensions, nonfully-connected, or with additional flavors (indices) pose greater challenges. While we believe the argument of Ref.~\cite{sa2022a} can be extended to this case, we were not able to do so
because of the large number of different possibilities that have to be checked.
In any case, as far as we know, there is no known example outside the 24-fold classification.

Having discussed the general classification of PT-symmetric quantum systems, in the remainder of the paper we consider a specific paradigmatic example, namely, a non-Hermitian two-site SYK model. By varying its parameters, we find that it realizes 14 of the 24 allowed symmetry classes, which are tabulated in Table~\ref{tab:RMT_realization} in Sec.~\ref{sec:level_statistics}.

\section{Paradigmatic example: the PT-symmetric two-site SYK model}
\label{sec:2siteSYK}

\subsection{Definition of the Hamiltonian}
\label{sec:model}

We restrict the general Hamiltonian Eq.~(\ref{eq:H_general_PT}) to have $q$-body intrasite and $2r$-body intersite interactions ($q$ and $2r$ are even natural numbers). The Hamiltonian reads
\begin{equation}
\label{eq:def_H}
H=H_L(\kappa) +\alpha(-1)^{q/2} H_R(\kappa) +\lambda H_I,
\end{equation}
with non-Hermiticity parameter $0\leq\kappa\leq1$, PT-symmetry-breaking parameter $\alpha$, and coupling constant $\lambda$. The two single-site $q$-body SYK Hamiltonians $H_{L,R}$ and the $2r$-body Hamiltonian $H_I$ coupling them
are given by
\begin{equation}
\begin{split}\label{eq:hami}
&H_L=-\i^{q/2} \sum_{i_1<\cdots<i_q}^{N}
\(J_{i_1\cdots i_q}+iM_{i_1\cdots i_q}\) \, \psi^L_{i_1} \cdots \, \psi^L_{i_q},
\\
&H_R=-\i^{q/2}\sum_{i_1<\cdots<i_q}^{N}
\(J_{i_1\cdots i_q}-iM_{i_1\cdots i_q}\) \, \psi^R_{i_1} \cdots \, \psi^R_{i_q},
\\
&H_I=\i^r \frac{N^{1-r}}{r} \sum_{i_1<\cdots<i_r}^{N}\psi^L_{i_1}\cdots \psi^L_{i_r}\psi^R_{i_1}\cdots \psi^R_{i_r},
\end{split}
\end{equation}
where the couplings $J_{i_1\cdots i_q}$ and $M_{i_1\cdots i_q}$ are real Gaussian random variables with zero mean and variance
\begin{equation}
\left\langle J^2_{i_1\cdots i_q}\right\rangle=
(1-\kappa)\frac{{2^{q-1}(q-1)!}}{q N^{q-1}}
\qquad \text{and} \qquad
\left\langle M^2_{i_1\cdots i_q}\right\rangle=
\kappa\,\frac{{2^{q-1}(q-1)!}}{q N^{q-1}}.
\end{equation}

For $\alpha=1$, the Hamiltonian Eq.~(\ref{eq:def_H}) is, by construction, PT-symmetric, with the PT symmetry implemented by the anti-unitary $\sT_+=QK$. In our study, we will also address the effect of weakly explicitly broken PT symmetry, controlled by a parameter $\alpha\approx 1$.
There are two special values of $\kappa$, for which $H$ has an additional symmetry. For $\kappa=0$~\cite{garcia2023b}, the uncoupled two-site Hamiltonian,
\begin{equation}\label{eq:h0}
	H_0=H_L +\alpha(-1)^{q/2} H_R,
\end{equation}
is Hermitian, and since the coupling Hamiltonian $H_I$ is always Hermitian, the full Hamiltonian $H$ is also Hermitian. $H$ can then be taken as a standard Hamiltonian of an isolated two-site quantum system. The PT symmetry reduces to an ordinary time-reversal symmetry that allows to establish a duality to a traversable wormhole in the low-temperature limit~\cite{maldacena2018,garcia2023b}, see Sec.~\ref{sec:examples_wormholes} for a more detailed discussion. For $\kappa=1$, $H_0$ is instead anti-Hermitian. If we interpret the L and R copies as acting, respectively, on the ket and bra of a density matrix (i.e., $H$ acting in the so-called vectorized Liouville space), then $H$ can be mapped~\cite{garcia2022} into a quantum Markovian generator of Lindblad type~\cite{belavin1969,gorini1976,lindblad1976}, see Sec.~\ref{sec:examples_open} for more details. For these reasons, in what follows we will dub the three regimes $\kappa=0$, $0<\kappa<1$, and $\kappa=1$ as Hermitian, general PT-symmetric, and Lindbladian, respectively.
In these three regimes, the symmetry classification proceeds differently. In the Hermitian case ($\kappa=0$), $H$ is Hermitian and, in Ref.~\cite{garcia2023b}, we found that it belongs to six out of the ten AZ classes, depending on the choice of parameters. When $\kappa\neq0$, the Hamiltonian is non-Hermitian and belongs to a non-Hermitian BL class. For the Lindbladian case ($\kappa=1$), the set of classes realized by $H$ is enlarged compared to the general non-Hermitian situation ($0<\kappa<1$) because of the anti-Hermiticity of $H_0$. For this special case, a symmetry classification was put forward in Ref.~\cite{kawabata2022a}, but we shall see that some classes were missed as it was not noticed that  different blocks of $H$ can belong to different classes. The classification depends not only on $\kappa$ but also on $N\mod 4$~\cite{kawabata2022a,garcia2023b}, the
parities of $q/2$ and $r$\footnote{
	Since the classification depends on the parity of $r$ only, it coincides with the one of the
	slightly different Hamiltonian of Ref.~\cite{garcia2023b}, where $H_I$ was given by (up to an overall constant) $\(i \sum_{i=1}^N \psi^L_i \psi^R_i\)^r$. The two interacting Hamiltonians differ by $s$-body terms, where $s<r$ has the same parity of $r$ and, thus, belong to the same class. Moreover, in the large-$N$ limit, these extra terms are subleading in $1/N$ (they are suppressed as $1/N^{r-s}$) and the resulting phenomenology also coincides for both models. We choose the current model because of its clearer connection to the Lindbladian, see Sec.~\ref{sec:examples_open}, and the Keldysh wormhole model of Ref.~\cite{garcia2022e}.
}, and whether $\alpha = 1$ or $\alpha\neq 1$.
Below, we develop a classification scheme that allows for a unified treatment of all three regimes of the PT-symmetric two-site SYK models.

\subsection{Action of symmetry operators}
As discussed in Sec.~\ref{sec:Majorana}, the symmetry properties of $H$ are based on the action of the complex-conjugation operator $K$ and the unitary operators $Q$ [defined in Eq.~(\ref{eq:def_Q})], $S$ Eq.~(\ref{eq:def_S}), and $S_{L,R}$ Eq.~(\ref{eq:def_SLR}). Using Eqs.~(\ref{eq:action_K_psi})--(\ref{eq:action_Q_psi}) and (\ref{eq:action_SL_psi})--(\ref{eq:action_S_psi}), we find the transformation relations of the Hamiltonian Eq.~(\ref{eq:def_H}) under the unitary and anti-unitary transformations. Since the intersite coupling Hamiltonian $H_I$ in Eq.~(\ref{eq:def_H}) is independent of $\kappa$, it always has the transformation properties:
\begin{align}
\label{eq:action_SL_HI}
&S_{L,R} H_I S_{L,R}^{-1} = (-1)^r H_I,
\\
\label{eq:action_Q_HI}
&Q H_I Q^{-1} = H_I,
\\
\label{eq:action_K_HI}
&K H_I K^{-1} = H_I,
\\
\label{eq:action_dg_HI}
&H_I^\dagger = H_I.
\end{align}

On the other hand, the transformation properties of the decoupled Hamiltonian $H_0$ depend on $\kappa$. For the three cases discussed above ($\kappa=0$, $\kappa=1$, and $0<\kappa<1$), they are:
\begin{align}
\nonumber
&\bm{\kappa=0}:
&&\bm{\kappa=1}:
&&\bm{0<\kappa<1}:
\\
\label{eq:action_SL_H}
&S_{L,R} H_0 S_{L,R}^{-1} = H_0,
&&S_{L,R} H_0 S_{L,R}^{-1} = H_0,
&&S_{L,R} H_0 S_{L,R}^{-1} = H_0,
\\
\label{eq:action_Q_H}
&Q H_0 Q^{-1} = (-1)^{q/2} H_0,
&&Q H_0 Q^{-1} = -(-1)^{q/2} H_0,
&&Q H_0 Q^{-1} = (-1)^{q/2} H_0^\dagger,
\\
\label{eq:action_K_H}
&K H_0 K^{-1} = (-1)^{q/2} H_0,
&&K H_0 K^{-1} = -(-1)^{q/2} H_0,
&&K H_0 K^{-1} = (-1)^{q/2} H_0^\dagger,
\\
\label{eq:action_dg_H}
& H_0^\dagger  = H_0,
&& H_0^\dagger  = -H_0,
\end{align}
where the transformation under $Q$ Eq.~(\ref{eq:action_Q_H}) holds only in the left-right symmetric point $\alpha=1$.\footnote{
	Strictly speaking, it is also satisfied for the antisymmetric point $\alpha=-1$, but this only exchanges the role of even and odd $q/2$. As such, the classification for $\alpha=-1$ follows straightforwardly from the one we give for $\alpha=1$.
}
In the following sections, we shall show that the sought symmetry classification follows naturally from a systematic consideration of Eqs.~(\ref{eq:comm_K_SLSR})--(\ref{eq:comm_QS}) and (\ref{eq:action_SL_HI})--(\ref{eq:action_dg_H}).
We first search for unitary symmetries commuting with the Hamiltonian. These do not specify the symmetry class, but they define the block structure of the Hamiltonian, within which anti-unitary symmetries and unitary involutions must act. Later, we define projector operators that enable us to investigate the symmetries of each block. With this information at hand, we finally identify all symmetries of the Hamiltonian, separating the cases of $\kappa = 0$, already investigated in Ref.~\cite{garcia2023b}, $\kappa = 1$, and $0 < \kappa < 1$.

\subsection{Commuting unitary symmetries and block structure of the Hamiltonian}
\label{sec:blocks}

The irreducible blocks of the Hamiltonian (to which we also refer as symmetry sectors of the theory) are defined by the maximal set of mutually commuting unitary operators that also commute with the Hamiltonian.
\begin{itemize}
	\item For all values of the parameters $N$, $q$, $r$, and $\kappa$, $S$ commutes with $H$, i.e., the total fermionic parity is always conserved. Since $S$ has eigenvalues $s=\pm1$, the Hamiltonian $H$ has always at least two blocks, indexed by $s$.
	\item For even $r$, $H$ commutes with both $S_L$ and $S_R$ independently. Since $S_L$ and $S_R$ always commute with each other (for even $N$), in this case, there are four symmetry blocks, indexed by the eigenvalues $s_{L,R}=\pm1$.
	\item For the general PT-symmetric situation, $0<\kappa<1$, there are no further unitary symmetries, as the remaining unitary operator $Q$ maps $H_0$ into $H_0^\dagger$.
	\item For the left-right symmetric case ($\alpha=1$), in the Hermitian ($\kappa=0$) and Lindbladian ($\kappa=1$) regimes, $Q$ maps $H_0$ into $\pm H_0$. Therefore, $Q$ is an additional unitary symmetry when the plus sign holds. In this case, the existing blocks of the Hamiltonian corresponding to fixed $S$ or $S_{L,R}$~\cite{garcia2023b} may be further split.
	\item For $\kappa=0$, $QH_0Q^{-1}=H_0$ when $q/2$ is even.
	\begin{enumerate}[(i)]
		\item If, $r$ is odd, since $S$ and $Q$ always commute, Eq.~(\ref{eq:comm_QS}), $Q$ splits the two existing blocks (labeled by $s=\pm1$) into two subblocks each, leading to four blocks indexed by the eigenvalues of $Q$, $k=\pm1,\pm i$.
		\item If, instead $r$ is even, it follows from Eq.~(\ref{eq:comm_QSLR}) that $Q$ and $S_{L,R}$ commute only in sectors with $s=+1$, i.e., $s_L=s_R$. The blocks $s_L=s_R=\pm1$ thus get split into two subblocks indexed by $k=\pm1$. The two blocks with $s_L=-s_R$ do not get split by $Q$. For our purposes, it is more convenient to equivalently label the latter two blocks by the eigenvalue $k=\pm\i$. In total, we thus have six blocks.
	\end{enumerate}
	For odd $q/2$, $H$ and $Q$ do not commute and the block structure is the same as in the generic case ($0<\kappa<1$).
	\item For $\kappa=1$, $QH_0Q^{-1}=H_0$ when $q/2$ is odd. The block structure for even and odd $q/2$ is thus interchanged with respect to $\kappa=0$ case.
	\item For $\alpha\neq 1$, left-right asymmetric case, $Q$ is not a symmetry of the Hamiltonian for any $\kappa$ since Eq.~(\ref{eq:action_Q_H}) holds only for $\alpha=1$. The block structure is, therefore, the same as in the $0<\kappa<1$ regime.
\end{itemize}

The number of blocks and the eigenvalues by which they are indexed for each case is summarized in Table~\ref{tab:block_structure}.

\begin{table}[t]
	\caption{Block structure of the Hamiltonian, for different parities of $q/2$ and $r$. The block structure differs for the following three cases: $\kappa=0$ and $\alpha=1$; $\kappa=1$ and $\alpha=1$; and $0<\kappa<1$ or $\alpha\neq1$. For each possibility and set of parities, we list the number of blocks and, in parentheses, the conserved quantum numbers (eigenvalues of commuting unitary operators) labeling them.}
	\label{tab:block_structure}
	\begin{tabular}{@{}Sc Sc Sc Sc Sc Sc@{}}
		\toprule
		\multirow{2}{*}{$(-1)^{q/2}$} & \multirow{2}{*}{$(-1)^r$} & $\kappa=0$                         & $\kappa=1$                         & $0<\kappa<1$        \\
		                              &                           & \multicolumn{1}{c}{and $\alpha=1$} & \multicolumn{1}{c}{and $\alpha=1$} & \multicolumn{1}{c}{or $\alpha\neq 1$} \\ \midrule
		$+1$                          & $+1$                      & $6$ ($s_L,s_R,k$)                  & $4$ ($s_L,s_R$)                    & $4$ ($s_L,s_R$)                       \\
		$+1$                          & $-1$                      & $4$ ($k$)                          & $2$ ($s$)                          & $2$ ($s$)                             \\
		$-1$                          & $+1$                      & $4$ ($s_L,s_R$)                    & $6$ ($s_L,s_R,k$)                  & $4$ ($s_L,s_R$)                       \\
		$-1$                          & $-1$                      & $2$ ($s$)                          & $4$ ($k$)                          & $2$ ($s$)              \\
	 \bottomrule
	\end{tabular}
\end{table}

\subsection{Anti-unitary symmetries and projectors into symmetry sectors}
\label{sec:proj}

We now proceed to characterize the anti-unitary symmetries inside the blocks identified previously, which will result in the full symmetry classification of the Hamiltonian.
Since $S$ is always a conserved quantity of $H$, the relevant anti-unitary operators are taken to be $K$, $QK$, $S_LK$, and $QS_LK$. Using Eqs.~(\ref{eq:comm_K_SLSR})--(\ref{eq:comm_QS}), we find their squares to be
\begin{equation}
\label{eq:square_antiunitaries}
K^2=(QK)^2=+1,\qquad
(S_L K)^2=(-1)^{N/2},\qquad
(QS_LK)^2=(-1)^{N/2}S.
\end{equation}
The action of these operators on the Hamiltonian will determine the type of symmetry ($\sT_\pm$ or $\sC_\pm$) they represent, which we shall show depends on the value of $\kappa$ ($\kappa=0$, $\kappa=1$, or $0<\kappa<1$) under consideration, see Sec.~\ref{sec:classification}.

As mentioned above, in order to define a universality class by a set of anti-unitary symmetries, they must act within a single block of the Hamiltonian.
To check this, we need the commutation relation of the unitary operators and the complex-conjugation operator with the projectors into sectors of fixed eigenvalues of $S_{L,R}$, $S$, and $Q$, defined, respectively, as
\begin{align}
\label{eq:PLR}
\mathbb{P}_{L,R}^{s_{L,R}}&=\frac{1}{2}\(\id+s_{L,R}S_{L,R}\),
\\
\label{eq:PS}
\mathbb{P}_{S}^{s}&=\frac{1}{2}\(\id+sS\),
\\
\label{eq:PQ}
\mathbb{P}_{Q}^{k}&=\frac{1}{4}\(\id+\frac{Q}{k}+\frac{Q^2}{k^2}+\frac{Q^3}{k^3}\),
\end{align}
where the subscript denotes the symmetry being projected and the superscript the eigenvalue labeling the block.

Using the commutation relations of the unitary operators, Eqs.~(\ref{eq:comm_SLSR})--(\ref{eq:comm_QS}), we find:
\begin{align}
\label{eq:comm_SLR_P}
& S_{L,R} \mathbb{P}_S^s = \mathbb{P}_S^{s} S_{L,R},
&&S_{L,R} \mathbb{P}_Q^k = \mathbb{P}_Q^{sk} S_{L,R},
\\
\label{eq:comm_S_P}
& S \mathbb{P}_{L,R}^{s_{L,R}} = \mathbb{P}_{L,R}^{s_{L,R}} S,
&&S \mathbb{P}_Q^k = \mathbb{P}_Q^k S,
\\
\label{eq:comm_Q_P}
& Q \mathbb{P}_{L,R}^{s_{L,R}} = \mathbb{P}_{L,R}^{s s_{L,R}} Q,
&&Q \mathbb{P}_S^s = \mathbb{P}_S^s Q,
\end{align}
where used that $S$ always commutes with the Hamiltonian and we can, therefore, replace it by its eigenvalue $s$.
We conclude that $S_L$ and $S_R$ always act within blocks of fixed $S$ (for even $N$), but within blocks of fixed $Q$ only when $s=+1$ (i.e., $k=\pm1$).
On the other hand, $S$ always acts within blocks of fixed $S_{L,R}$ or $Q$.
Lastly, $Q$ always acts within blocks of fixed $S$ but only inside blocks of fixed $S_{L,R}$ when $s=+1$ (i.e., $s_L=s_R$).

The commutation relations of the projectors with the complex conjugation operators depend on the reality of both the unitary operators themselves and their eigenvalues. Since $S$ is real with real eigenvalues, the projector $\mathbb{P}_S$ is also real and, accordingly,
\begin{equation}
\label{eq:comm_K_PS}
K \mathbb{P}_{S}^{s}=\mathbb{P}_{S}^{s} K.
\end{equation}
$\mathbb{P}_Q$ also commutes with $K$,
\be
\label{eq:comm_K_PQ}
K \mathbb{P}_{Q}^{k}=\frac 14 \(\id +\frac{Q^{-1}}{k^*}+\frac{Q^{-2}}{{k^*}^2}
+\frac{Q^{-3}}{{k^*}^3} \)K=\mathbb{P}_{Q}^{k} K,
\ee
where we have used $ K Q = Q^{-1} K$ [Eq.~(\ref{eq:comm_K_S_Q})], $Q^4=1$ [Eq.~(\ref{eq:square_unitaries})], and $k^4=1$.
We see that $K$ always acts within blocks of fixed $S$ or fixed $Q$.
Finally, the projector $\mathbb{P}_{L,R}$ commutes with $K$ only when $S_{L,R}$ is real. Using Eq.~(\ref{eq:comm_K_SLSR}) we have
\begin{align}
\label{eq:comm_K_PLR}
&K \mathbb{P}_{S_{L,R}}^{s_{L,R}}
=\frac{1}{2}\(\id+(-1)^{N/2}s_{L,R}S_{L,R}\) K
=\mathbb{P}_{S_{L,R}}^{(-1)^{N/2}s_{L,R}} K,
\end{align}
i.e., $K$ acts within blocks of fixed $S_{L,R}$ only when $N\mod4=0$.

Finally, combining Eqs.~(\ref{eq:comm_SLR_P})--(\ref{eq:comm_K_PLR}), the commutation relations of the anti-unitary operators with the projectors are:
\begin{align}
\label{eq:comm_QK_P}
&
QK \mathbb{P}_{Q}^k = \mathbb{P}_{Q}^k QK,&&
QK \mathbb{P}_{S}^s = \mathbb{P}_{S}^s QK,&&
QK \mathbb{P}_{L,R}^{s_{L,R}} = \mathbb{P}_{L,R}^{(-1)^{N/2}ss_{L,R}} QK
\\
\label{eq:comm_SLK_P}
&
S_LK \mathbb{P}_{Q}^k = \mathbb{P}_{Q}^{(-1)^{N/2}s k} S_LK,&&
S_LK \mathbb{P}_{S}^s = \mathbb{P}_{S}^{s} S_LK,&&
S_LK \mathbb{P}_{L,R}^{s_{L,R}} = \mathbb{P}_{L,R}^{(-1)^{N/2}s_{L,R}} S_LK,
\\
\label{eq:comm_QSLK_P}
&
QS_LK \mathbb{P}_{Q}^k = \mathbb{P}_{Q}^{s k} QS_LK,&&
QS_LK \mathbb{P}_{S}^s = \mathbb{P}_{S}^{s} QS_LK,&&
QS_LK \mathbb{P}_{L,R}^{s_{L,R}} = \mathbb{P}_{L,R}^{ss_{L,R}} QS_LK.
\end{align}
From these equations, we can read off immediately whether the anti-unitary symmetries act with a symmetry sector or not.

We now have all the ingredients to determine the symmetries of the Hamiltonian Eq.~(\ref{eq:def_H}) depending on the value of $\kappa$.

\section{Symmetry classification of the PT-symmetric SYK model}
\label{sec:classification}

\subsection{Hermitian regime ($\kappa=0$)}
\label{sec:Hermitian}

\begin{table}[t]
	\caption{Symmetry classification of the two-site SYK
		Hamiltonian Eq.~(\ref{eq:def_H}) in the Hermitian regime ($\kappa=0$), for even $q/2$ and even $r$. Each line corresponds to a block of the Hamiltonian, labeled by the eigenvalues of the conserved quantities: $S_{L,R}$ and $Q$
		($\alpha=1$) or $S_{L,R}$ ($\alpha\neq1$). For each block, we give the Hermitian random matrix symmetry class as a function of $N\mod4$ and $\alpha$.
	}
	\label{tab:classification_Hermitian_even_even}
	\begin{tabular}[t]{@{}Sc Sc Sc Sc Sc Sc Sc@{}}
		\toprule
		\multirow{2}{*}{$S_L$} & \multirow{2}{*}{$S_R$} & \multirow{2}{*}{$Q$} & \multicolumn{2}{c}{$\ N\mod4=0\ $}   & \multicolumn{2}{c}{$\ N\mod4=2\ $}    \\
		&                      &                       & $\ \alpha=1$ & $\alpha\neq1$       & $\ \alpha=1$ & $\alpha\neq1$       \\ \midrule
		\multirow{2}{*}{$+1$}  & \multirow{2}{*}{$+1$}  & $+1$                & AI         & \multirow{2}{*}{AI} & A          & \multirow{2}{*}{A} \\
		&                        & $-1$                & AI         &                     & A          &                     \\ \midrule
		\multirow{2}{*}{$-1$}  & \multirow{2}{*}{$-1$}  & $+1$                & AI         & \multirow{2}{*}{AI} & A          & \multirow{2}{*}{A} \\
		&                        & $-1$               & AI         &                     & A          &                     \\ \midrule
		$+1$                   & $-1$                   & ---                  & AI         & AI                  & AI         & A                  \\ \midrule
		$-1$                   & $+1$                   & ---                  & AI         & AI                  & AI         & A                  \\ \bottomrule
	\end{tabular}
	
	\caption{Same as Table~\ref{tab:classification_Hermitian_even_even}, but for even $q/2$ and odd $r$. The blocks are labeled by the eigenvalues of $S$ and $Q$ ($\alpha=1$) or $S$ ($\alpha\neq1$).}
	\label{tab:classification_Hermitian_even_odd}
	\begin{tabular}{@{}Sc Sc Sc Sc Sc Sc@{}}
		\toprule
		\multirow{2}{*}{$S$}  & \multirow{2}{*}{$Q$} & \multicolumn{2}{c}{$\ N\mod4=0\ $}   & \multicolumn{2}{c}{$\ N\mod4=2\ $} \\
		&                      & $\ \alpha=1$   &  $\alpha\neq1$      & $\ \alpha=1$ & $\alpha\neq1$       \\ \midrule
		\multirow{2}{*}{$+1$} & $+1$                 & AI             & \multirow{2}{*}{AI} & AI           & \multirow{2}{*}{AI} \\
		& $-1$                 & AI             &                     & AI           &                     \\ \midrule
		\multirow{2}{*}{$-1$} & $+\i$                & AI             & \multirow{2}{*}{AI} & AI           & \multirow{2}{*}{AI} \\
		& $-\i$                & AI             &                     & AI           &                     \\ \bottomrule
	\end{tabular}	
	
	\caption{Same as Table~\ref{tab:classification_Hermitian_even_even}, but for odd $q/2$ and even $r$. The blocks are labeled by the eigenvalues of $S_{L,R}$.}
	\label{tab:classification_Hermitian_odd_even}
	\begin{tabular}{@{}Sc Sc Sc Sc Sc Sc@{}}
		\toprule
		\multirow{2}{*}{$S_L$} & \multirow{2}{*}{$S_R$} & \multicolumn{2}{c}{$\ N\mod4=0\ $}  & \multicolumn{2}{c}{$\ N\mod4=2\ $} \\
		&                        & $\ \alpha=1$   & $\alpha\neq1$      & $\ \alpha=1$ &  $\alpha\neq1$      \\ \midrule
		$+1$                   & $+1$                   & AI             & A                  & A            & A                   \\
		$-1$                   & $-1$                   & AI             & A                  & A            & A                   \\ \midrule
		$+1$                   & $-1$                   & A              & A                  & AI           & A                   \\
		$-1$                   & $+1$                   & A              & A                  & AI           & A                   \\ \bottomrule
	\end{tabular}
	
	\caption{Same as Table~\ref{tab:classification_Hermitian_even_even}, but for odd $q/2$ and odd $r$. The blocks are labeled by the eigenvalues of $S$.}
	\label{tab:classification_Hermitian_odd_odd}
	\begin{tabular}{@{}Sc Sc Sc Sc Sc@{}}
		\toprule
		\multirow{2}{*}{$S$} & \multicolumn{2}{c}{$\ N\mod4=0\ $}   & \multicolumn{2}{c}{$\ N\mod4=2\ $}  \\
		& $\ \alpha=1$   & $\alpha\neq1$     & $\ \alpha=1$ &  $\alpha\neq1$   \\ \midrule
		$+1$ & BDI  & D & CI & C  \\
		$-1$ & BDI  & D & CI & C  \\ \bottomrule
	\end{tabular}
\end{table}

The symmetry classification of the two-site SYK Hamiltonian Eq.~(\ref{eq:def_H}) in the Hermitian regime ($\kappa=0$) was performed in Ref.~\cite{garcia2023b}. Here, we reproduce it for pedagogical purposes, in preparation of the more challenging non-Hermitian case ($\kappa > 0$).

As mentioned above, for Hermitian matrices, the two transformations $\sT_+$ and $\sC_+$ coincide and are denoted as $T$---time reversal symmetry (TRS)---and the coinciding symmetries $\sT_-$ and $\sC_-$ are denoted as $C$---particle-hole symmetry (PHS). If $\alpha=1$, the PT-symmetry operator plays the role of TRS, $T=QK$. However, if the left-right symmetry is broken ($\alpha\neq1$), there is still a residual TRS $\widetilde{T}=K$ for the case of odd $q/2$. Moreover, if both $q/2$ and $r$ are odd, there is an anti-unitary PHS, given by $C=S_L K$ (regardless of $\alpha$). From Eq.~(\ref{eq:square_antiunitaries}), we have that $T^2=\widetilde{T}^2=+1$, while $C^2=(-1)^{N/2}$. In order
to determine the symmetry class, it remains to check whether the anti-unitary symmetries act within a single block of the Hamiltonian, resorting to Eqs.~(\ref{eq:comm_K_PS})--(\ref{eq:comm_QSLK_P}).
By evaluating them for the different block structures of the Hamiltonian in the first ($\alpha=1$) and third ($\alpha\neq1$) columns of Table~\ref{tab:block_structure}, we fill out Tables~\ref{tab:classification_Hermitian_even_even}--\ref{tab:classification_Hermitian_odd_odd}.

An in-depth analysis of the classification is given in Ref.~\cite{garcia2023b}. Let us highlight the most salient features. Of the ten classes in the Hermitian RMT  classification~\cite{altland1997,verbaarschot1993a,verbaarschot1994}, only six are realized in the two-site model (A, AI, BDI, CI, D, and C). As a general result of left-right symmetric two-site models, there are no classes with $T^2=-1$ (AII, CII, and DIII). This restriction prevents both the occurrence of Kramers degeneracy inside a fixed block of the Hamiltonian and the independence of a state and its time-reversal. We have seen in Sec.~\ref{sec:class_general_PT} that a similar statement is believed to hold also for non-Hermitian systems~\cite{sa2022a}. The absence of the remaining class (AIII) is not fundamental and could, in principle, be realized in some modified two-site model. Moreover, remarkably, different blocks of the same Hamiltonian can belong to different symmetry classes, a result that also carries over to the non-Hermitian setting.

\subsection{Lindbladian regime ($\kappa=1$)}
\label{sec:Lindbladian}

Next, we turn to the classification of the two-site SYK Hamiltonian Eq.~(\ref{eq:def_H}) in the Lindbladian regime ($\kappa=1$) where, at least for $\alpha = 1$, the Hamiltonian can be interpreted as the vectorized Lindbladian describing a Hermitian system coupled to a Markovian bath.

\begin{table}[t]
	\caption{Symmetry classification of the two-site SYK Hamiltonian Eq.~(\ref{eq:def_H}) in the Lindbladian regime ($\kappa=1$), for even $q/2$ and even $r$. Each line corresponds to a block of the Hamiltonian, labeled by the eigenvalues of the conserved quantities, $S_{L,R}$.
		For each block, we give the non-Hermitian Bernard-LeClair class as a function of $N\mod4$ and $\alpha$. The subscript $\nu$ denotes the index $\nu = \Tr \mathbb{P}_L\mathbb{P}_{R} Q$ of AIII$_\nu$ and BDI$^\dagger_\nu$, which we have recently shown \cite{garcia2023c} to also be a topological invariant.}
	\label{tab:classification_Lindbladian_even_even}
	\begin{tabular}{@{}Sc Sc Sc Sc Sc Sc Sc Sc Sc Sc@{}}
		\toprule
		\multirow{2}{*}{$S_L$} & \multirow{2}{*}{$S_R$} & \multicolumn{2}{c}{$\ N\mod4=0\ $}  &  \multicolumn{2}{c}{$\ N\mod4=2\ $} \\
		&                        & $\ \alpha=1$   & $\alpha\neq1$      & $\ \alpha=1$ &  $\alpha\neq1$       \\ \midrule
		$+1$                   & $+1$                   & BDI$^\dagger_\nu$ & AI$^\dagger$ & AIII$_\nu$  & A \\
		$-1$ & $-1$ & BDI$^\dagger_\nu$ & AI$^\dagger$   & AIII$_\nu$  & A        \\ \midrule
		$+1$ & $-1$ & AI$^\dagger$  & AI$^\dagger$  & AI    & A   \\
		$-1$ & $+1$ & AI$^\dagger$  & AI$^\dagger$  & AI    & A  \\ \bottomrule
	\end{tabular}
	
	\caption{Same as Table~\ref{tab:classification_Lindbladian_even_even}, but for even $q/2$ and odd $r$. The blocks are labeled by the eigenvalues of $S$. In this case, the index of classes BDI$_{++\nu}$ and CI$_{--\nu}$ is given by $\nu=\Tr \mathbb{P}_S Q$.}
	\label{tab:classification_Lindbladian_even_odd}
	\begin{tabular}{@{}Sc Sc Sc Sc Sc@{}}
		\toprule
		\multirow{2}{*}{$S$} & \multicolumn{2}{c}{$\ N\mod4=0\ $}   & \multicolumn{2}{c}{$\ N\mod4=2\ $} \\
		& $\ \alpha=1$   & $\alpha\neq1$     & $\ \alpha=1$ &  $\alpha\neq1$   \\ \midrule
		$+1$ & BDI$_{++\nu}$ & BDI$^\dagger$  & CI$_{--\nu}$  & CI$^\dagger$ \\
		$-1$ & CI$_{+-}$  & BDI$^\dagger$ & BDI$_{-+}$ & CI$^\dagger$ \\ \bottomrule
	\end{tabular}
	
	\caption{Same as Table~\ref{tab:classification_Lindbladian_even_even}, but for odd $q/2$ and even $r$. The blocks are labeled by the eigenvalues of $S_{L,R}$ and $Q$ ($\alpha=1$) or $S_{L,R}$ ($\alpha\neq1$).}
	\label{tab:classification_Lindbladian_odd_even}
	\begin{tabular}[t]{@{}Sc Sc Sc Sc Sc Sc Sc@{}}
		\toprule
		\multirow{2}{*}{$S_L$} & \multirow{2}{*}{$S_R$} & \multirow{2}{*}{$Q$} & \multicolumn{2}{c}{$\ N\mod4=0\ $}   & \multicolumn{2}{c}{$\ N\mod4=2\ $}    \\
		&                      &                       & $\ \alpha=1$ & $\alpha\neq1$       & $\ \alpha=1$ & $\alpha\neq1$       \\ \midrule
		\multirow{2}{*}{$+1$}  & \multirow{2}{*}{$+1$}  & $+1$                & AI         & \multirow{2}{*}{AI} & A          & \multirow{2}{*}{A} \\
		&                        & $-1$                & AI         &                     & A          &                     \\ \midrule
		\multirow{2}{*}{$-1$}  & \multirow{2}{*}{$-1$}  & $+1$                & AI         & \multirow{2}{*}{AI} & A          & \multirow{2}{*}{A} \\
		&                        & $-1$               & AI         &                     & A          &                     \\ \midrule
		$+1$                   & $+1$                   & ---                  & AI         & AI                  & AI         & A                  \\ \midrule
		$-1$                   & $-1$                   & ---                  & AI         & AI                  & AI         & A                  \\ \bottomrule
	\end{tabular}

	\caption{Same as Table~\ref{tab:classification_Lindbladian_even_even}, but for odd $q/2$ and odd $r$. The blocks are labeled by the eigenvalues of $S$ and $Q$ ($\alpha=1$) or $S$ ($\alpha\neq1$).}
	\label{tab:classification_Lindbladian_odd_odd}
	\begin{tabular}{@{}Sc Sc Sc Sc Sc Sc@{}}
		\toprule
		\multirow{2}{*}{$S$} & \multirow{2}{*}{$Q$} & \multicolumn{2}{c}{$\ N\mod4=0\ $}   &  \multicolumn{2}{c}{$\ N\mod4=2\ $}  \\
		&      & $\ \alpha=1$   & $\alpha\neq1$  & $\ \alpha=1$ &  $\alpha\neq1$     \\ \midrule
		\multirow{2}{*}{$+1$} & $+1$  & BDI  & \multirow{2}{*}{BDI} & CI & \multirow{2}{*}{CI} \\
		& $-1$  & BDI  &                       & CI   \\ \midrule
		\multirow{2}{*}{$-1$} & $+\i$ & AI   & \multirow{2}{*}{BDI} & AI & \multirow{2}{*}{CI}  \\
		& $-\i$ & AI   &                       & AI &                        \\ \bottomrule
	\end{tabular}	
\end{table}

Since this model is non-Hermitian, we must consider the four anti-unitary transformations listed in Table~\ref{eq:nHsym}.
Let us start with the explicitly PT-symmetric case, $\alpha=1$. We have the following anti-unitary symmetries. $\sT_+$ is always present and is implemented by the PT-symmetry operator $\sT_+=QK$. If $q/2$ is even, there is a $\sC_+$ symmetry implemented by $\sC_+=K$. If $r$ is odd, there is a $\sC_-$ symmetry implemented by $\sC_-=QS_LK$. Naturally, if $q/2$ is even and $r$ is odd, there is also a $\sT_-$ symmetry, given by the composition of the previous three symmetries, $\sT_-=S_L K$ (up to a factor $S$, which is constant inside each block). If PT symmetry is explicitly broken, $\alpha\neq1$, the anti-unitary transformations involving the $Q$ operator ($\sT_+$ and $\sC_+$) cease to be symmetry transformations. There is still $\sC_+=K$ for even $q/2$, and $\sT_-=S_LK$ for even $q/2$ and odd $r$. There are also residual anti-unitary symmetries, namely, $\widetilde{\sT}_+=K$ for odd $q/2$, and $\widetilde{\sC}_-=S_LK$ for odd $q/2$ and odd $r$. The squares of the anti-unitary symmetries are given by Eq.~(\ref{eq:square_antiunitaries}):
\begin{equation}
\label{eq:Lindbladian_square_antiunitaries}
\sT_+^2=\widetilde{\sT}_+^2=\sC_+^2=+1,
\qquad
\sC_-^2=(-1)^{N/2}S,
\qquad \text{and} \qquad
\sT_-^2=\widetilde{\sC}_-^2=(-1)^{N/2}.
\end{equation}
Note that the square of $\sC_-$ depends explicitly on $S$, giving the first hint that different parity sectors can belong to distinct symmetry classes. In the Hermitian regime, this was solely due to symmetries being broken in some sectors and not in others. However, in this case, the square of one of the anti-unitaries depends explicitly on the sector.
By evaluating Eqs.~(\ref{eq:comm_K_PS})--(\ref{eq:Lindbladian_square_antiunitaries}) for the different block structures of the Hamiltonian in the second ($\alpha=1$) and third ($\alpha\neq1$) columns of Table~\ref{tab:block_structure}, we can fill out Tables~\ref{tab:classification_Lindbladian_even_even}--\ref{tab:classification_Lindbladian_odd_odd}.

We note that if either (i) $\alpha=1$, $r$ is even, $N\mod4=2$, and $s=+1$, or (ii) $\alpha\neq1$, $q/2$ is odd, $r$ is even, and $N\mod4=2$, then none of the anti-unitary symmetries act within a single block. However, before assigning these cases to non-Hermitian class A, we must check for the unitary involutions (compositions of
anti-unitary symmetries, $\sP$ and $\sQ_\pm$), which could have nontrivial actions on $H$ despite the anti-unitary symmetries being individually broken. Indeed, for one case, namely, $\alpha=1$, even $q/2$ and $r$, and $N\mod4=2$, for the two sectors with $s=+1$, there is a unitary involution $\sQ_+=Q$. As such, this particular block of the Hamiltonian belongs to non-Hermitian class AIII. The remaining cases belong to class A.

In the Hermitian regime, there were no classes with $T^2=-1$. Likewise, in the Lindbladian regime, there are no classes with $\sT_+^2=-1$ or $\sC_+^2=-1$.
The absence of classes with $\sC_+^2=-1$, in particular, precludes the existence of non-Hermitian Kramers degeneracy inside fixed sectors of the Hamiltonian. In order to illustrate our remarks about the generality of this result (made in Sec.~\ref{sec:general_classification}) and to understand more explicitly the absence of these classes in the concrete model of this paper, we note that the only anti-unitary symmetries that can square to $-1$ are $S_LK$ and $QS_LK$ and these always implement only type $\sT_-$ or $\sC_-$ symmetries.

Some classes in Tables \ref{tab:classification_Lindbladian_even_even} and \ref{tab:classification_Lindbladian_even_odd} (even $q/2$) come with an index $\nu=\Tr\mathbb{P}Q$, where $\mathbb{P}Q$ denotes an appropriate projection of $Q$, since the left-right exchange operator $Q$, while not being a symmetry of $H$, still induces a finer substructure of its blocks. To understand this, we note that, in the sectors with $s=+1$ (i.e., $k=\pm1$), $Q$ acquires an anomalous trace:
\begin{align}
\label{eq:nu_S}
&\Tr \mathbb{P}^{(+1)}_{S} Q = 2^{N/2},
\\
\label{eq:nu_LR}
&\Tr \mathbb{P}^{(+1)}_{L}\mathbb{P}^{(+1)}_R Q =
\Tr \mathbb{P}^{(-1)}_{L}\mathbb{P}^{(-1)}_R Q = 2^{N/2}/2.
\end{align}
As a consequence, a block of $H$ with $s=+1$ acquires itself a substructure of the form
\begin{equation}
\begin{pmatrix}
 A & B \\ C & D
\end{pmatrix}.
\end{equation}
Here, $A$ contains the matrix elements of the block of $H$ that connect eigenstates of $Q$ with $k=+1$, $D$ connects two eigenstates with $k=-1$, while $B$ and $C$ connect eigenstates with $k=1$ to eigenstates with $k=-1$. Since there are $\nu$ more eigenstates with $k=+1$ than with $k=-1$, $A$ and $D$ have different dimensions and $B$ and $C$ are rectangular. In Sec.~\ref{topo}, we will discuss in detail how this anomalous trace results in $\nu$ exactly real eigenvalues of $H$, which are robust and cannot become pairs of complex-conjugated eigenvalues by a collision at an exceptional point.
On the other hand, $Q$ has the same number of eigenstates with eigenvalues $k = \pm i$ and, hence, $\nu=0$ for sectors with $s=-1$. Moreover, for odd $q/2$, $Q$ is a symmetry of $H$ and hence, determines the block structure of $H$ itself and not the finer substructure of the blocks. In these two cases there are, therefore, no classes with a nontrivial block substructure.
We have shown recently \cite{garcia2023c} that the rectangular block structure and the anomalous trace are signatures of topological behavior characterized by the class dependent topological invariant $\nu(N)$.
Therefore, the combination of these two facts points to the existence of an index theorem in this case equating the topological invariant $\nu(N)$ and the analytic index $\Tr \mathbb{P} Q$, with $\mathbb{P} $
a projection on a subspace as discussed earlier in this section.
Likewise, as mentioned in Ref.~\cite{garcia2023c}, the anomalous trace is then a signature of a pseudo-Hermitian quantum anomaly.

As shall be discussed in detail in Sec.~\ref{sec:examples_open}, the PT-symmetric SYK Hamiltonian Eq.~(\ref{eq:def_H}) for $\kappa=1$ and $\alpha=1$ can be mapped onto an SYK Lindbladian with a Hermitian dissipator. This particular case of the model coincides with the one employed in Ref.~\cite{kawabata2022a} to classify dissipative quantum chaos. Although the two classifications agree in many cases, there are still important differences stemming from the fact that the classification of Ref.~\cite{kawabata2022a} did account neither for the modifications to the block structure due to the $Q$ symmetry (leading, for example, to the mentioned topological features \cite{garcia2023c}) nor for the possibility that blocks of the Hamiltonian with opposite parity $S$ can, in some cases, belong to different symmetry classes (either because the square of an anti-unitary depends explicitly on $S$ or because a symmetry is broken in one of the parity blocks but not in the other). These two results are some of the main novel features of our work and, arguably, the most interesting characteristics of the model.
More specifically, the classification of Ref.~\cite{kawabata2022a} should be corrected in the following cases ($\alpha=\kappa=1$ in all instances):
\begin{enumerate}
	\item $q/2$ even, $r$ even, $N\mod4=0$. In the two sectors with $s=-1$, $\sT_+$ is broken. These blocks belong to class AI$^\dagger$ (instead of BDI$^\dagger$).
	\item $q/2$ even, $r$ even, $N\mod4=2$. In the two sectors with $s=-1$, $\sT_+$ is unbroken and $\sQ_+$ is broken. These blocks belong to class AI (instead of AIII).
	\item $q/2$ even, $r$ odd. The square of $\sC_-$ is $\sC_-^2=(-1)^{N/2}S$, while Ref.~\cite{kawabata2022a} incorrectly uses $\sC_-^2=+1$, resulting in the misidentification of the classes when $(-1)^{N/2}s=-1$. For $N\mod4=0$, the block with $s=-1$ thus belongs to class CI$_{+-}$ (instead of BDI$_{++}$). For $N\mod4=2$, the $s=+1$ block belongs to class CI$_{--\nu}$ (instead of BDI$_{-+}$).
	\item $q/2$ odd, $r$ even, $N\mod4=2$. In the two sectors with $s=-1$ (i.e., $k=\pm\i$), $\sT_+$ is unbroken, because these blocks are not labeled by $s_{L,R}$ but instead by $k$. These blocks belong to class AI (instead of A).
	\item $q/2$ odd, $r$ odd. In the two sectors with $s=-1$ (i.e., $k=\pm i$), $\sC_-$ is broken, because it does not commute with the projector $\mathbb{P}_Q$. These blocks belong to class AI (instead of BDI for $N\mod4=0$ and CI for $N\mod4=2$).
\end{enumerate}
The numerical checks of the classification in Ref.~\cite{kawabata2022a} were performed only for $q=4$ (i.e., even $q/2$), for which the block structure is not modified by the operator $Q$. Moreover, for the cases 1., 2., and 3. enumerated above with even $q/2$, the numerical indicators used in Ref.~\cite{kawabata2022a} are unable to distinguish the different classes. In Sec.~\ref{sec:level_statistics}, we show explicitly that these different universality classes in different blocks of the Hamiltonian can indeed be distinguished by employing a more complete set of spectral observables.

In summary, the PT-symmetric Hamiltonian Eq.~(\ref{eq:def_H}) in the Lindbladian regime ($\kappa = 1$) realizes 12 non-Hermitian symmetry classes (A, AIII, AI, BDI, CI, AI$^\dagger$, BDI$^\dagger$, CI$^\dagger$, BDI$_{++}$, BDI$_{-+}$, CI$_{+-}$, and CI$_{--}$), three of which (AIII, BDI$_{++}$, and CI$_{--}$) always come with an index $\nu >0$, and one (BDI$^\dagger$) for which the index $\nu$ is nonzero in some cases only.
We have found \cite{garcia2023c} that this index $\nu(N)> 0$ is indeed a topological invariant, so these four classes are topological since they describe systems with topologically nontrivial features. Remarkably, this list of 12 classes includes eight classes of the Lindbladian tenfold way with unbroken $\sT_+$ symmetry~\cite{sa2022a} realized by a single model (only classes AI$_+$ and AI$_-$ are missing), further illustrating the richness of the SYK Lindbladian~\cite{sa2022,kulkarni2022,garcia2022e,kawabata2022,kawabata2022a}.

\subsection{General PT-symmetric regime ($0<\kappa<1$)}
\label{sec:general}

Finally, we consider the classification of the two-site SYK Hamiltonian Eq.~(\ref{eq:def_H}) in the general PT-symmetric regime ($0<\kappa<1$). The model is still non-Hermitian and the classification proceeds similarly to the previous case. However, because the decoupled Hamiltonian $H_0$ is no longer anti-Hermitian, the anti-unitary symmetries are more restricted. The block structure is also simpler, since $Q$ is not a unitary symmetry for any choice of $q$, see Sec.~\ref{sec:blocks}.

\begin{table}[t]
	\caption{Symmetry classification of the two-site SYK Hamiltonian Eq.~(\ref{eq:def_H}) in the general PT-symmetric regime ($0<\kappa<1$), for even $q/2$ and even $r$. Each line corresponds to a block of the Hamiltonian, labeled by the eigenvalues of the conserved quantities, $S_{L,R}$. For each block, we give the non-Hermitian Bernard-LeClair class as a function of $N\mod4$ and $\alpha$. The subscript $\nu$ denotes the index $\nu = \Tr \mathbb{P}_L\mathbb{P}_{R} Q$ of AIII$_\nu$ and BDI$^\dagger_\nu$, which we have recently shown \cite{garcia2023c} to also be a topological invariant.}
	\label{tab:classification_general_even_even}
	\begin{tabular}{@{}Sc Sc Sc Sc Sc Sc@{}}
		\toprule
		\multirow{2}{*}{$S_L$} & \multirow{2}{*}{$S_R$} & \multicolumn{2}{c}{$\ N\mod4=0\ $}   & \multicolumn{2}{c}{$\ N\mod4=2\ $}  \\
		&      & $\ \alpha=1$   & $\alpha\neq1$  & $\ \alpha=1$ &  $\alpha\neq1$  \\ \midrule
		$+1$ & $+1$ & BDI$^\dagger_{\nu}$ & AI$^\dagger$ & AIII$_{\nu}$  & A  \\
		$-1$ & $-1$ & BDI$^\dagger_{\nu}$ & AI$^\dagger$ & AIII$_{\nu}$  & A  \\ \midrule
		$+1$ & $-1$ & AI$^\dagger$  & AI$^\dagger$ & AI    & A  \\
		$-1$ & $+1$ & AI$^\dagger$  & AI$^\dagger$ & AI    & A  \\ \bottomrule
	\end{tabular}
	
	\caption{Same as Table~\ref{tab:classification_general_even_even}, but for even $q/2$ and odd $r$. The blocks are labeled by the eigenvalues of $S$. The index of class BDI$^\dagger_\nu$ is given by $\nu=\Tr \mathbb{P}_S Q$.}
	\label{tab:classification_general_even_odd}
	\begin{tabular}{@{}Sc Sc Sc Sc Sc@{}}
		\toprule
		\multirow{2}{*}{$S$} & \multicolumn{2}{c}{$\ N\mod4=0\ $} & \multicolumn{2}{c}{$\ N\mod4=2\ $} \\
		& $\ \alpha=1$   & $\alpha\neq1$     & $\ \alpha=1$ &  $\alpha\neq1$  \\ \midrule
		$+1$ & BDI$^\dagger_\nu$ & AI$^\dagger$ & BDI$^\dagger_\nu$ & AI$^\dagger$  \\
		$-1$ & BDI$^\dagger$ & AI$^\dagger$ & BDI$^\dagger$ & AI$^\dagger$  \\ \bottomrule
	\end{tabular}
	
	\caption{Same as Table~\ref{tab:classification_general_even_even}, but for odd $q/2$ and even $r$. The blocks are labeled by the eigenvalues of $S_{L,R}$.}
	\label{tab:classification_general_odd_even}
	\begin{tabular}{@{}Sc Sc Sc Sc Sc Scc@{}}
		\toprule
		\multirow{2}{*}{$S_L$} & \multirow{2}{*}{$S_R$} & \multicolumn{2}{c}{$\ N\mod4=0\ $}  & \multicolumn{2}{c}{$\ N\mod4=2\ $}  \\
		&      & $\ \alpha=1$   & $\alpha\neq1$  & $\ \alpha=1$ &  $\alpha\neq1$  \\ \midrule
		$+1$ & $+1$ & AI & A & A  & A \\
		$-1$ & $-1$ & AI & A & A  & A \\ \midrule
		$+1$ & $-1$ & A  & A & AI & A \\
		$-1$ & $+1$ & A  & A & AI & A \\ \bottomrule
	\end{tabular}

	\caption{Same as Table~\ref{tab:classification_general_even_even}, but for odd $q/2$ and odd $r$. The blocks are labeled by the eigenvalues of $S$.}
	\label{tab:classification_general_odd_odd}
	\begin{tabular}{@{}Sc Sc Sc Sc Sc@{}}
		\toprule
		\multirow{2}{*}{$S$} & \multicolumn{2}{c}{$\ N\mod4=0\ $} & \multicolumn{2}{c}{$\ N\mod4=2\ $} \\
		& $\ \alpha=1$   & $\alpha\neq1$     & $\ \alpha=1$ &  $\alpha\neq1$  \\ \midrule
		$+1$ & BDI & D & CI & C \\
		$-1$ & BDI & D & CI & C \\ \bottomrule
	\end{tabular}
\end{table}

We start again with the explicitly PT-symmetric case, $\alpha=1$. As before, there is always a $\sT_+$ (PT) symmetry implemented by $\sT_+=QK$ and a $\sC_+$ symmetry for even $q/2$ implemented by $\sC_+=K$. The anti-unitary operator $QS_LK$ is no longer a symmetry operation and a $\sC_-$ symmetry is instead implemented by $\sC_-=S_LK$, but only for simultaneously odd $q/2$ and $r$. Finally, there is no $\sT_-$ symmetry for any choice of parameters. When the PT symmetry is explicitly broken by $\alpha\neq1$, the $\sT_+=QK$ anti-unitary operator ceases to be a symmetry operation because it involves the operator $Q$. However, the $\sC_\pm$ symmetries continue to hold. Using Eq.~(\ref{eq:square_antiunitaries}), the squares of the anti-unitary symmetries are
\begin{equation}
\label{eq:general_square_antiunitaries}
\sT_+^2=\sC_+^2=+1\qquad\text{and}\qquad
\sC_-=(-1)^{N/2}.
\end{equation}
By evaluating Eqs.~(\ref{eq:comm_K_PS})--(\ref{eq:comm_SLK_P}) and (\ref{eq:general_square_antiunitaries}) for the different block structures of the Hamiltonian in the third column of Table~\ref{tab:block_structure}, we can fill out Tables~\ref{tab:classification_general_even_even}--\ref{tab:classification_general_odd_odd}.

As was the case in the Lindbladian regime, once again there are no classes with $\sT_+^2=-1$ or $\sC_+^2=-1$.
Moreover, there are again classes with a nontrivial topological structure. For even $q/2$ and $r$, BDI$^\dagger_\nu$ and AIII$_\nu$ retain their topological features from the more symmetric point $\kappa=1$. For even $q/2$ and odd $r$, classes BDI$_{++\nu}$ and CI$_{--\nu}$ for $\kappa=1$ become BDI$^\dagger_\nu$ here. Therefore, no further topological classes arise.

In summary, the Hamiltonian Eq.~(\ref{eq:def_H}) in the general PT-symmetric case realizes nine non-Hermitian symmetry classes (A, AIII, AI, BDI, CI, D, C, AI$^\dagger$ and BDI$^\dagger$), of which seven are also realized in the Lindbladian regime and two (D and C) are exclusive to the general PT-symmetric regime. Of these nine, two (AIII and BDI$^\dagger$) can have, for some parameters, a rectangular block representation which we have recently found to be related to topologically nontrivial features~\cite{garcia2023c}. Combining this with the results of previous section, we conclude that the two-site SYK Hamiltonian Eq.~(\ref{eq:def_H}) realizes, in total, six Hermitian and 14 non-Hermitian symmetry classes.

The confirmation of this symmetry classification by a spectral and eigenvector analysis using exact diagonalization techniques is discussed in Sec.~\ref{sec:level_statistics}, while applications of the classification are addressed in Sec.~\ref{sec:examples}. Before that, we discuss in more detail the relation alluded above between the observed rectangularization of the SYK Hamiltonian for classes AIII$_\nu$, BDI$^\dagger_\nu$, BDI$_{++\nu}$, and CI$_{--\nu}$ and the existence of robust $\nu(N)$ purely real modes with distinct level statistics that agree with the predictions for an \emph{Hermitian} random matrix ensemble with the corresponding symmetry.

\section{Rectangular blocks and real eigenvalues in non-Hermitian random matrix theory and in the PT-symmetric SYK model}
\label{topo}

In the previous section, we have shown that the symmetry of classes AIII, BDI$^\dagger$, BDI$_{++}$, and CI$_{--}$ can be altered by the existence of rectangular blocks in the Hamiltonian characterized by $\Tr Q=\nu(N) > 0$. We have recently shown \cite{garcia2023c} that $\nu(N)$ is a topological invariant so that these four classes are topological. In this section, we will give a perturbative characterization of these symmetry classes and the differences from the $\nu = 0$ case. We will first do this in detail for class AIII$_\nu$  and then show how to adapt the discussion to the remaining classes. The presence of rectangular blocks in non-Hermitian random matrix
theory was first
found in the context of QCD at nonzero chemical potential
\cite{Osborn:2004rf,Akemann:2004dr} and for the QCD Dirac operator of Wilson fermions
\cite{Damgaard:2010cz,Akemann:2010em,Kieburg:2011uf,Akemann:2011kj,Kieburg:2013xta,Kieburg:2015vqa}.

\subsection{Class AIII$_\nu$}
The simplest example is AIII, which is well known from the analysis of Wilson fermions in QCD~\cite{Itoh:1987iy,Narayanan:1994gw,Gattringer:1997ci,Damgaard:2010cz,Akemann:2010em,Kieburg:2011uf,Akemann:2011kj,Kieburg:2013xta,Kieburg:2015vqa}.
Its block structure is
\be
H = \left( \begin{array}{cc}a A & C \\-C^\dagger & a B   \end{array}\right ),
\ee
where $A$ and $B$ are Hermitian matrices of dimension $n$ and $n+\nu$, respectively, with $\nu$ the difference between the number of rows and columns of the matrix that we term the index, $C$ is a complex $n \times(n+\nu)$ matrix, and $a$ is a real parameter that in lattice QCD is the lattice spacing. For $a=0$, the matrix $H$ is anti-Hermitian and has
$\nu$ zero modes. This ensemble will be denoted by AIII$_\nu$.

If we define the operator
\be
\gamma_c ={\rm diag}(\underbrace{1,\cdots, 1}_n,\underbrace{-1,\cdots -1}_{n+\nu})
\ee
then $\gamma_c H=H^\dagger \gamma_c$ because
\be
\left [ \gamma_c,  \left( \begin{array}{cc}a A & 0 \\0 & a B   \end{array}
  \right )\right ] =0
  \qquad \text{and}\qquad
\left \{ \gamma_c,  \left( \begin{array}{cc} 0 & C \\-C^\dagger & 0
  \end{array}
  \right )\right \} =0.
\ee
In a chiral basis $\{\phi_{j+}, \phi_{j-}\}_j$ with $\gamma_c \phi_{j\pm} = \pm
\phi_{j\pm} $, the matrix $C$ only connects states of opposite chirality (eigenvalues of $\gamma_c$). Note that $H$ has rectangular blocks when the index
$\nu=\Tr\gamma_c \ne 0$.

The operator $\gamma_c(H +m\id) $ with $m$ real is Hermitian with eigenvalue $E_j$. For $a=0$,
the spectral flow of its zero modes as a function of $m$
goes as $m$, while the nonzero modes flow as $\pm \sqrt {E_j^2 +m^2}$.
Interestingly, for $a \neq 0$, the flow lines will generically intersect the real axis at $m_j \ne  0$. As
  $a$ increases, flow lines may have one or more pairs of additional
  intersections with the real axis.
 The intersection points $m_j$ satisfy
  \be
  \gamma_c(H+m_j\id) \phi_j =0.
  \ee
  Therefore,
  \be
H \phi_j = -m_j \phi_j,
\ee
so that the intersection points of the spectral flow lines of $\gamma_c(H +m\id) $ with the real axis correspond to purely real eigenvalues of the non-Hermitian (but pseudo-Hermitian) operator $H$. Therefore, the zero modes at $a = 0$ become purely real eigenvalues at finite $a$. In addition,
it has been shown analytically in the $a \ll 1$ region that the level statistics of the $\nu > 0$ real eigenvalues originating from the zero modes at $a = 0$ have GUE statistics, while for
$\nu = 0$ the statistics of real eigenvalues is close to Poisson.
The full joint probability distribution of AIII$_\nu$ random matrices, worked out
in Ref.~\cite{Kieburg:2011uf,Akemann:2011kj,Kieburg:2013xta}, is $\nu$-dependent.

Next, we show that, to leading order in $a$, the characteristic polynomial factorizes into the product of the characteristic polynomial of an $2n\times 2n$ non-Hermitian matrix with the same anti-unitary symmetries of $H$ and the characteristic polynomial of an \emph{Hermitian} $\nu\times \nu$ matrix, which, hence, determines the statistics of the $\nu$ real eigenvalues.
In order to prove this result, we decompose
the complex matrix $C$ as
\be
\label{eq:topo_AIII_rotation_C}
C= U_1^\dagger \left (C'\  0_{n\times \nu}\right ) U_2,
\ee
where $C'$ is an $n\times n$ square complex matrix, $0_{n\times\nu}$ is a $n\times\nu$ matrix with all entries equal to zero, and $U_1$ and $U_2$ are unitary matrices of dimension $n$ and $n+ \nu$, respectively. Writing also $A=U_1^\dagger A' U_1$ and
\begin{equation}
\label{eq:topo_AIII_rotation_B}
aB=U_2^\dagger \begin{pmatrix}
a B' & a f \\ a f^\dagger & a b
\end{pmatrix} U_2,
\end{equation}
where $A'$, $B'$, and $b$ are Hermitian matrices of dimension $n$, $n+\nu$, and $\nu$, respectively, and $f$ is an $n\times \nu$ complex matrix, the Hamiltonian acquires the form
\be
H = \begin{pmatrix}
	U_1^\dagger & 0 \\ 0 & U_2^\dagger
\end{pmatrix}
\left(\begin{array} {ccc}
	aA' & C' & 0 \\
	-{C'}^\dagger & a B' &af\\
	0 & af^\dagger& ab
\end{array}\right )
\begin{pmatrix}
	U_1 & 0 \\ 0 & U_2
\end{pmatrix}.
\ee
To order $a$, the secular equation factorizes as
\be
\label{eq:topo_AIII_factorization}
\det( H - \zeta \id) = \det\left[
\left( \begin{array}{cc}a A' & C' \\{-C'}^\dagger & a B'   \end{array}\right )
-\zeta \id \right ] \det[ a b- \zeta \id].
\ee
After the unitary rotation \eref{eq:topo_AIII_rotation_B}, the Hermitian matrix $b$ is a block of the matrix $B$.  Its eigenvalues are continuously connected
to the zero modes of $\gamma_c H$ at $a=0$.
The matrix $b$  has no symmetries beyond being Hermitian, and thus belongs to Hermitian class A (GUE). The local level statistics of its $\nu$ eigenvalues (the $\nu$ real eigenvalues of $H$) are those of the GUE, both in the bulk and at the origin.
If $B$ is taken to be a random matrix, then so is $b$, and we expect the real eigenvalues to have a semicircular distribution.
The matrix in the first determinant in Eq.~\eref{eq:topo_AIII_factorization}
is a $2n\times2n$ non-Hermitian matrix with the same symmetries of $H$ but only square blocks. In general, this matrix will also have real eigenvalues, but with different spectral correlations, which were studied in Ref.~\cite{Shindou2022}.

Next, we show that, for the parameters corresponding to the AIII$_\nu$ universality class, these RMT results are also a feature of the SYK model studied in this paper. As a first step, we split $H_0$ Eq.~(\ref{eq:h0}) into its Hermitian and anti-Hermitian parts, $H_{0J}$ and $H_{0M}$, respectively:
\be
H_0 = H_{0J}+H_{0M}.
\ee
The operator $Q$ takes the role of $\gamma_c$ in RMT, since (recall that $q/2$ is even for an SYK in class AIII$_\nu$)
\be
[Q, H_{0J}+ H_I]=0  \qquad \text{and}\qquad \{ Q,H_{0M} \} = 0.
\ee
The operator $ H_{0J}+ H_I$ does not mix states with different $Q$ quantum numbers (i.e., eigenvalues $k$), while $H_{0K}$ only couples states with opposite $k$.
The rectangular structure arises when the index
\be
  \nu(N) =\Tr \mathbb{P}_S Q \ne 0.
  \ee
This is the case in the sector $s=1$ (when $k=\pm1$), for both odd and even $r$. In the former case, $S_L$ and $S_R$ are individually conserved and an additional projection, say with $\mathbb{P}_L$, is required.
  By a careful analysis of the states, $\nu$ is given by Eqs.~(\ref{eq:nu_S}) and (\ref{eq:nu_LR}).
  Using this result, where $Q$ plays the role of $\gamma_c$ in the AIII$_\nu$ random matrix ensemble above, the SYK Hamiltonian with
  $\kappa=1$
and $\lambda=0$ has $\nu(N)=2^{N/2}/2$ zero modes. At small but nonzero $\lambda$ and $1-\kappa$, equivalent to small $a$ in the AIII$_\nu$  random matrix above, these zero modes become real modes that have GUE statistics.

As was mentioned earlier, in the analysis of the PT-symmetric SYK models
we have found three more non-Hermitian ensembles
with an index $\nu$: BDI$^\dagger_\nu$, BDI$_{++\nu}$, and C$_{--\nu}$.
We have shown recently~\cite{garcia2023c} that this index $\nu$ is indeed topological.
 In the following, we adapt the analysis of AIII$_\nu$ to each of the three remaining topological classes. The mapping to the SYK results proceeds as before.

\subsection{Class BDI$^\dagger_\nu$}
\label{sec:topo_BDIdag}

For class BDI$^\dagger_\nu$, the block structure is still given by
\begin{equation}
H=\begin{pmatrix}
a A & C \\ -C^\dagger & a B
\end{pmatrix},
\end{equation}
but with $A$, $B$, and $C$ real ($A$ and $B$ are also still Hermitian, and all matrices have the same dimension as before). Performing exactly the same procedure as for AIII$_\nu$, we arrive again at Eq.~(\ref{eq:topo_AIII_factorization}), but where $b$ is now a real symmetric matrix, with no further symmetries, i.e., it belongs to Hermitian class AI (GOE). The $\nu$ real eigenvalues are thus correlated according to the GOE, both in the bulk and near the origin. If $B$ is a random matrix, $b$ belongs to the GOE and the eigenvalues are distributed according to the semicircle distribution.

\subsection{Class BDI$_{++\nu}$}

For class BDI$_{++\nu}$, the block
structure is given by
\begin{equation}
H=\begin{pmatrix}
& & a A & B \\
& & C & a D \\
aA^\dagger & -C^\dagger & & \\
-B^\dagger & aD^\dagger & &
\end{pmatrix}
\equiv
\begin{pmatrix}
& \mathcal{H}_1\\
\mathcal{H}_2
\end{pmatrix}
\end{equation}
where the matrices $A$, $B$, $C$, and $D$ are real with size
$n\times n$,  $n\times (n+\nu)$, $(n+\nu)\times n$ and
$(n+\nu)\times (n+\nu)$, respectively. The Hamiltonian  $H$ has a chiral structure and its characteristic polynomial reads:
\begin{equation}
\begin{split}
  \det\left( H-\zeta\id \right)
&=\det\( \zeta^2-\mathcal{H}_1\mathcal{H}_2\)
\\
&=(-1)^\nu \det\(\mathcal{H}_1\mathcal{H}_2-\zeta^2\id\).
\end{split}
\end{equation}
We have
\begin{equation}
\mathcal{H}_1\mathcal{H}_2 = \begin{pmatrix}
a^2 A A^\dagger-B B^\dagger & -a (A C^\dagger-B D^\dagger)\\
a (A C^\dagger-B D^\dagger)^\dagger & a^2 D D^\dagger-C C^\dagger
\end{pmatrix},
\end{equation}
where $AA^\dagger$ and $BB^\dagger$
are $n\times n$ matrices, $AC^\dagger$ and $B D^\dagger$ are $n\times(n+\nu)$ matrices, and $DD^\dagger$ and $CC^\dagger$ are $(n+\nu)\times (n+\nu)$ matrices.
We now modify the procedure of AIII$_\nu$ as follows. First, we introduce an $(n+\nu)\times (n+\nu)$ unitary $V$ such that
\begin{equation}
CC^\dagger=V^\dagger\begin{pmatrix}
C' C'^\dagger & 0_{n\times \nu} \\
0_{\nu\times n} &0_{\nu\times \nu}
\end{pmatrix} V,
\end{equation}
where $C'$ is an $n\times n$ matrix, and then, as in the AIII$_\nu$ case,
an  $n\times n$ unitary matrix $U_1$ and an $(n+\nu)\times (n+\nu)$
unitary matrix $U_2$ such that
\begin{equation}
-(AC^\dagger-BD^\dagger)V^\dagger= U_1^\dagger \left (G\  0_{n\times \nu}\right ) U_2,
\end{equation}
where $G$ is an $n\times n$ matrix.
We further write $a^2 AA^\dagger -BB^\dagger=U_1^\dagger F U_1$ and
\begin{equation}
U_2VD=\begin{pmatrix}
d_1 & d_2 \\ d_3 & d_4
\end{pmatrix}
\implies
U_2VDD^\dagger V^\dagger U_2^\dagger =\begin{pmatrix}
d_1d_1^\dagger +d_2 d_2^\dagger & d_1 d_3^\dagger + d_2 d_4^\dagger \\
d_3 d_1^\dagger +d_4 d_2^\dagger & d_3 d_3^\dagger+d_4 d_4^\dagger
\end{pmatrix}
\equiv
\begin{pmatrix}
D'D'^\dagger & f \\ f^\dagger & dd^\dagger
\end{pmatrix},
\end{equation}
where $d_1$, $d_2$, $d_3$, and $d_4$ are
$n\times n$, $n\times \nu$, $\nu\times n$,
$\nu\times\nu$ matrices, respectively.
Putting everything together, we obtain
\begin{equation}
\mathcal{H}_1\mathcal{H}_2 = \begin{pmatrix}
U_1^\dagger & 0 \\ 0 & V^\dagger U_2^\dagger
\end{pmatrix}
\begin{pmatrix}
F & a G & 0 \\
-a {G}^\dagger & a^2 D'D'^\dagger - C'C'^\dagger  & a^2 f\\
0 & a^2 f^\dagger & a^2 dd^\dagger
\end{pmatrix}
\begin{pmatrix}
U_1 & 0 \\ 0 & U_2 V
\end{pmatrix},
\end{equation}
and, hence, to leading order in $a$, we have the factorization of the characteristic polynomial
\begin{equation}
\label{eq:topo_BDI++_factorization}
\det\left( H-\zeta\id \right)=
(-1)^\nu\det\left[\begin{pmatrix}
F & a G \\
-a {G}^\dagger & a^2 D'D'^\dagger - C'C'^\dagger
\end{pmatrix} -\zeta^2\id\right]
\det\left[a^2 dd^\dagger - \zeta^2\id\right].
\end{equation}
Therefore, the squares of the topological real eigenvalues coincide with the eigenvalues of the matrix $dd^\dagger$ with $d$ real, which belongs to Hermitian class BDI (chGOE). In the bulk, its eigenvalues are correlated according to the GOE, but there is chGOE universality in the microscopic regime.

\subsection{Class CI$_{--\nu}$}
Finally, we consider the case of CI$_{--\nu}$, whose block structure is given by
\begin{equation}
H=\begin{pmatrix}
& & a A & B \\
& & -B^\top & a D \\
aA^\dagger & -B^* & & \\
-B^\dagger & aD^\dagger & &
\end{pmatrix},
\end{equation}
where the dimensions of $A$, $B$, and $D$ are the same as for BDI$_{++\nu}$, but $A$ and $D$ are now complex symmetric and $B$ is arbitrary. The computation follows exactly the same steps as for BDI$_{++}$, with the replacement $C\to-B^\top$. We arrive at the factorized characteristic polynomial Eq.~(\ref{eq:topo_BDI++_factorization}), but where $d$ is now a complex symmetric matrix. Correspondingly, $dd^\dagger$ belongs to the Hermitian class CI. The eigenvalues of $H$ have GOE correlations in the bulk and CI correlations near the origin, namely, eigenvalues close to zero.

\section{Comparison between SYK and RMT: level statistics and eigenvector overlaps}
\label{sec:level_statistics}

In order to confirm the symmetry classification proposed in Sec.~\ref{sec:classification}, we now carry out a comparison between the spectral and eigenvector properties of the two-site SYK model Eq.~(\ref{eq:def_H}) for the reported fourteen symmetries classes and the predictions of RMT in each case.
The block structure of the 14
random matrix ensembles is shown in Table~\ref{tab:RMT_realization}. This
allows for an easy comparison with the corresponding RMT,
which is obtained by replacing the blocks in Table~\ref{tab:RMT_realization} by random matrices with Gaussian distributed entries subjected to the constraints mentioned in the table. Each of the 14 classes is realized by the SYK model
for a suitable choice of its parameters, see Table~\ref{tab:full_classification}. In each of these cases, we confirm the predicted agreement with the level statistics of the RMT ensemble with the corresponding symmetry.

\newcommand\tsp{\rule{0pt}{2em}}         
\newcommand\bsp{\rule[0.9em]{0pt}{0pt}}   

\begin{table}[tbp]
	\caption{Matrix structure corresponding to the 14 symmetry classes realized in the PT-symmetric SYK model. The letters A, B, C, and D in the second column denote complex random matrices subject to the listed constraints.
		The anti-unitary and involutive operators used to
		construct the block structure are listed in the third column. In the case of four anti-unitary
		symmetries we give only three of them --- the fourth one is given by their product. We indicate with an index $\nu$ the classes that allow
		for a rectangular block structure.
		However, in some cases (BDI$_\nu$ and CI$_{+-\nu}$), while $\nu\neq0$ is allowed in principle,
		the  SYK model only realizes the  $\nu=0$ case.  }
	\label{tab:RMT_realization}
	\begin{tabular}{|c|c|c|}
		\hline
		Class & Matrix Realization& Symmetry operator \\
		\hline\tsp
		A & A &\\[1em]
		\hline\tsp
		AI$^\dag$ &A,  A$=$A$^\top$ &$\sC_+=K$ \\[1em]
		\hline\tsp
		AI & A,  A$=$A$^*$
		& $\sT_+=K$\\[1em]
		\hline
		\tsp
		AIII$_\nu$ &   $
		\begin{pmatrix}
		\mathrm{A} & \mathrm{C} \\
		\mathrm{-C^\dagger} & \mathrm{B}
		\end{pmatrix}$,
		\;  \dob $\mathrm{A}=\mathrm{A}^\dag$, $\mathrm{B}=\mathrm{B}^\dag$\\
		\edob
		& $\sQ_+ =\sigma_z $
		\\[1em] \hline
		\tsp
		D & \dob $\rA$,\; $\rA=-\rA^{\top}$ \\
		\edob  & $\sC_-=K$\\[1em]
		\hline
		\tsp
		C &$
		
		\begin{pmatrix} \rA & \rB \\ \rC & -\rA^{\top} \end{pmatrix} $,
		
		\; \dob $\rB=\rB^{\top},\;\rC=\rC^{\top}$ \\
		\edob &
		$\sC_-=i\sigma_y K$
		\\[1em]
		\hline \tsp
		BDI$_\nu$ & $\begin{pmatrix}
		\mathrm{A} & \mathrm{B} \\
		\mathrm{B^\dagger} & \mathrm{C}
		\end{pmatrix}$,\;  \dob  $\rA= -\rA^\top=-\rA^\dagger,\;\rB^*=-\rB,\;
		\rC=-\rC^\top=-\rC^\dagger$,\;\\
		
		\edob  &$\left . \begin{array}{l}\sT_+=\sigma_z K,\\ \sC_-= K \end{array} \right.$
		\\[1em]\hline
		\tsp
		CI & $ \begin{pmatrix}
		\mathrm{A} & \mathrm{B} \\
		\mathrm{C} & -\mathrm{A}^\dagger
		\end{pmatrix}$,\; \dob $\rA=\rA^*,\;\rB=-\rB^*=-\rB^\dagger,\;
		\rC=-\rC^* =-\rC^\dagger $
		\\
		
		\edob
		&$\left . \begin{array}{l}\sT_+=\sigma_z K,\\ \sC_-=i\sigma_y K\end{array} \right .$  \\[1em]
		\hline \tsp
		BDI$^{\dag}_{\nu}$ & $
		\begin{pmatrix}
		\mathrm{A} & \mathrm{B} \\
		\mathrm{-B}^\dagger & \mathrm{C}^*
		\end{pmatrix}$,\; \dob  $\rA= \rA^\top=\rA^\dagger,\;
		\rB=\rB^*,\; \rC= \rC^\top=\rC^\dagger$\; \\
		
		\edob &$\left . \begin{array}{l}\sT_+=K,\\ \sC_+=\sigma_z K \end{array} \right .$  \\[1em]
		\hline \tsp
		CI$^\dag$ & $
		\begin{pmatrix}
		\mathrm{\rA} & \mathrm{B} \\
		\mathrm{B}^* & -\rA^*
		\end{pmatrix}$,\; \dob $ \rA=\rA^\top,\; \rB= \rB^\dagger$,\; \\
		\edob
		&$\left . \begin{array}{l}\sT_- =i\sigma_y K, \\\sC_+= K\\ \end{array} \right .$ \\[1em]
		\hline\tsp
		BDI$_{-+}$ & $\begin{pmatrix}
		\mathrm{0} & \mathrm{B} \\
		\mathrm{B}^\top & \mathrm{0}
		\end{pmatrix}$, \; \dob $\rB=\rB^\dagger$ \\
		\edob
		&$\left . \begin{array}{l}\sT_+=\sigma_x K,\\ \sT_-=i\sigma_y K, \\
		\sC_+=K\end{array} \right .$   \\[1em]
		\hline \tsp
		BDI$_{++\nu}$ & $
		\left (  \begin{array}{cccc}
		0&0&          \rA & \rB \\
		0&0&    \rC& \rD\\
		\rA^\top & \rC^\top & 0 &0\\
		\rB^\top & \rD^\top &0&0\end{array} \right )$,\;  \dob $\rA=\rA^*,\;\rB =-\rB^*,\;\rC=-\rC^*,\;\rD=\rD^*$,\\
		\edob
		&$ \left . \begin{array}{l}  \sT_+=  (\unit \otimes \sigma_z) K, \\
		\sC_-=(\sigma^z\otimes\unit)K, \\
		\sC_+=K\end{array} \right . $
		\\[1em]\hline
		\tsp
		CI$_{+-\nu}$ & $\begin{pmatrix}
		0 & \mathrm{A} \\
		\mathrm{B} & 0
		\end{pmatrix}$,\; \dob $\rA=\rA^*=\rA^\top,\; \rB=\rB^*=\rB^\top$ \\
		\edob   &
		$\left . \begin{array}{l}\sT_+=K,\\ \sC_-=i\sigma_y K,\\ \sC_+=\sigma_x K \end{array} \right .$   \\[1em]
		\hline \tsp
		CI$_{--\nu}$ & $ \left (\begin{array}{cccc}
		0&0&          \rA & \rB \\
		0&0&   -\rB^\top & \rC\\
		\rA^* & \rB^* & 0 &0\\
		-\rB^\dagger & \rC^* &0&0
		\end{array} \right )$,\; \dob $ \rA^\top = \rA, \;  \rC^\top = \rC $ \\
		\edob
		& $
		\left . \begin{array}{l}  \sT_+= (\sigma_x \otimes \unit) K,\\
		\sT_-=(i\sigma_y\otimes \unit) K,\\
		\sC_+=(\sigma_x \otimes \sigma_z)  K \end{array} \right . $
		\\[1em]\hline
	\end{tabular}
\end{table}

\begin{table}[htbp!]
	\caption{Numerical confirmation of the 14 universal classes realized in the PT-symmetric non-Hermitian SYK model Eq.~(\ref{eq:def_H}).
		The entries marked with ``$/$'' indicate that we have checked that quantity, but it does not display special properties.
		We first classify all cases into three classes using $\sC_+^2$, using the average complex spacing ratio $\langle z\rangle$.
		Second, we use the global symmetry of the spectrum to distinguish different reflection symmetries.
		Third, we compute the eigenvector overlaps $\mathcal{O}_{ab}$ and $\omega_{\z\z}$,
		where $a,\; b\in \{ \z, \z^*, -\z, -\z^* \}$ and $\z$ are complex eigenvalues of $H$, to uniquely determine the symmetry class. We note that the last eight lines of the table correspond to only four SYK Hamiltonians thus illustrating that different sectors of a given Hamiltonian, in this case labeled by the eigenvalues of the parity operator, can have different symmetries.}
	\label{tab:full_classification}
	\begin{tabular}{|c|c|c|c|c|c|c|c|c|}
		\hline Example &Class & $\sT_{+}^2$ & $\sC_{-}^2$ & $\sC_{+}^2$ & $\sT_{-}^2$& \makecell[c]{CSR\\$\langle z\rangle$}&\makecell[l]{Spectrum\\ Symmetry} & \makecell[c]{Eigenvector \\Overlap}\\
		\hline \makecell[l]{$N=14, q=6, r=2, \alpha=1.1$,\\  $S_L=-S_R=1, \lambda=0.424, \kappa=0.5$}&$\mathrm{A}$ & 0 & 0 & 0 & 0 &\makecell[c]{$\mathrm{A}$\\0.738}&/&/\\
		\hline \makecell[l]{$N=14, q=8, r=2, \alpha=1$,\\  $S_L=S_R=-1, \lambda=0.141, \kappa=0.5$}&$\mathrm{AIII}_\nu$ & 0 & 0 & 0 & 0 &\makecell[c]{$\mathrm{A}$\\0.738}&$D_1$(Re)&$\omega_{\z\z}\ne 0$ \\
		\hline \makecell[l]{$N=16, q=6, r=1, \alpha=1.1$,\\  $S=-1, \lambda=0.113, \kappa=0.5$}&$\mathrm{D}$ & 0 & $+1$ & 0 & 0 &\makecell[c]{$\mathrm{A}$\\0.738}&$C_\pi$&\makecell[l]{$\mathrm{Im}(\mathcal{O}_{-\z \z})=0$\\ $\mathrm{Re}(\mathcal{O}_{-\z \z})>0$}\\
		\hline \makecell[l]{$N=14, q=6, r=1, \alpha=1.1$,\\  $S=-1, \lambda=0.0424, \kappa=0.5$}&$\mathrm{C}$ & 0 & $-1$ & 0 & 0 &\makecell[c]{$\mathrm{A}$\\0.738}&$C_\pi$&\makecell[l]{$\mathrm{Im}(\mathcal{O}_{-\z \z})=0$\\ $\mathrm{Re}(\mathcal{O}_{-\z \z})<0$}\\
		\hline \makecell[l]{$N=16, q=6, r=1, \alpha=1$,\\  $Q=-1, \lambda=0.12, \kappa=1$}&$\mathrm{BDI}$ & $+1$ & $+1$ & 0 & 0 &\makecell[c]{$\mathrm{A}$\\0.738}&$D_2$&\makecell[l]{$\mathrm{Im}(\mathcal{O}_{-\z \z})=0$\\ $\mathrm{Re}(\mathcal{O}_{-\z \z})>0$} \\
		\hline \makecell[l]{$N=14, q=4, r=1, \alpha=1.1$,\\  $S=-1, \lambda=0.15, \kappa=1$}&$\mathrm{CI}^{\dagger}$ & 0 & 0 & $+1$ & $-1$ &\makecell[c]{$\mathrm{AI^\dag}$\\0.719}&$D_1$(\Im)&$\mathcal{O}_{-\z^* \z}=0$\\
		\hline \makecell[l]{$N=16, q=8, r=2, \alpha=1$,\\  $S_L=-S_R=1, \lambda=0.32, \kappa=1$}&$\mathrm{AI}^{\dagger}$ & 0 & 0 & $+1$ & 0 &\makecell[c]{$\mathrm{AI}^\dag$\\0.720}&/&/\\
		\hline \makecell[l]{$N=16, q=8, r=2, \alpha=1$,\\  $S_L=S_R=-1, \lambda=0.32, \kappa=1$}&$\mathrm{BDI}^{\dagger}_\nu$ & $+1$ & 0 & $+1$ & 0 &\makecell[c]{$\mathrm{AI^\dag}$\\0.720}&$D_1$(Re)&/\\
		\hline \makecell[l]{$N=14, q=6, r=1, \alpha=1$,\\  $Q=i, \lambda=0.06, \kappa=1$}&$\mathrm{AI}$ & $+1$ & 0 & 0 & 0&
		\makecell[c]{$\mathrm{A}$\\0.737}&$D_1$(Re)&$  \omega_{\z\z}=0$\\
		\hline \makecell[l]{$N=14, q=6, r=1, \alpha=1$,\\  $Q=-1, \lambda=0.1, \kappa=1$}&$\mathrm{CI}$ & $+1$ & $-1$ & 0 & 0 &\makecell[c]{$\mathrm{A}$\\0.737}&$D_2$&\makecell[l]{$\mathrm{Im}(\mathcal{O}_{-\z \z})=0$\\ $\mathrm{Re}(\mathcal{O}_{-\z \z})<0$}\\
		\hline \makecell[l]{$N=16, q=4, r=1, \alpha=1$,\\  $S=1, \lambda=0.06, \kappa=1$}&$\mathrm{BDI}_{++\nu}$ & $+1$ & $+1$ & $+1$ & $+1$ &\makecell[c]{$\mathrm{AI^\dag}$\\0.719}&$D_2$&\makecell[l]{$\mathrm{Im}(\mathcal{O}_{-\z \z})=0$\\ $\mathrm{Re}
			(\mathcal{O}_{-\z \z})>0$}\\
		\hline \makecell[l]{$N=16, q=4, r=1, \alpha=1$,\\  $S=-1, \lambda=0.06, \kappa=1$}&$\mathrm{CI}_{+-}$ & $+1$ & $-1$ & $+1$ & $+1$ &\makecell[c]{$\mathrm{AI^\dag}$\\0.718}&$D_2$&\makecell[l]{$\mathrm{Im}(\mathcal{O}_{-\z \z})=0$\\ $\mathrm{Re}
			(\mathcal{O}_{-\z \z})<0$}\\
		\hline \makecell[l]{$N=14, q=4, r=1, \alpha=1$,\\  $S=1, \lambda=0.12, \kappa=1$}&$\mathrm{CI}_{--\nu}$ & $+1$ & $-1$ & $+1$ & $-1$ &\makecell[c]{$\mathrm{AI^\dag}$\\0.720}&$D_2$&\makecell[l]{$\mathrm{Im}(\mathcal{O}_{-\z \z})=0$\\
			$\mathrm{Re}(\mathcal{O}_{-\z \z})<0$
			\\ $\mathcal{O}_{-\z^* \z}=0$}\\
		\hline \makecell[l]{$N=14, q=4, r=1, \alpha=1$,\\  $S=-1, \lambda=0.2, \kappa=1$}&$\mathrm{BDI}_{-+}$ & $+1$ & $+1$ & $+1$ & $-1$ &\makecell[c]{$\mathrm{AI^\dag}$\\ 0.720}&$D_2$&\makecell[l]{$\mathrm{Im}(\mathcal{O}_{-\z \z})=0$\\
			$\mathrm{Re}(\mathcal{O}_{-\z \z})>0$
			\\ $\mathcal{O}_{-\z^* \z}=0$}\\
		\hline
	\end{tabular}
\end{table}

\subsection{Bulk level fluctuations}
We first consider the level statistics in the bulk of the complex spectrum, which is only sensitive~\cite{garcia2022d,sa2022a} to the sign of $\sC_+^2$, namely, statistics of class A (no $\sC_+$), AI$^\dag$ ($\sC_+^2=1$), and AII$^\dag$ ($\sC_+^2=-1$). In our classification scheme, we do not have classes with $\sC_+^2=-1$ and, hence, no AII$^\dagger$ statistics, so the 14 universality classes are split into two groups, A and AI$^\dagger$, with seven elements each:
\par (a) Broken $\sC_+$: A, AIII, AI, D, C, BDI, and CI;
\par (b) $\sC_+^2=1$: AI$^\dagger$, BDI$^{\dagger}$, CI$^{\dagger}$, BDI$_{++}$, BDI$_{-+}$, CI$_{+-}$, and CI$_{--}$.

\begin{figure}[t!]
	\centering
	\subfigure[]{\includegraphics[width=7cm]{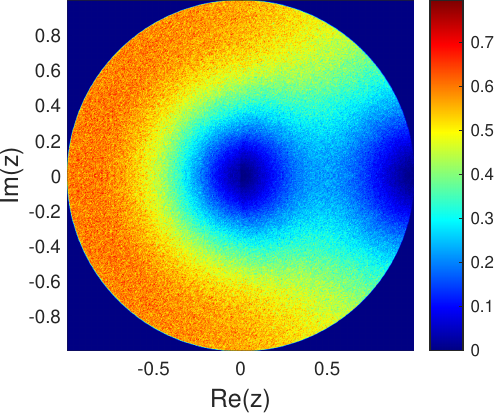}}
	\subfigure[]{\includegraphics[width=7cm]{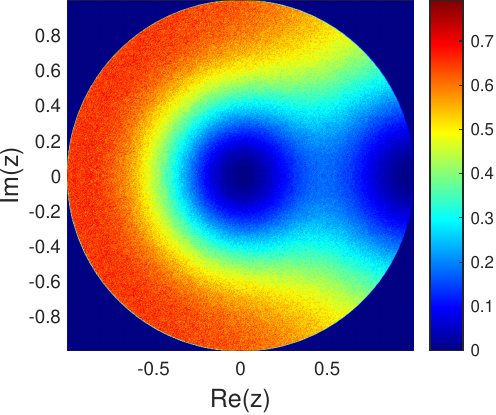}}
	\\
	\subfigure[]{\includegraphics[width=7cm]{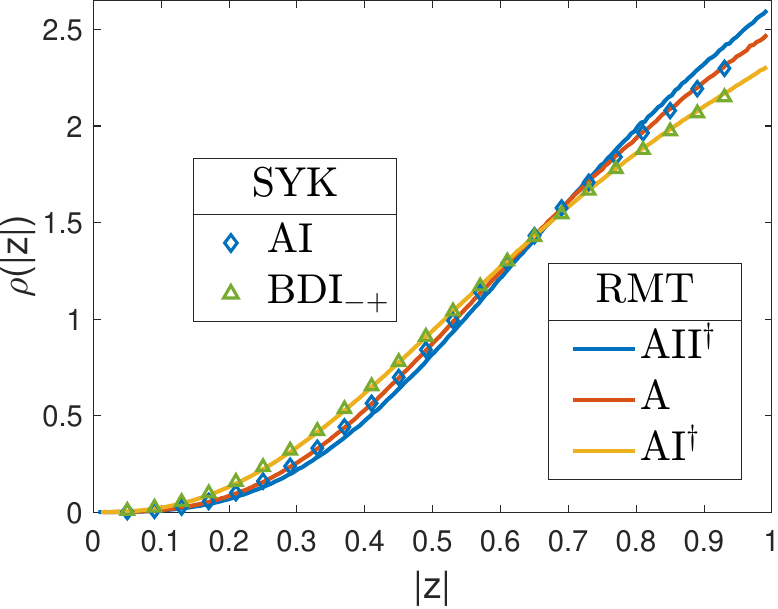}}
	\subfigure[]{\includegraphics[width=7cm]{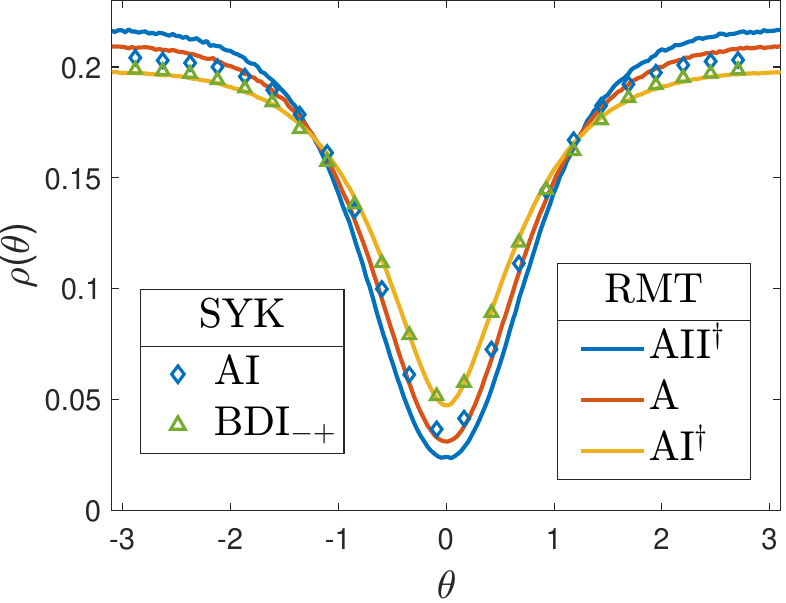}}
	\caption{Complex spacing ratios, Eq.~(\ref{eq:csr1}), of  the spectrum of the PT symmetric SYK model Eq.~(\ref{eq:def_H}). In the upper row, we show the distribution of the complex spacing ratio $z$ in the complex unit disk for $N=14,\; \kappa=1,\;r=1,\;\alpha=1$, and $q=4$ (a)
		or $q=6$ (b). From Table~\ref{tab:full_classification}, when $q/2$ is odd, there exists no $\sC_+$ symmetry, corresponding to class AI (bulk class A); when $q/2$ is even, $\sC_+^2=1$, corresponding to class BDI$_{-+}$ (bulk class AI$^\dagger$). In the lower row, we show
		the angular (c) and radial (d) distributions of $z$ for cases (a) and (b). We compare the SYK results (points) to the RMT predictions
		(curves) of bulk classes $\mathrm{A}$, $\mathrm{AI}^\dag$, and $\mathrm{AII}^\dag$. The agreement with the random matrix result of the predicted class is excellent.
		For $q=4$, we employ $5\times 10^3$ disorder realizations, $ \lambda=0.2$, and $s=-1$, while for $q=6$ we have $10^4$ realizations, $\lambda=0.06$ and $k=\i$.
	}
	\label{fig:Gapz}
\end{figure}

We probe the bulk spectral correlations by computing the complex spacing ratios (CSRs)~\cite{sa2019PRX}:
\begin{equation}
	z_i=\frac{\zeta_i^{\mathrm{NN}}-\zeta_i}{\zeta_i^{\mathrm{NNN}}-\zeta_i}, \label{eq:csr1}
\end{equation}
 where $\zeta_i$ denotes the $i$th complex eigenvalue of the Hamiltonian Eq.~(\ref{eq:def_H}) and NN/NNN refers to the nearest and next-to-nearest neighboring eigenvalues with respect to $\zeta_i$.
The complex eigenvalues were obtained numerically by exact diagonalization techniques. For a given set of parameters, we obtained a minimum of $5\times 10^5$ eigenvalues.
By definition, the CSRs are distributed inside the complex unit circle.
For uncorrelated eigenvalues, corresponding to Poisson statistics, the distribution is uniform. By contrast, as shown in Fig.~\ref{fig:Gapz}, the CSR distribution has a characteristic half-eaten doughnut shape
with a radial and angular distribution that fully characterizes the statistics of classes A and AI$^\dag$.

In order to proceed, we first compare the average CSR, $\langle z\rangle$,
with the RMT predictions $\langle z\rangle_\mathrm{A} = 0.7384$, $\langle z\rangle_{\mathrm{AI}^\dag} = 0.7222$, $\langle z\rangle_{\mathrm{AII}^\dag}= 0.7486$, and $\langle z\rangle_\mathrm{Poisson} = 2/3$~\cite{garcia2022d}.
As is shown in Table~\ref{tab:full_classification}, the numerical average of the bulk CSR agrees with the RMT prediction of either A or AI for all 14 classes.
Second, we also compare the RMT predictions with the full radial and angular distributions of the CSR, $\rho(|z|)$ and $\rho(\theta(z))$,  where $z=|z|e^{i\theta(z)}$, see Figs.~\ref{fig:Gapz}(c) and (d), respectively.
The agreement is also excellent in both cases with only small deviations near $\theta(z)=\pi$, likely due to finite-size effects.
We note that
 the coupling strength $\lambda$ cannot be set too small because the spectral degeneracy at $\lambda = 0$ is not fully lifted for finite system sizes, but not too large either because
 the dynamics is then controlled by the nonrandom intersite interaction.
 We also note that in the analysis of the bulk spectral correlations
we have excluded the spectral region on the real axis because, as we show later, real eigenvalues have qualitatively different spectral correlations.

In summary, in all cases, the bulk spectral correlations agree with the RMT prediction according to the predicted symmetry classification, summarized in Table \ref{tab:full_classification}.
Therefore, assuming that the BGS conjecture applies to this non-Hermitian setting in all cases, something which is not yet entirely clear \cite{garcia2023a}, the dissipative dynamics is quantum chaotic.
However, because bulk spectral correlations only distinguish two universality classes, this does not fully confirm the validity of our classification, which has 14 classes. In the next section, we will characterize these additional universality classes by a combination of spectral symmetries
and eigenvector overlaps.

\subsection{Spectral symmetry and eigenvector overlaps}

\begin{figure}[t!]
	\centering
	\subfigure[]{\includegraphics[width=7cm]{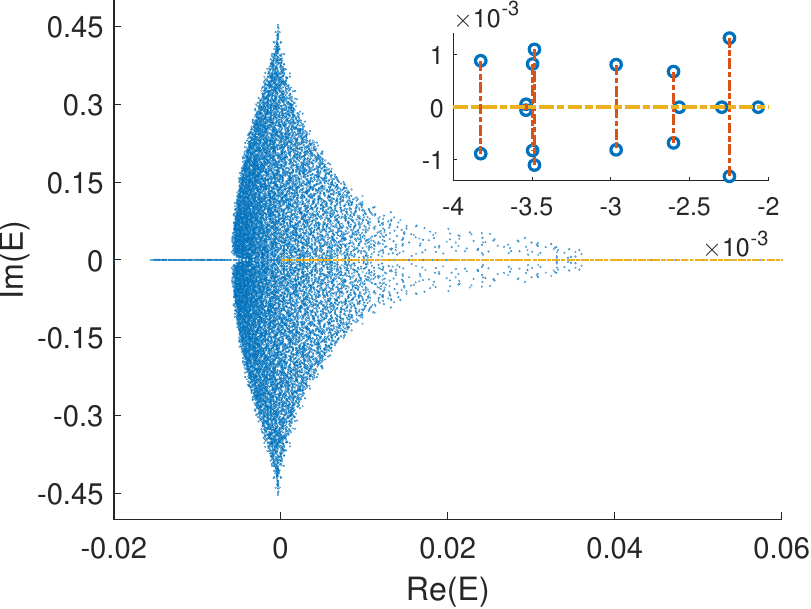}}
	\subfigure[]{\includegraphics[width=7cm]{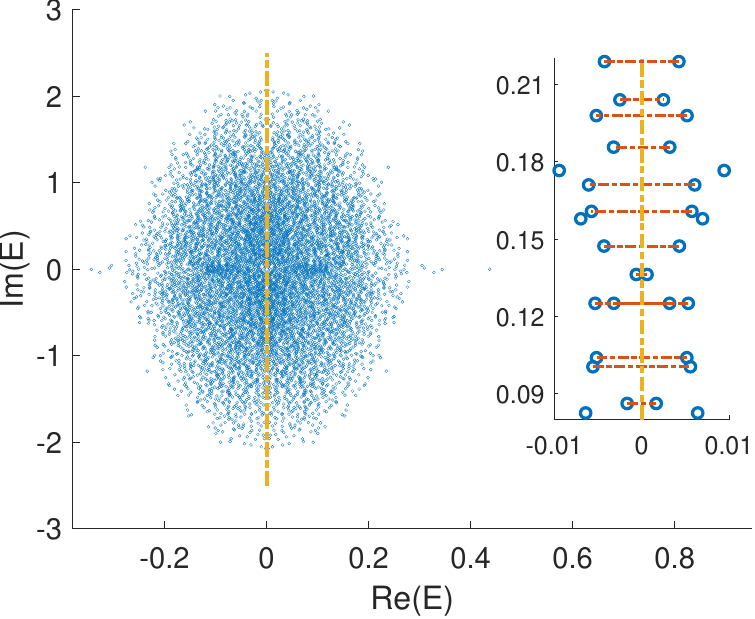}}
	\\
	\subfigure[]{\includegraphics[width=7cm]{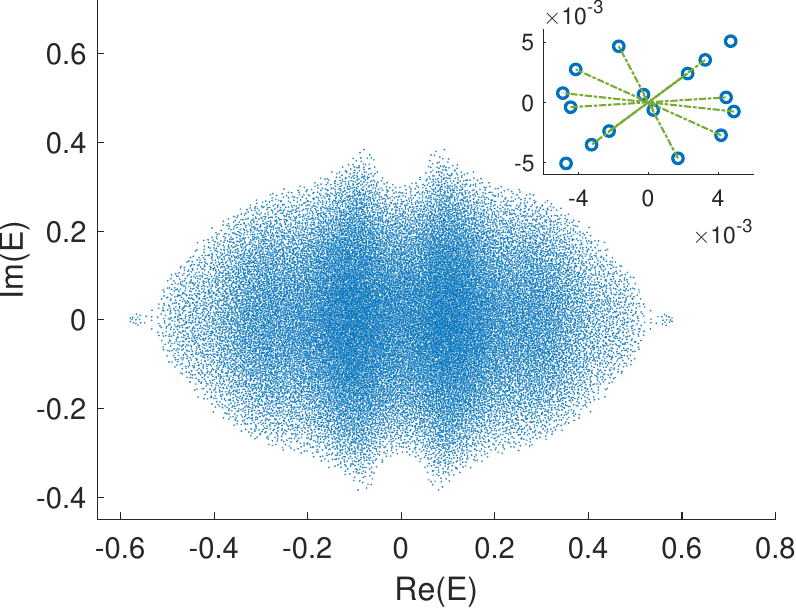}}
	\subfigure[]{\includegraphics[width=7cm]{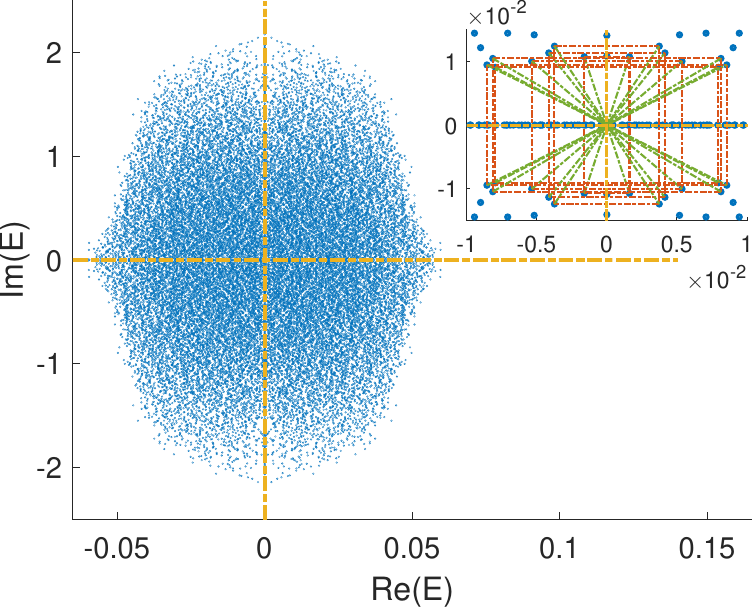}}
	\caption{Global spectral symmetry for a single disorder realization. The yellow dashed-dotted lines in the main plots denote the axes of the reflection symmetry and the insets show a magnification around these axes. In every inset, pairs of levels related by reflection and point symmetry are connected by dashed-dotted lines of orange and green respectively. The parameters of the Hamiltonian Eq.~(\ref{eq:def_H}) for the
		four figures are:
		(a) $N=16$, $q=8$, $r=2$, $\kappa=1$, $\lambda=0.32$, $S_L=S_R=-1$, $\alpha=1$ (BDI$^\dagger$),
		(b) $N=14$, $q=4$, $r=1$, $\kappa=1$, $\lambda=0.15$, $S=-1$, $\alpha=1.1$ (CI$^\dagger$),
		(c) $N=16$, $q=6$, $r=1$, $\kappa=0.5$, $\lambda=0.113$, $S=-1$, $\alpha=1.1$ (D),
		(d) $N=16$, $q=4$, $r=1$, $\kappa=1$, $\lambda=0.06$, $r=1$, $S=1$, $\alpha=1$ (BDI$_{++}$).}
	\label{fig:SpSym}
\end{figure}

In order to fully identify each of the universality classes, we have to go beyond the bulk level statistics studied in the previous section.
For that purpose, we first use the global spectral symmetry~\cite{sa2022a} to help identify the symmetry in each case. The spectral symmetries given in the fourth column
of Table \ref{eq:nHsym} are shown in Table~\ref{tab:full_classification} in terms of the corresponding
discrete symmetry group. The spectrum can either have a reflection symmetry across the real axis ($\sT_+$ and $\sQ_+$), denoted as $D_1(\Re)$; the imaginary axis ($\sT_-$ and $\sQ_-$), denoted $D_1(\Im)$; or the origin ($\sC_-$), denoted as $C_\pi$. If any two of these symmetries is present, the third automatically also is, and the spectrum has dihedral symmetry $D_2$.
In Fig.~\ref{fig:SpSym}, we give four examples of SYK models belonging
to classes BDI$^\dag$, CI$^\dag$, D, and BDI$_{++}$, see
Table~\ref{tab:full_classification} for the corresponding symmetries.
In the inset of each figure, we highlight the symmetry relation by drawing lines between eigenvalues connected by the global symmetries.

Since the bulk correlations and the spectral symmetries do not yet fully resolve the 14 classes, we use the eigenvector overlap method~\cite{sa2022a} to further distinguish them.
For non-Hermitian operators,
the left and right eigenvectors are defined by
\begin{equation}
\begin{aligned}
  H \phi_\z^R=\z \phi_\z^R, \qquad \text{and} \qquad
  {\phi_\z^L}^* H =\z {\phi_\z^L}^*,
\end{aligned}
\end{equation}
respectively.
The left eigenvector can also be expressed as a right eigenvector of $H^\dagger$
corresponding to eigenvalue $\z^*$:
\be
H^\dagger {\phi_\z^L} = \z^* {\phi_\z^L}.
\ee
In the case of a symmetry that relates $H$ and $H^\dagger$ ($\sC_\pm$ and $\sQ_\pm$), we can relate left and right
eigenvectors. For anti-unitary symmetries $\sC_\pm$, we have that
\be
\phi^L_\z &=& \sC_+\phi^R_{\z},\\
\phi^L_\z &=& \sC_-\phi^R_{-\z}.
\ee
Similarly, for the involutive symmetries $\sQ_\pm$, the left and right eigenvectors are related by
\be
\phi^L_\z &=& \sQ_+\phi^R_{\z^*},\\
\phi^L_\z &=& \sQ_-\phi^R_{-\z^*}.
\ee
On the other hand, a $\sT_\pm$ symmetry relates right eigenvectors of $H$ according to
\be
\phi^R_{\z}  &=& \sT_+ \phi_{\z^*}^R,\\
\phi^R_{\z}  &=& \sT_- \phi_{-\z^*}^R,
\ee
and equivalent relations apply to left eigenvectors.
Note that, for an anti-unitary operator $A$ and vectors $\phi$ and $\psi$, we have that
\begin{equation}
\phi^* A\psi=(A^2\psi)^*A\phi =A^2 \psi^*A\phi.
\end{equation}
In particular, for $A$ that squares to $-1$, the vectors $\phi$ and $A\phi$ are linearly independent, so that $\phi^* A \phi =0$.

We consider the normalized overlap matrix~\cite{chalker1998}
 \be
 \sO_{ab} = \frac{\phi_a^{L\;*} \cdot  \phi_{b}^L\phi_{b}^{R\;*}\cdot \phi_{a}^R}
{\phi_a^{L\;*} \cdot  \phi_{b}^R\phi_{b}^{R\;*}\cdot \phi_{a}^L}
 ,
\ee
where $\phi_a^{L}$ and $ \phi_b^{R}$ are the left and right eigenvectors
corresponding to $a$ and $b$, and $a, b\in \{ \z, \z^*, -\z, -\z^* \}$,
are eigenvalues of $H$ Eq.~(\ref{eq:def_H}).
These overlaps were discussed in detail in Ref.~\cite{sa2022a}, but for completeness, we summarize the main points here.

For a $\sT_-$ symmetry, we have that $\sT_- \phi_\z^R$ is a right eigenvector
with eigenvalue $-\z^*$ and we consider the overlap
 \be
\sO_{\z\; -\z^*} =\frac{\phi^{L\;*}_\z\cdot \phi^L_{-\z^*} \phi^{R\;*} _{-\z^*} \cdot\phi^R_\z}{\phi^{L\;*}_\z\cdot \phi^R_{-\z^*} \phi^{R\;*} _{-\z^*} \cdot\phi^L_\z},
 \label{overlap}
\ee
which vanishes for $\sT_-^2 =-1$ because $\phi^{R} _{-\z^*}\sim \sT_- \phi^R_\z$ and these two states are orthogonal as discussed above.
When $\sT_\pm^2 =+1$, the overlaps are complex and no useful information can be extracted.
The case $\sT_+^2 =-1$ does
not occur in the coupled SYK model.

For a symmetry of the $\sC_-$ type, we consider the overlap
\be
\sO_{\z\;-\z} =\frac{\phi^{L\;*}_\z\cdot \phi^L_{-\z} \phi^{R\;*} _{-\z} \cdot\phi^R_\z}{\phi^{L\;*}_\z\cdot \phi^R_{-\z} \phi^{R\;*} _{-\z} \cdot\phi^L_\z},
\label{ovcm}
\ee
and use that $\phi^{L}_{-\z} = \sC_- \phi^R_\z$
(up to an irrelevant constant).
The numerator of the normalized overlap~\eref{ovcm} can be written as
\be
\phi^{L\;*}_\z\cdot \phi^L_{-\z} \phi^{R\;*} _{-\z} \cdot\phi^R_\z &=&
\phi^{L\;*}_\z\cdot \sC_- \phi^{R}_{\z}
\sC_-^2 (\sC_- \phi^{L} _{\z})^* \cdot\phi^R_\z\nn\\
&=&
\phi^{L\;*}_\z\cdot \sC_- \phi^{R}_{\z}
\sC_-^2( (\sC_-^2 \phi^{L} _{\z})^* \sC_-\cdot\phi^R_\z)^*\nn\\
&=&\sC_-^4\, |\phi^{L\;*}_\z\cdot \sC_- \phi^{R}_{\z}|^2.
\ee
and for the denominator of \eref{ovcm} we obtain
 \be
\phi^{L\;*}_\z\cdot \phi^R_{-\z} \phi^{R\;*} _{-\z} \cdot\phi^L_\z
&=&\sC_-^2 \phi^{L\;*}_\z\cdot \sC_- \phi^L_{\z} \phi^{R\;*} _{-\z} \cdot\phi^L_\z\nn\\
&=&\sC_-^2 (\sC_-\phi^{R}_{-\z})^*\cdot \sC_- \phi^L_{\z} \phi^{R\;*} _{-\z} \cdot\phi^L_\z\nn\\
&=&\sC_-^2\, |\phi^{R\;*}_{-\z}\cdot \phi^L_{\z}|^2.
\ee
This shows that the overlap \eref{ovcm} is real with sign equal to $\sC_-^2$.
In Table \ref{tab:overlap}, we enumerate the overlap matrix elements $\sO_{ab}$ for eigenvalues with the same
absolute value of $\zeta$. The sign of the overlap matrix elements on the
diagonal and antidiagonal give the value of $\sC_+^2$ and $\sC_-^2$, respectively. The matrix elements with $\sT_-$ are zero when $\sT_-^2=-1$.

\begin{figure}[t!]
	\centering
	\subfigure[]{\includegraphics[width=7.5cm]{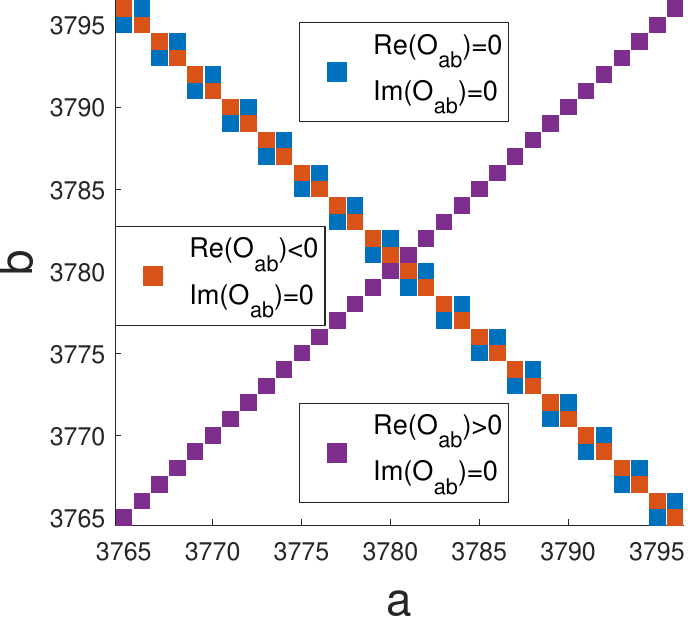}}
	\subfigure[]{\includegraphics[width=7.5cm]{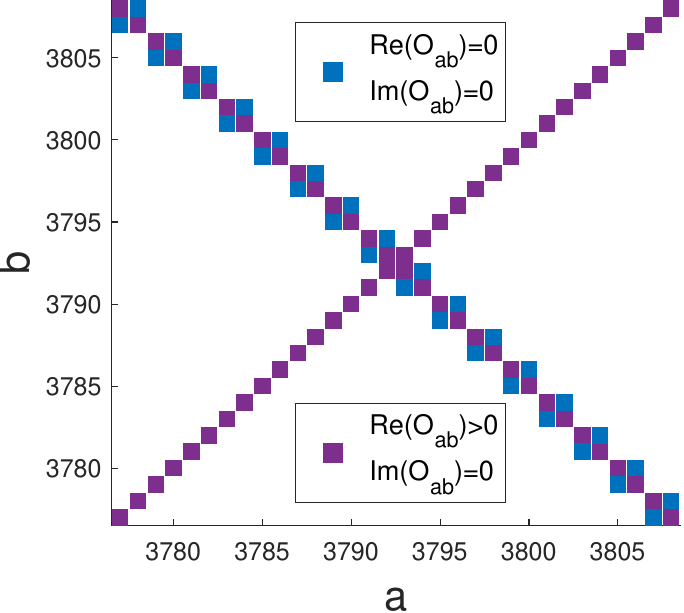}}
	\caption{Central matrix elements (a $30\times 30$ matrix around the center of $\mathcal{O}_{ab}$) of the eigenvector overlap matrix $\mathcal{O}_{ab}$ for a single realization of the Hamiltonian with parameters
		$N=14,\;\kappa=1,\; \lambda=0.2,\; \alpha=1,\;q=4$ and $r=1$.
		In this case, the $S=1$ block (left) belongs to class CI$_{--}$, while the $S=-1$ block (right) has BDI$_{-+}$ symmetry.
		In both cases, the spectrum has dihedral symmetry $D_2$. We rule out the purely real eigenvalues since they are degenerated.
		The eigenvalues are ordered, with increasing order, first by the values of their real parts, then by the values of their imaginary parts. The signatures of the colored elements in matrix $\mathcal{O}_{ab}$ correspond to the symmetries given in Table \ref{tab:overlap}.
	}
	\label{fig:Overlap}
\end{figure}

\begin{table}[t!]
  \caption{Matrix elements of the overlap $\sO_{ab }$. We give the operator that determines the
    sign of the overlap ($\sC_\pm^2$)  or its value in case it vanishes ($\sT_-$).}
  \label{tab:overlap}
\centerline{\begin{tabular}{c|cccc}
$\sO_{ab}$ &$-\z$ & $-\z^*$ & $\z^*$ &$\z$\\
\hline
$-\z$ &\;$\sC_+^2$  &    & $\sT_- $ &$\sC_-^2$ \\
$-\z^*$&\;  & $\sC_+^2$ &$\sC_-^2$  &$\sT_- $\\
$\z^*$&\;$\sT_- $ & $\sC_-^2$ &$\sC_+^2$ &   \\
$\z$&\; $\sC_-^2$ &$\sT_- $ &   & $\sC_+^2$
\end{tabular}}
\end{table}
As an example of how the universality classes can be distinguished by a combination of bulk level statistics, dihedral/point symmetries and eigenvector overlap, we consider the case $N=14, \kappa=1, \lambda=0.2,
\alpha=1,\; q=4,\; r=1$ (see Table~\ref{tab:full_classification} for the global symmetries and the expected symmetry classes).
An interesting feature for this choice, not noted in a recent classification of dissipative quantum chaotic systems~\cite{kawabata2022a}, is that two symmetry classes can coexist in different blocks of the same Hamiltonian, which are characterized by different eigenvalues $s=\pm 1$ of the unitary operator $S$ that commutes with the Hamiltonian.
The $S=1$ and $S = -1$ blocks share the same bulk level statistics
(because they have the same $\sC_+^2$), and the same dihedral symmetry $D_2$ of the global spectrum.
The only difference is the value of $\sC_-^2$. For $S=1$, we have $\sC_-^2=-1$, while for $S=-1$, $\sC_-^2=+1$.
This can be seen from the values of the overlaps of states with eigenvalue
$\zeta$ and $-\zeta$, as depicted by the colors in Fig.~\ref{fig:Overlap} for $S=1$ (left) and $S=-1$ (right).
In these figures, we also show the diagonal overlaps, which are relevant for the value of $\sC_+^2=1$, and overlaps of states with eigenvalues $\z$ and $-\z^*$, which vanish in both cases because of a $\sT_-$ symmetry with $\sT_-^2=-1$. This is enough to identify twelve universality classes which so far fully confirms the symmetry classification of Table~\ref{tab:full_classification}, namely, class CI$_{--}$ for $S=1$, and class BDI$_{-+}$ for $S=-1$.

More specifically, we have distinguished classes with the same
$\sC_+^2$, determined by the CSR,
through the use of global spectral symmetry and overlaps of eigenvectors.
Following this procedure, we were able to uniquely distinguish 12 of the 14 classes realized in the PT-symmetric SYK model, see Table~\ref{tab:full_classification}: A, AI$^\dagger$, D, C, BDI, CI, BDI$^\dagger$, CI$^\dagger$, BDI$_{++}$, BDI$_{-+}$, CI$_{+-}$, and CI$_{--}$. Only classes AIII and AI have not yet been differentiated. For this purpose we consider
the overlap
\be
\omega_{\z\z}=\phi^L_\z \cdot \phi^R_{\z},
\ee
which vanishes for class AI but not for class AIII and can, thus, be used to distinguish the two classes. To see this, we consider the expectation value
\be
   \z\phi^L_{\z} \cdot \phi^R_\z=  \phi^L_{\z}\cdot  H  \phi^R_\z.
\ee
For matrices in class AI, $H^\dagger = H^\top$, and the left eigenvectors are given by
\be
   H^\top \phi^L_\z= \z^* \phi^L_\z,
\ee
so that
\be
\z \phi^L_{\z} \cdot  \phi^R_\z = \phi^L_{\z} \cdot H  \phi^R_\z
= (H^\top\phi^L_{\z}) \cdot \phi^R_\z = \z^*\phi^L_{\z^*}\cdot \phi^R_\z.
\ee
This shows that $\omega_{\z\z}=\phi^L_{\z}\cdot \phi^R_\z =0$ if $\z^* \ne \z$ for class AI. This argument fails for AIII because in that case $H^\dagger \ne H^\top$.

Combining the overlap $\omega_{\z\z}$ with CSR, global spectral symmetry, and eigenvector overlaps, in Table~\ref{tab:full_classification} we list the full classification scheme according to anti-unitary and involutive symmetries of the SYK Hamiltonian Eq.~(\ref{eq:def_H}), confirming the existence of fourteen different universality classes.

As pointed out in Sec.~\ref{sec:classification}, in four of these classes, the Hamiltonian has a finer structure, parameterized by the index $\nu(N)>0$. The observables employed so far are not able to detect this additional structure. Indeed, in these cases, the Hamiltonian Eq.~(\ref{eq:def_H}) contains rectangular blocks with $\nu$ being the difference between the number of rows and columns, due to the nonvanishing of the trace of the operator $Q$, $\Tr Q = \nu$ which we have recently shown \cite{garcia2023c} to have its origin in the existence to a topological invariant of the same value. As discussed in Sec.~\ref{topo}, the consequence of the rectangularization is the existence of $\nu(N)$ purely real eigenvalues.
We shall see in the next section that the level statistics of these real eigenvalues characterizes the subclass by comparing it with the RMT prediction for the corresponding
$Hermitian$ class.

\subsection{Level statistics of real eigenvalues}
\label{sec:levreal}

Nine out of the 14 symmetry classes in our SYK model have
purely real and/or purely imaginary eigenvalues (those with $\sT_+$ type symmetry with $\sT_+^2=+1$ or a $\sQ_+$ type symmetry).
It was recently reported~\cite{Shindou2022} that the level statistics of real eigenvalues in non-Hermitian random matrices have distinct properties that could be exploited to fully characterize non-Hermitian universality classes without having to analyze the complex eigenvalues.
However, as discussed in Sec.~\ref{topo}, for some classes, we have two sources of real eigenvalues: $\nu(N)$ of them
coming from the rectangular structure of the Hamiltonian, which are robust to changes in $\lambda$; plus those, sensitive to $\lambda$, that appear when two complex-conjugated eigenvalues join the real axis, and which can exit it upon an additional variation of $\lambda$.
At least in some cases, this feature is expected to complicate the spectral characterization of non-Hermitian universality classes based only on the analysis of level statistics of purely real eigenvalues.
We study next to what extent this is possible, starting with the Lindbladian regime $\kappa =1$.

\subsubsection{Lindbladian regime ($\kappa=1$)}

In order to illustrate this point, we begin our study of real eigenvalues by comparing a case (AI) with $\nu(N) = 0$ --- corresponding to the absence of a topological invariant \cite{garcia2023c}, so that real eigenvalues are never topological --- and a case (AIII$_\nu$) with $\nu(N) > 0$ real eigenvalues of topological origin.
We consider the SYK Hamiltonian Eq.~(\ref{eq:def_H}) with $N\mod 4=2,\;\kappa=1,\;\alpha=1,\;q=4,\;r=2$, for $S=1$ (AIII$_\nu$) and $S=-1$ (AI).
For AI symmetry, we find that the level statistics does not depend on $\lambda$ if $\lambda$ is not too small or too large and $\kappa \ne 1$. However, this is not the case for the AIII$_\nu$ symmetry class.

\begin{figure}[t!]
	\centerline{
		\includegraphics[width=8cm]{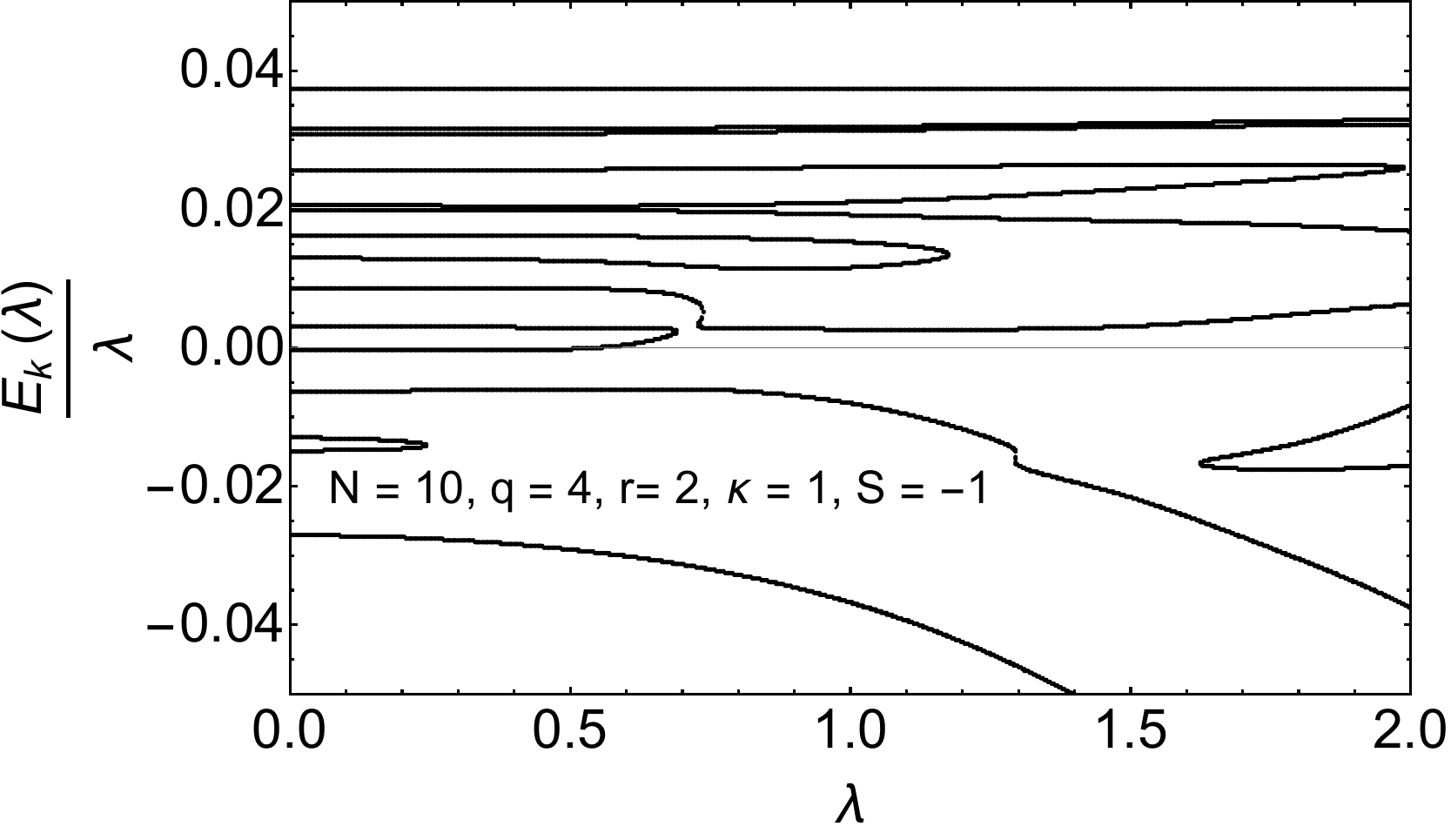}
		\includegraphics[width=8cm]{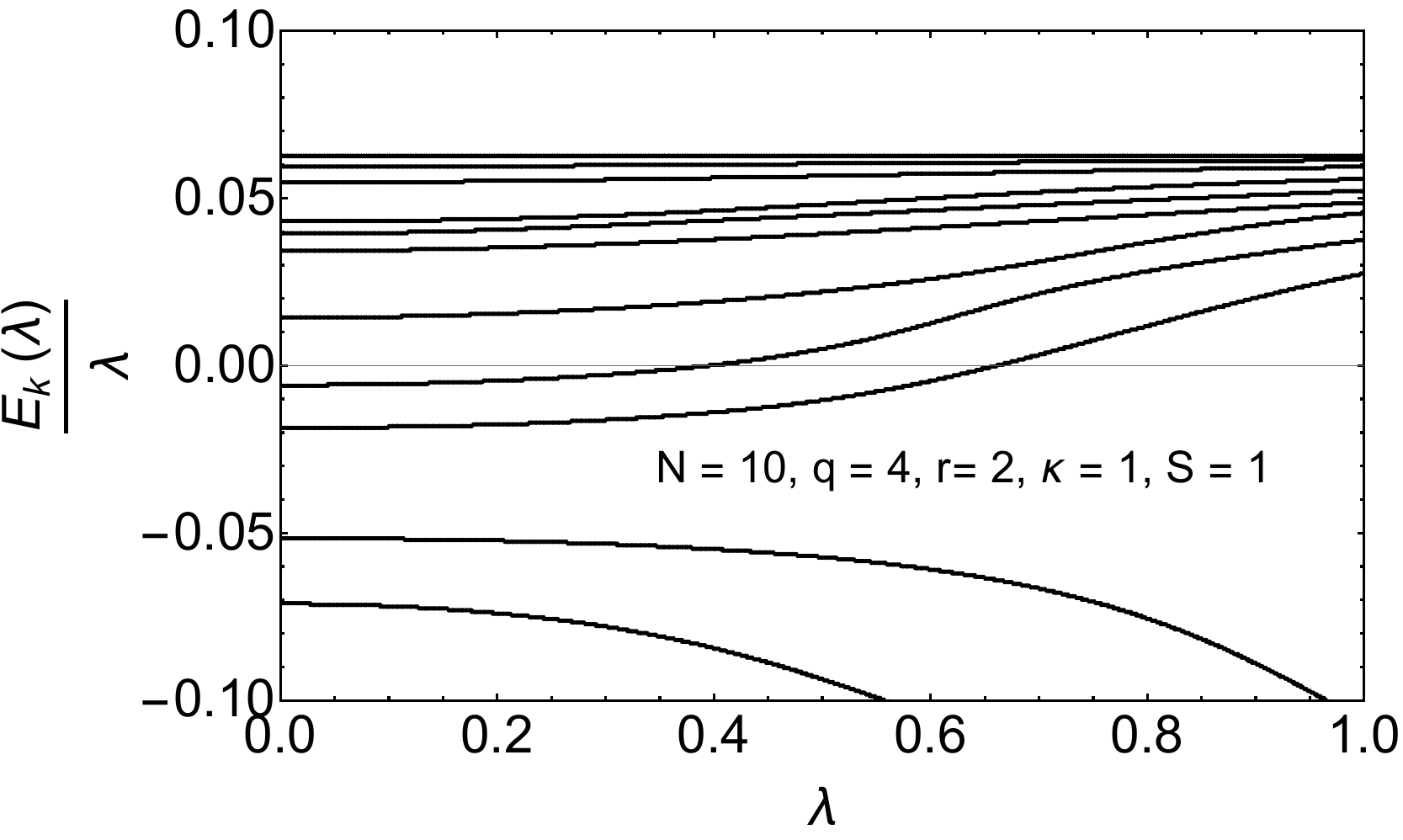}
	}
	\caption{Spectral flow of purely real eigenvalues of the SYK Hamiltonian Eq.~(\ref{eq:def_H}) versus $\lambda$ for $N=10$, $\kappa=1$, $\alpha =1$, $q=4$, and $r=2$. Left: $S_L= -S_R=1$ blocks corresponding to class AI. We observe the creation and annihilation of pairs of real eigenvalues for $\lambda > 0.2$. Right: $S_L= S_R=1$ blocks corresponding to class AIII$_\nu$. The number of real eigenvalues does not depend on $\lambda$, up to relatively large values of $\lambda$, resulting in GUE statistics in this region, see Fig.~\ref{fig:Realratio}.
	}
	\label{fig:flowk1}
\end{figure}

To understand the difference between the two cases, we show in Fig.~\ref{fig:flowk1} the spectral flow of the real eigenvalues of $H$ as a function of $\lambda$ (we set $N=10$ in order to have only a small number of flow lines). For the AI class, see the left plot of Fig.~\ref{fig:flowk1}, we observe that for very small $\lambda < 0.2$, the flow is laminar, so we expect agreement with the Hermitian symmetry class (GOE) in this region. For larger values of $\lambda$, spectral flow lines start to disappear, resulting in a gap in the spectrum, and pairs of complex eigenvalues join the real axis, leading to two very close eigenvalues.
Therefore, level statistics are expected to deviate strongly from the Hermitian class and approach the non-Hermitian RMT prediction (AI) ~\cite{Shindou2022} characterized by stronger spectral fluctuations.
The situation is different for the flow lines corresponding to AIII$_\nu$, see Fig.~\ref{fig:flowk1} (right). For sufficiently small $\lambda <0.5$, the flow lines do not coalesce, and no new pairs of flow lines appear. Therefore, we find a spectrum with no large gaps or close pairs, and, as discussed in Sec.~\ref{topo}, we expect to observe \emph{Hermitian} GUE statistics. The reason for that behavior is the recently shown \cite{garcia2023c} topological nature, for a not too strong $\lambda$,  of all real eigenvalues in this symmetry class.

\begin{figure}[t!]
	\centering
	{\includegraphics[width=7.5cm]{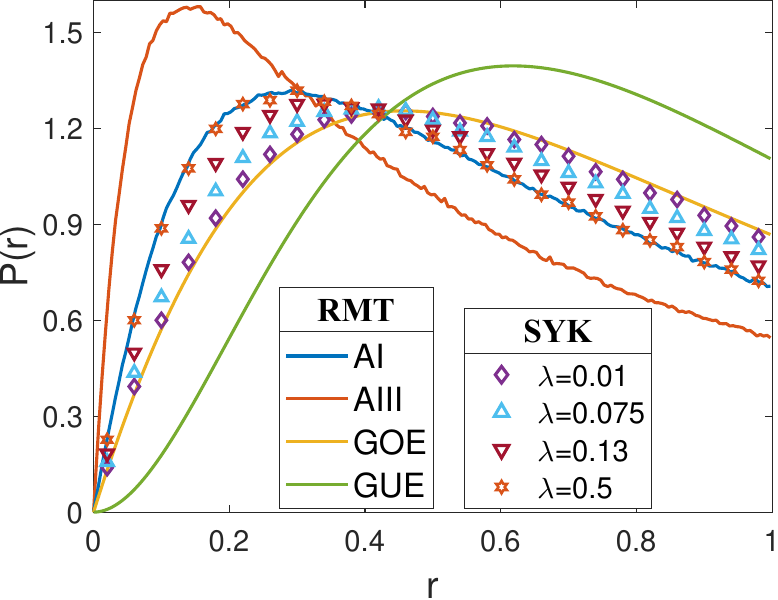}}
	{\includegraphics[width=7.5cm]{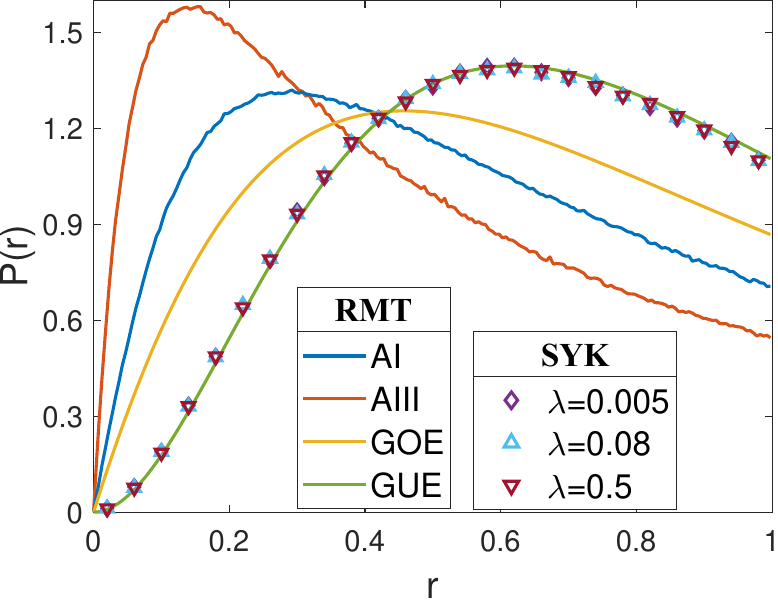}}
	\caption{Gap ratio distribution \cite{atas2016}, $P(r)$,
		for the real eigenvalues of the SYK Hamiltonian Eq.~(\ref{eq:def_H}) with $N=14,\; \kappa=1,\; \alpha=1,\; q=4$, $r=2 $, and $2\times10^4$ disorder realizations. For the sake of clarity, we stress that $r$ in the SYK Hamiltonian and in $P(r)$ are completely unrelated. In the left plot, we show results for class AI ($S_L= -S_R=1$) and in the right plot, we give the results for class AIII$_\nu$ ($S_L= S_R=1$). The values of $\lambda$ are shown in the legend of the figures. As predicted, for class AI (left panel), we observe a crossover from GOE to AI statistics as $\lambda$ increases, while for AIII$_\nu$ (right panel), $P(r)$ fits well the GUE result in the broad range of $\lambda \in[0.005,0.5]$ considered.
	}\label{fig:Realratio}
\end{figure}

To make the previous statements more quantitative, we consider the gap ratio distribution $P(r)$~\cite{oganesyan2007,atas2016},
where $r_i=\mathrm{min}(s_{i+1}/s_i,s_i/s_{i+1})$ and $s_{i}=E_i-E_{i-1}$ are the spacings of the ordered real eigenvalues of $H$, Eq.~(\ref{eq:def_H}).
 To eliminate finite size effects,
 we now choose $N = 14$, which is in the same universality class. We also remove
 one or two eigenvalues at the edge of the spectrum, which have nonuniversal
 properties (see, e.g., Fig.~\ref{fig:flowk1}, where the lowest eigenvalues are separated
 from the rest of the spectrum by a gap).
 As predicted, since the SYK model in the
 AI class does not have eigenvalues of topological origin \cite{garcia2023c},
 we find good agreement with the non-Hermitian AI random matrix prediction for the distribution of the gap spacing ratio $P(r)$ when $\lambda \geq 0.3$, see Fig.~\ref{fig:Realratio} (left).
 For very small $\lambda$, the distribution is close to GOE statistics. This can be seen as follows.
 For $\lambda = 0$, the zero eigenvalues of $H_L+ H_R$ are  due to eigenstates of
 the form $|k\rangle \otimes |k\rangle$ with $|k\rangle$
 the eigenstates of a single Hermitian SYK.
  At nonzero
$\lambda$, by degenerate perturbation theory, the zero eigenvalues flow into the eigenvalues of the
  matrix  $\langle kk |H_I|ll \rangle$. This is a Hermitian matrix, and because it is in class AI, it must
  be real. Therefore, its eigenvalues are correlated according to the GOE.
 The situation is different for class AIII$_\nu$, see Fig.~\ref{fig:Realratio} (right),
 for which we find good agreement with GUE statistics for a wide range of $\lambda \in [0.005,0.5]$, consistent with our previous discussion in Sec.~\ref{topo} on the AIII$_\nu$ university class.
 The crossover between Hermitian and non-Hermitian level statistics seen for the nontopological eigenvalues of AI is absent in this case.
 It is worth noticing that, as predicted, the number of real eigenvalues in each realization of AIII$_\nu$ is always $\nu = 2^{N/2}/2 = 64$ for not too large $\lambda$, exactly the same as the number of zero modes for $\kappa=1, \lambda=0$.

\begin{figure}[t!]
	\centering
	\includegraphics[width=8cm]{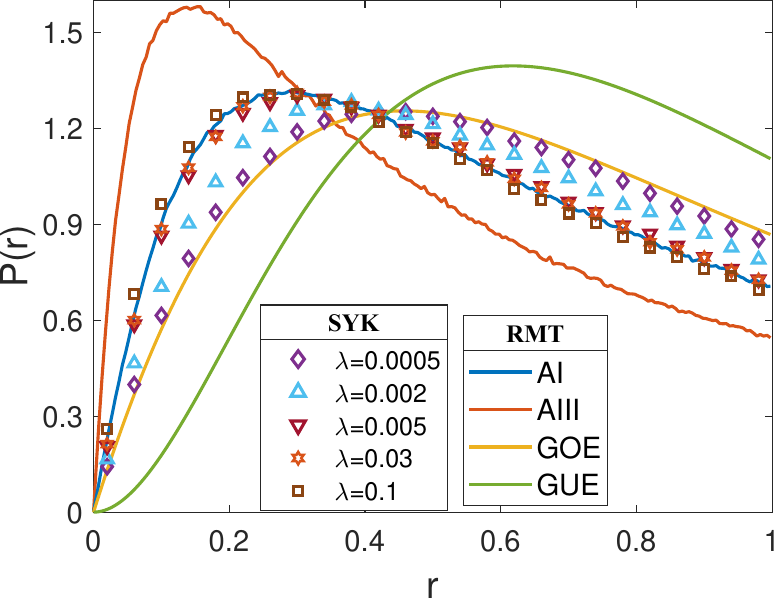}
	\caption{Gap ratio distribution, $P(r)$ for the real eigenvalues of the SYK Hamiltonian Eq.~(\ref{eq:def_H}) for $N=14,\;\kappa=1,\;\alpha=1,\;q=6,\;r=1,\; Q=i$, $2\times10^4$ realizations, and different values of $\lambda$.  We observe a crossover in $P(r)$, from GOE to AI as $\lambda$ increases. This is consistent with the results of Fig.~\ref{fig:Realratio} (left), also for class AI, but with a different choice of parameters. Notably, the minimum $\lambda$ for which agreement with AI class is observed is much smaller.}
	\label{fig:ai}
\end{figure}

In order to better understand the crossover between the level statistics of Hermitian and non-Hermitian RMT for real eigenvalues of nontopological origin, we discuss another example of class AI corresponding to $N=14$, $q=6$, $r=1$, $ \kappa =1$, $\alpha =1$, in the sector with $Q=i$. Using the
 same perturbative argument as in the previous paragraph, it is clear that we also expect GOE statistics in $\lambda \ll 1$ limit and a crossover to AI statistics for sufficiently large $\lambda$. The results for the gap spacing distribution $P(r)$ depicted
 in Fig.~\ref{fig:ai} confirm this prediction. However, the crossover scale in $\lambda$ is one order of magnitude different from the previous case. An analytical estimation of the crossover point clarifies the origin of this difference.
The crossover point is approximately given by
 \be
 \lambda \lVert H_{I} \rVert = \Delta E,
 \ee
 where $\Delta E$ is the level spacing of $H_0$ and $\lVert \ldots \rVert$ stands for the sup norm. For $r = 1$, we find that
 \be
 \lVert H_{I} \rVert = \frac N2,   \qquad r=1,
 \ee
 while for $r=2$ the result is
\be
 \lVert H_{I} \rVert = \frac 1{4N}\left ( \left(\frac N2\right)^2-\frac N2\right ),   \qquad r=2.
 \ee
 For $N = 14$, the norm for $r = 1$ is a factor 8.6 larger than the norm for $r = 2$.
 On the other had, the $q$ dependence of the norm of $H_0$ is given
 by
 \be
 \lVert H_{0} \rVert \sim 2 N \frac {2^{-{q/2}}}{q^{3/2}}.
\ee
Therefore, the level spacing for $q = 6$ is about a factor 4 smaller than the level spacing for $q = 4$. Altogether, this results in
a crossover point which for $q = 6$, $r = 1$ is about thirty times smaller than for $q = 4$, $r = 2$, in agreement with Fig.~\ref{fig:Realratio} (left) and Fig.~\ref{fig:ai}.

\begin{figure}
	\centering
	\includegraphics[width=8cm]{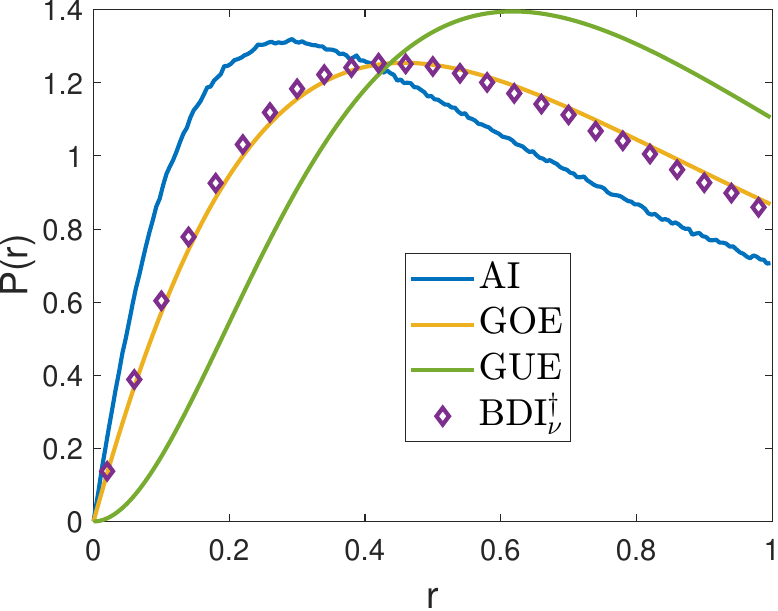}
	\includegraphics[width=8cm]{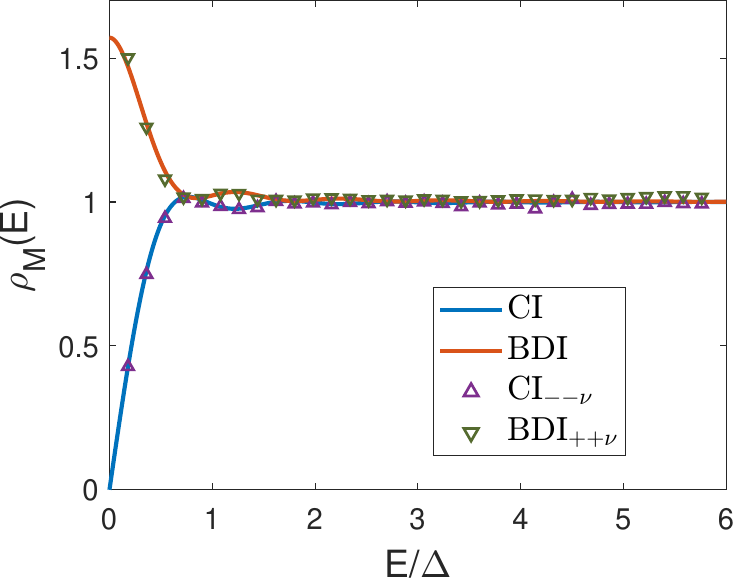}
	\caption{Left: Gap ratio distribution \cite{atas2016}, $P(r)$, for the real eigenvalues of the SYK Hamiltonian Eq.~(\ref{eq:def_H}) for $N = 12,\; \kappa=1,\; \lambda=0.02,\; \alpha=1,\; q=4$, $r=2$, $S_L=S_R=1$, and $8 \times 10^4$ realizations corresponding to the BDI$^\dagger_\nu$ class. Each realization has 64 exactly real modes, including 16 nonrandom ones which are not affected by the random SYK couplings. We only take the remaining 48 modes for the $P(r)$ computation. We find excellent agreement with the predicted RMT result (GOE), see Sec.~\ref{sec:topo_BDIdag}. Right:  Microscopic spectral density $\rho_M(E)$ of the real modes close to $E=0$ in units of the mean level spacing $\Delta$ for $\kappa=1,\; \alpha=1,\; q=4$, $r=1 $, $S=1$. For $N=12, \; \lambda=0.07, \;10^6$ realizations (BDI$_{++\nu}$ symmetry class) and also for $N=14, \; \lambda=0.02, \;2.5\times 10^5$ realizations  (CI$_{--\nu}$ symmetry class). In both cases, we find very good agreement with the BDI and CI random matrix predictions, respectively.}
	\label{fig:othertopo}
\end{figure}

 We now turn to the study of the level statistics of the real eigenvalues
 of the remaining three classes, CI$_{--\nu}$, BDI$_{++\nu}$, and BDI$^\dagger_\nu$ which in Sec.~\ref{topo} we predicted, by a RMT analytic calculation, to belong to the CI, BDI, and GOE Hermitian symmetry classes, respectively. Results for the distribution of the gap ratio $P(r)$, depicted in Fig.~\ref{fig:othertopo}, show an excellent agreement with the RMT prediction for the corresponding class. We expect this agreement is robust in a broad range of not too large values of $\lambda$ because we have recently shown these real eigenvalues have topological origin~\cite{garcia2023c}.

\subsubsection{General PT-symmetric regime ($0<\kappa<1$)}
 We now move to the study of real modes for the general PT-symmetric case $0 < \kappa < 1$.
 For the sake of concision, we again focus on the two classes AIII$_\nu$ (with two sources of real modes since $\nu(N) > 0$) and AI (for which there is only one source, as the blocks of the Hamiltonian are never rectangular, so $\nu(N) = 0$).

\begin{figure}[t!]
\centerline{
	\includegraphics[width=8cm]{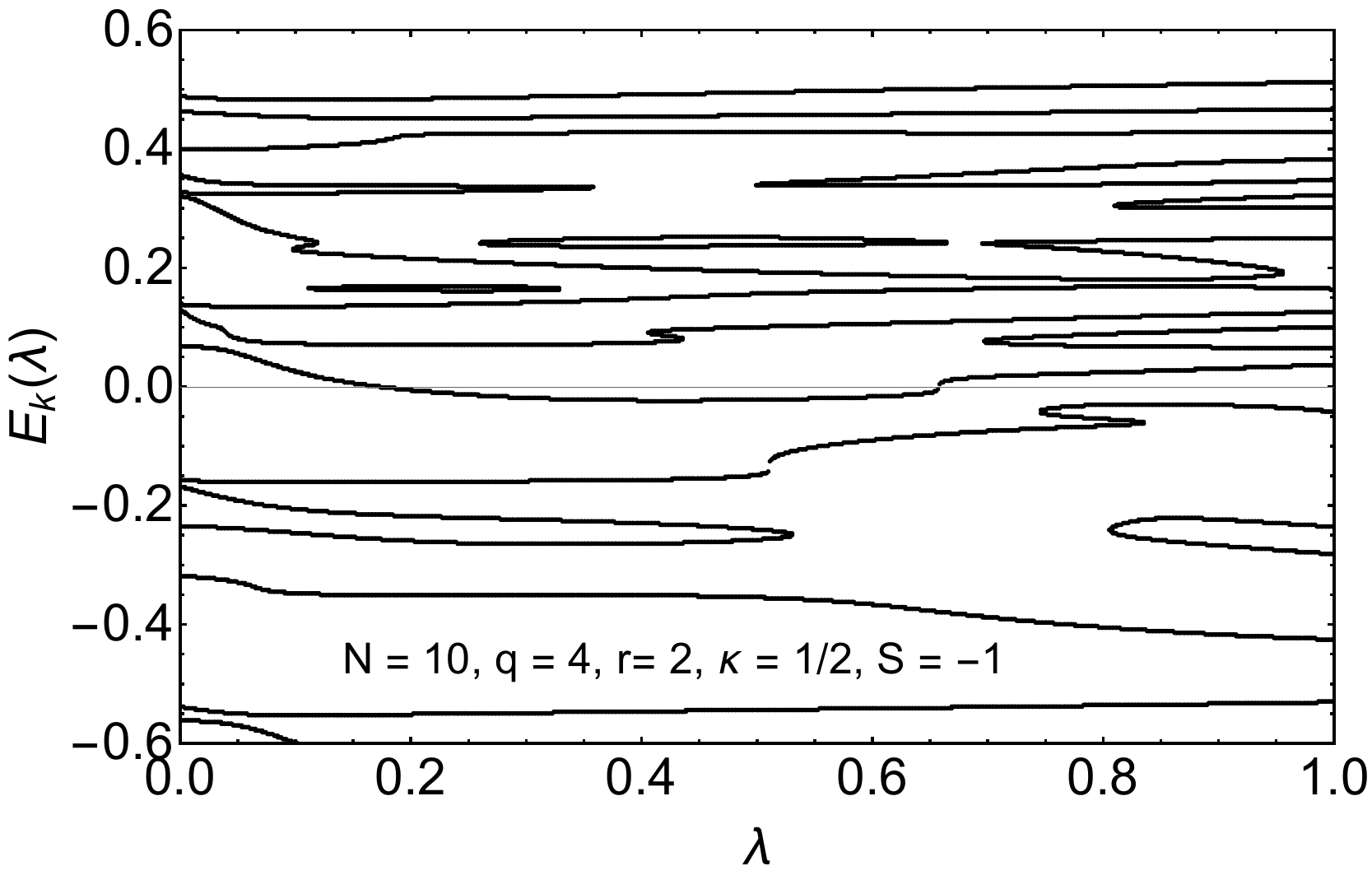}
    \includegraphics[width=8cm]{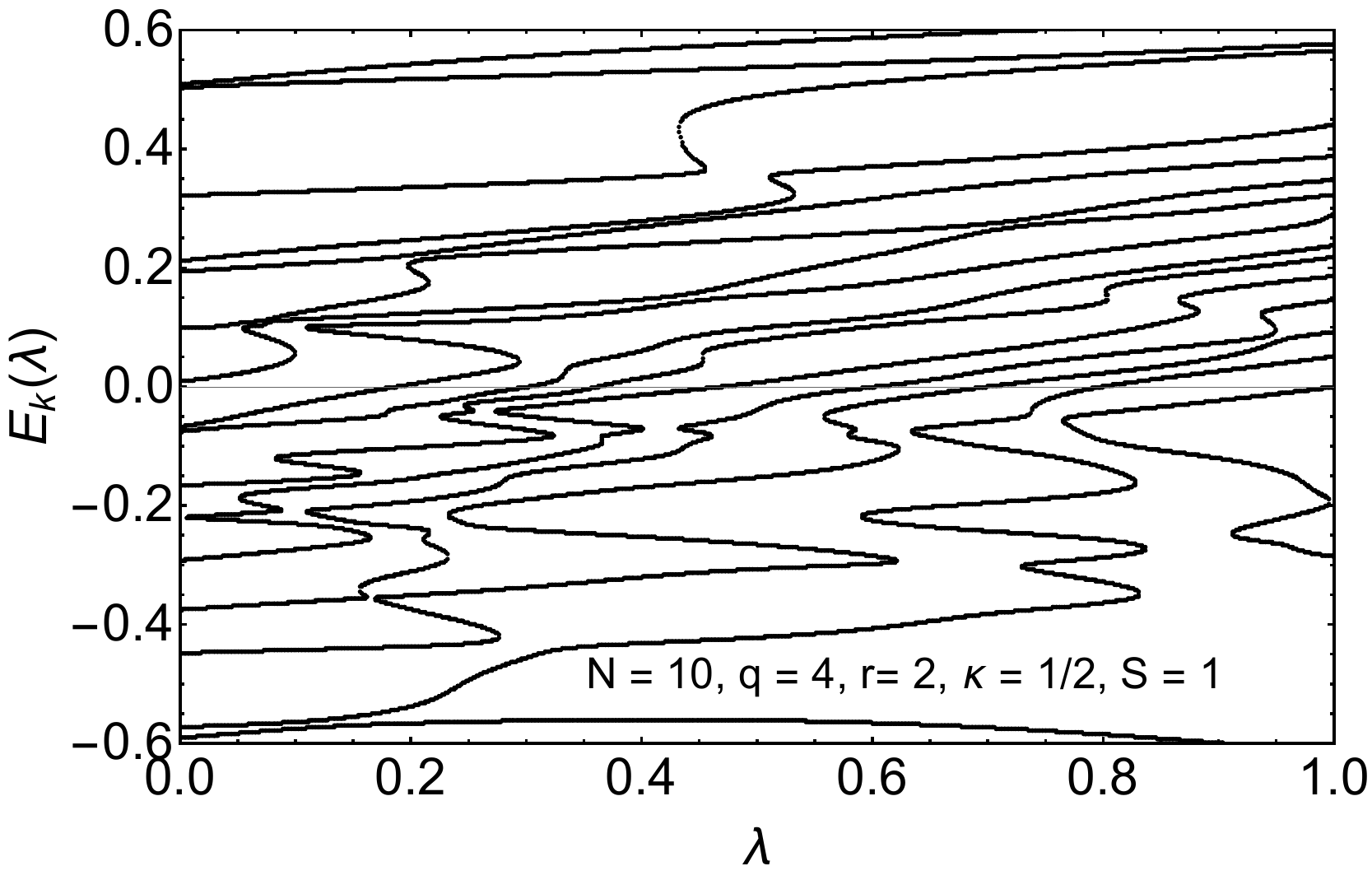}
}
\caption{Spectral flow of the real eigenvalues of the Hamiltonian Eq.~(\ref{eq:def_H}) versus $\lambda$ for $N=10$, $\kappa=1/2$, $\alpha =1$, $q=4$, and $r=2$. Left: $S_L= -S_R=1$ corresponding to class AI. We observe the creation and annihilation of pairs of real eigenvalues all over the spectrum and uniformly in $\lambda$. Right: $S_L= S_R=1$ corresponding to class AIII$_\nu$. We observe an asymmetry between positive and negative real eigenvalues and fewer pairs with positive energy appear or disappear at larger values of $\lambda$.
}
\label{fig:flowhlf}
\end{figure}

For $0<\kappa<1$, we never have exact zero modes for $\lambda=0$.
 In this case, we can split $H_0$ into its Hermitian and anti-Hermitian parts as
\be
H_0 = H_{0J}+ H_{0M}.
\ee
For even $q$, both $H_I$ and $H_{0J}$ commute with $Q$ while $H_{0M}$
anticommutes with $Q$. This implies that no symmetries are restored
in the small $\lambda$ limit, and we are always in a parameter
range with both $\nu(N)$ real eigenvalues and additional real eigenvalues
due to pairs of complex eigenvalues coalescing with the real axis. Therefore, we expect that, in this case, the spectral statistics
for class AIII$_\nu$ depend
on the mixture of both types of real eigenvalues which changes as a function of $\lambda$.
By contrast, in class AI, $\nu(N) = 0$, so we expect to observe the predicted AI level statistics, at least for sufficiently large $\lambda$, as real eigenvalues come only from
 pairs of complex eigenvalues that diffuse in and out the real axis already for small
 $\lambda$. Therefore, two new neighboring real
 eigenvalues occur when a pair just joins the
real axis, while a large gap is observed when a pair leaves the real axis.
This is illustrated in Fig.~\ref{fig:flowhlf}
for $N=10,\;$ $\kappa=1/2$, \;$\alpha =1$,\; $q=4$, and $r=2$, where we show the spectral flow as a function of $\lambda$ for symmetry class AI (left), corresponding to the sector $S=-1$, and for symmetry class AIII$_\nu$ (right), corresponding to the sector $S=+1$. For class AI, we do not observe the laminar flow of the $\kappa = 1$ case for $\lambda \ll 1$. Therefore, we do not expect a crossover from Hermitian to non-Hermitian RMT as $\lambda$ increases. The existence of real random couplings prevents us from carrying out the degenerate perturbation theory that, for $\kappa  = 1$, resulted in Hermitian level statistics in this $\lambda \ll 1$ region. Indeed, for $\lambda \ll 1$, the spectral flow shows quasidegeneracies that are gradually lifted as $\lambda$ increases. In the region of larger $\lambda$,
all eigenvalues diffuse in and out of the real axis which is the expectation for a non-Hermitian RMT.

The spectral flow for AIII$_\nu$, Fig.~\ref{fig:flowhlf} (right), also shows a similar gradual lifting of (quasi) degeneracies as $\lambda \ll 1$ increases. Unlike the AI flow, however, the region of eigenvalues diffusing in and out of the real axis is
mostly restricted to relatively small values of $\lambda$, while for large $\lambda$ the flow is laminar, at least for positive eigenvalues. Therefore, it may be possible that, for sufficiently large $\lambda$, in the aforementioned laminar region mostly controlled by the $\nu(N)$ eigenvalues, the level statistics are well described by the Hermitian RMT prediction, GUE in this case.

\begin{figure}[t!]
	\centering
	\includegraphics[width=8cm]{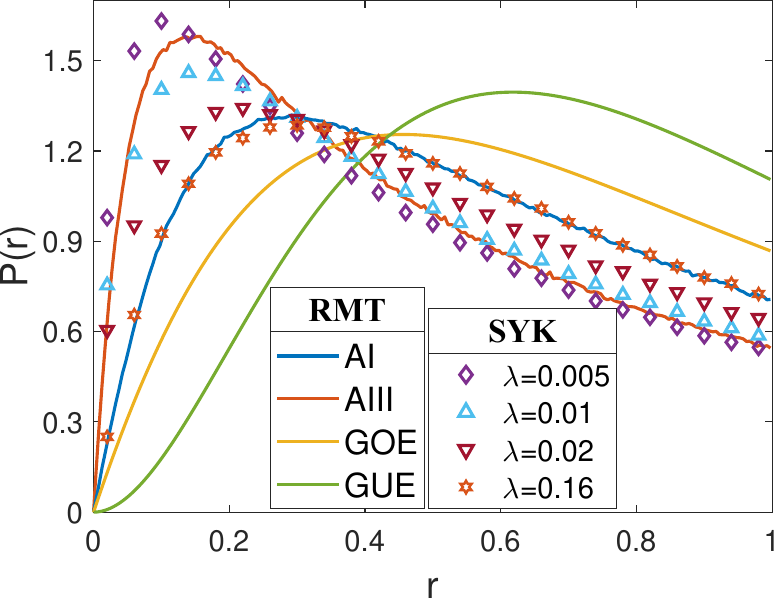}
	\includegraphics[width=8cm]{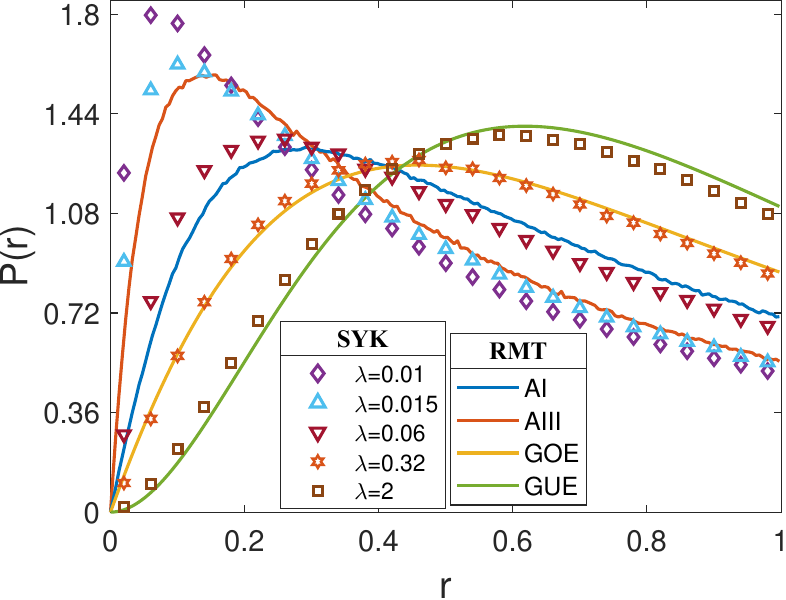}
        \caption{Gap ratio distribution, $P(r)$, for the real eigenvalues of the SYK Hamiltonian Eq.~(\ref{eq:def_H}) with $N=14,\;\kappa=0.5,\;\alpha=1,\;q=4,\;r=2$, and $2\times 10^4$ realizations. The values of $\lambda$ are given in the legend of the figure. In the left plot, we show results for class AI ($S_L= -S_R=1$) and, in the right plot, we present results for class AIII$_\nu$ ($S_L= S_R=1$). We find that for sufficiently large $\lambda$, $P(r)$ agrees well with the RMT prediction for the non-Hermitian AI class (left) and GUE (right) thus confirming that, also for $0 < \kappa < 1$, different blocks of the same Hamiltonian can have different symmetries with distinct levels statistics properties.
  }
    \label{fig:non_stationary}
\end{figure}

The numerical results depicted in Fig.~\ref{fig:non_stationary} confirm the above prediction stemming from the qualitative analysis of the spectral flow.
In both cases, we observe that the spacing ratio
distribution $P(r)$ changes gradually from close to Poisson in the $\lambda \to 0$ limit to close to GUE (non-Hermitian AI) for class AIII$_\nu$ (AI) for sufficiently large $\lambda$.
For class AIII$_\nu$, the persistence of deviations from the GUE until large values of $\lambda \gtrsim 1$ is consistent with the mentioned coexistence of real eigenvalues of different origin: $\nu(N)$ of them topological, recently reported in Ref.~\cite{garcia2023c}, and additional $\lambda$-dependent ones coming from interactions that push complex eigenvalues in and out of the real axis. We note that the agreement with GUE statistics starts much earlier for positive eigenvalues, a feature consistent with the observed laminar flow in this spectral region. Although we do not have a complete understanding of the origin of this asymmetry, we have observed that spectral density in this region develops a (likely nonuniversal)  mouth  that prevents eigenvalues from moving in and out of the real axis.

As a final remark, we note other spectral observables could also have been used to characterize most of the universality classes. Some of them may give useful complementary information but others may require a much greater computational effort. It is often a matter of trial
and error to find out what works best.

\section{Examples and applications of PT-symmetric SYK configurations}
\label{sec:examples}

In this section, we apply the classification scheme to specific physical problems: gravitational wormholes --- solutions of the
Einstein equations representing shortcuts in space-time --- and open quantum many-body systems coupled to a bath or subjected to continuous monitoring~\cite{schlosshauer2019}, i.e., a situation where unitary time evolution is interrupted by the repeated process of measurement modeled by the application of a projection
operator, representing the observable which  is being measured, on the wave function that describes the time evolution of the quantum system.

\subsection{Wormholes}
\label{sec:examples_wormholes}
The physics of wormholes has attracted a lot of recent interest after it became possible to construct~\cite{gao2016} traversable wormhole solutions
that did not violate the null-energy condition --- and were, therefore, legitimate physical solutions of the gravity equations ---
by adding a double-trace coupling between the boundaries in global AdS.
As an example, wormholes are now believed~\cite{penington2020,almheiri2020} to be indispensable to render the process of black-hole evaporation unitary, which is central for the resolution of the information paradox and also important in the so-called factorization problem \cite{maldacena2004,harlow2018} which poses a challenge for the application of the holographic principle beyond tree-level.

In the context of Jackiw-Teitelboim (JT)  gravity~\cite{jackiw1985,teitelboim1983}, a near-(A)dS$_2$ background~\cite{jackiw1985,teitelboim1983,almheiri2015,maldacena2016a}, different types of wormholes configurations have been intensively studied in the past few years.
Eternal traversable wormholes solutions where found  \cite{maldacena2018} by perturbing global AdS$_2$ with a weak double trace deformation coupling the boundaries only.
Euclidean wormholes solutions were found in this context by solving the Einstein's equations with an additional massless scalar with complex sources at the two boundaries of global AdS$_2$.
Similarly, it has been speculated that an analogous perturbation in a near-global de Sitter background~\cite{turiaci2019,cotler2020} may lead to the so-called Keldysh wormholes~\cite{garcia2022e} relevant for the description of late-time features of the dynamics.

The SYK field theory analogue of these different types of wormholes in Jackiw-Teitelboim gravity has already been found to correspond to different regions of parameters of the PT-symmetric SYK model Eq.~(\ref{eq:def_H}).
For traversable wormholes ~\cite{maldacena2018}, the field theory analogue is the low-temperature limit of a two-site Hermitian SYK model with a weak explicit coupling between the two sites, which corresponds to the $\kappa =  0, \lambda \ll 1$ limit of Eq.~(\ref{eq:def_H}). This weak intersite coupling in the field theory mimics the gravitational double-trace deformation mentioned earlier.
Euclidean wormholes have been related~\cite{maldacena2018,garcia2021}
to the low-temperature limit of a two-site non-Hermitian SYK model with no intersite coupling and complex couplings with complex conjugate symmetry and the requirements that the variance of the real part of the couplings must be larger than the one corresponding to the imaginary part, which corresponds to the $\lambda = 0$, $\kappa \leq 1/2$
region of our model.

In the gravitational setting, it is possible~\cite{garcia2022c} to study a transition from Euclidean to traversable wormholes by increasing the strength of the double-trace couplings for a fixed strength of the complex sources, which in the SYK Hamiltonian Eq.~(\ref{eq:def_H}) corresponds to an increase of $\lambda$ for a fixed $\kappa \leq 1/2$.
The latter condition is required since for $\kappa > 1/2$ the thermodynamic properties are pathological~\cite{maldacena2018,garcia2022a,garcia2021}, and not related to Euclidean wormholes configurations.
Analytical~\cite{garcia2022c} results for the gap on the gravity side are available since the low-energy (boundary) action is a generalized Schwarzian and they are in agreement with the SYK prediction resulting from a saddle point analysis.
In both Euclidean and traversable configurations, the free energy undergoes a thermal first-order phase transition that separates
the low-temperature wormhole phase from the high-temperature two-black hole phase. In the strict Euclidean case ($\lambda = 0$),
it is possible to show that the low-temperature wormhole phase is a consequence of the dominance of off-diagonal replica configurations~\cite{garcia2022a,garcia2022b}.

Keldysh wormholes~\cite{garcia2022e}, namely, wormholes configurations in two-dimensional de Sitter spaces~\cite{turiaci2019,cotler2020} have been conjectured to play an important role in the path integrals that govern the late time dynamics of a single site SYK coupled \cite{garcia2022e} to a Markovian bath. The dynamics of this model is described by a vectorized Liouvillian which is formally identical \cite{garcia2022e} to the limit $\kappa = 1$ and small $\lambda \ll 1$ of the SYK Hamiltonian Eq.~(\ref{eq:def_H}).

In summary, a large variety of gravity settings have been related to the two-site PT-symmetric SYK model Eq.~(\ref{eq:def_H}) we have employed for the symmetry classification, or small variations of it. Therefore, our classification scheme has the potential to stimulate research
on the role of symmetry in wormhole physics.

\subsection{Open quantum systems: Driven-dissipative dynamics and continuous monitoring}
\label{sec:examples_open}

An interesting point of the phase diagram of our model is $\kappa = 1$ with $\lambda > 0$.
In this case, using either path-integral~\cite{sieberer2016keldysh,kamenev2011field} or operator~\cite{prosen2008,garcia2022e} methods, $H$ can be mapped to a vectorized Liouvillian $\mathcal{L}$ that governs the time-evolution of the density matrix, $\partial_t \rho=\sL{\rho}$, of a single-site Hermitian SYK coupled to a Markovian environment characterized by jump operators $L_i \sim \sqrt{\lambda}\psi_i$. Indeed, this is essentially the
model considered in Ref.~\cite{kawabata2022a}
to classify dissipative quantum chaotic systems by its global symmetries.
In the vectorized representation (often called Liouville space), operators in the left copy correspond to operators acting from the left of the density matrix (i.e., on its ket), while operators in the right copy represent operators acting from the right (i.e., on its bra). The PT symmetry of the two-site model ensures that the dynamics generated by $\mathcal{L}$ preserves the Hermiticity of the density matrix~\cite{sa2022a}, while the vanishing of the real part of the intrasite couplings guarantees that it also preserves the trace of the density matrix.
A especially interesting case is that of sufficiently small $\lambda$ (and still $\kappa=1$), where the time-evolution is still quantum chaotic but the Keldysh path integral is dominated by the so-called Keldysh wormhole configuration~\cite{garcia2022e} at late times, see the preceding section for more details.
The dominance of these solutions leads to an anomalously large gap and, therefore, a much faster approach to infinite-temperature equilibrium than otherwise expected. In particular, even in the limit of vanishing coupling to the bath ($\lambda\to0$), there is a finite relaxation rate.

A dissipative system coupled to a bath is not the only type of system that is effectively described by the Lindbladian $\mathcal{L}$.
Indeed, the time evolution of a system whose Majorana operators are being continuously monitored is described~\cite{jacobs2006,jacobs2014,wiseman2009}, upon averaging over measurement outcomes (i.e., if they are not recorded), by the same Lindbladian and, hence, can be mapped to our PT-symmetric model. Other monitoring processes could also, under certain conditions, be mapped to a model with a different left-right coupling but the same global symmetries. This quantum information setting is, therefore, another area of direct relevance of our classification scheme.

Finally, we address the region $1/2 <\kappa < 1$. As we have seen above, real-time Lindblad dynamics of the density matrix can be interpreted as PT-symmetric dynamics in Liouville space. Following the same procedure, for the more generic situation $1/2 <\kappa < 1$, $H$ is mapped instead into a nontrace-preserving, so not of Lindbladian type, but still Hermiticity-preserving, vectorized Liouvillian.
Physically, this corresponds to a non-Hermitian SYK coupled to a Markovian bath where the non-Hermitian part has a relatively minor effect.

Moreover, following the previous interpretation in terms of continuous monitoring, this regime of the PT-symmetric Hamiltonian could potentially model a many-body system on which measurements are being performed and that, at the same time, is experiencing spontaneous decay or inelastic scattering, so that states acquire a finite lifetime. We stress once more that this interpretation may require the use of different jump operators from the ones we are considering but it could be possible to choose them so that the symmetry classification is unaltered.

\section{Outlook and conclusion}
\label{sec:conclusion}

We have proposed a classification of non-Hermitian but PT-symmetric systems. Putting some relatively mild restriction on the form of the Hamiltonian, which covers SYK models with Majoranas fermions, we have found that
the 38 universality classes~\cite{bernard2002,ueda2019} of general non-Hermitian systems are reduced to only 24 due to PT symmetry.
A simple two-site non-Hermitian but PT-symmetric SYK model with $N$ fermions, $q$-body interactions and a $r$-body intersite coupling has been employed to identify the different universality classes. By tuning $q$, $r$, and $N$, the strength of the imaginary couplings and a weak
asymmetry factor, we have found 14 different universality classes. Interestingly, among the symmetry classes not allowed by PT symmetry are those leading
to Kramers degeneracy in a fixed sector of the theory. Another salient result of our analysis is the identification of cases where different blocks of the Hamiltonian (for fixed parameters) have different symmetries. This feature had been missed in a recent classification of dissipative quantum chaotic systems~\cite{kawabata2022a}, which partially overlaps with ours. The symmetry classification has been confirmed by an exhaustive analysis of level statistics and eigenvector overlaps. An interesting direction for further work is to investigate modifications of the present two-site SYK model that realize the remaining classes from the 24-fold classification.

Another highlight of the PT symmetry classification is the identification of classes AIII$_\nu$, BDI$^\dagger_\nu$, BDI$_{++\nu}$, and CI$_{--\nu}$, characterized by $\nu(N)$ real eigenvalues
with level statistics that agree with the random matrix prediction for the Hermitian classes GUE, GOE, BDI, and CI, respectively. These spectral results are
robust to changes in the parameters of the models, such as the intersite coupling, that preserve the symmetry.
These real modes have their origin in the
existence of rectangular blocks in the Hamiltonian, characterized by $\nu$, which we have recently \cite{garcia2023c} found to be of topological origin and so we refer to these four classes as topological classes.
Since this is not the only source of purely real eigenvalues in non-Hermitian systems, this finding casts some doubts about the possibility, recently proposed in Ref.~\cite{Shindou2022}, of characterizing the universality class of certain non-Hermitian random ensembles by only looking at their real eigenvalues.

The proposed classification is of relevance to a broad range of problems: from wormhole configurations in gravity theories with a SYK field theory dual, which for nontraversable cases are related to quantum entanglement through the EPR$=$ER conjecture~\cite{maldacena2013}, to the time evolution of open systems coupled to a bath or subjected to continuous monitoring. As an example, we identify the range of parameters in our SYK for which its gravity analogue in the low-temperature limit may be a traversable, an Euclidean, or a Keldysh wormhole.
Likewise, we identify the range of parameters for which the SYK Hamiltonian can be mapped onto a vectorized Liouvillian that generates the dynamics, i.e., governs the dissipative dynamics of the SYK density matrix. We note that the PT symmetry of the Hamiltonian guarantees the preservation of Hermiticity in the dynamical problem~\cite{sa2022a}, although the form of the Liouvillian is not of Lindblad type unless the real part of the intrasite coupling of the SYK Hamiltonian vanishes.

We expect that this symmetry classification will stimulate the search of systems with these symmetries in the context of dissipative quantum chaos, quantum gravity, and quantum information.

\acknowledgments{
AMGG and YC were supported by NSFC Grant No.\ 11874259, the National Key R\&D Program of China (Project ID: 2019YFA0308603), and a Shanghai talent program. LS acknowledges was supported by a Research Fellowship from the Royal Commission for the Exhibition of 1851 and by Fundação para a Ciência e a Tecnologia (FCT-Portugal) through Grant No.\ SFRH/BD/147477/2019. JJMV acknowledges support from U.S.\ DOE Grant No.\ DE-FAG88FR40388.
}

\bibliography{librarynh_amg2.bib}

\end{document}